\newcommand{\ourtitle}{Envy-Free Division of Land}
\newcommand{\ourauthors}{Erel Segal-Halevi, Shmuel Nitzan, Avinatan Hassidim, Yonatan Aumann}

\def\ARXIV{}

\ifdefined\ARXIV

\documentclass{article}

\usepackage{amsthm}

\theoremstyle{plain}
\newtheorem{theorem}{Theorem}[section]
\newtheorem{lemma}[theorem]{Lemma}

\newtheorem{example}[theorem]{Example}
\newtheorem{definition}[theorem]{Definition}

\newtheorem{remark}[theorem]{Remark}
\newtheorem*{restatetheorem}{Theorem}

\newcommand{\ABSTRACT}[1]{\begin{abstract}#1\end{abstract}}
\newcommand{\KEYWORDS}[1]{~\\Keywords: #1\\}
\newcommand{\HISTORY}[1]{~\\History: #1\\}
\newcommand{\Halmos}{\qed}
\newcommand{\NatBibNumeric}{}

\else

\documentclass[moor]{informs1}              

\fi

\usepackage{amsmath,amssymb,amstext}

\usepackage{pstricks,pst-node,pst-tree}
\usepackage{graphics,graphicx}
\usepackage[overload]{textcase}

\usepackage{amsmath}

\usepackage[letterpaper]{geometry}
\usepackage{calc}
\usepackage{subscript}
\usepackage{framed}

\makeatletter

\usepackage{algorithm,algorithmicx}

\def\covernum(#1,#2){\operatorname{\textsc{CoverNum}}(#1,#2)}
\def\kcovernum(#1,#2){\operatorname{\textsc{KCoverNum}}(#1,#2)}
\def\prp(#1,#2){\operatorname{\textsc{Part}}(#1,#2)}
\def\prop(#1,#2,#3){\operatorname{\textsc{Prop}}(#1,#3,#2)}
\def\prope(#1,#2,#3){\operatorname{\textsc{PropEF}}(#1,#3,#2)}
\def\proprel(#1,#2,#3){\operatorname{\textsc{RelProp}}(#1,#3,#2)}
\def\properel(#1,#2,#3){\operatorname{\textsc{RelPropEF}}(#1,#3,#2)}
\newcommand{\East}{right }
\newcommand{\West}{left }

\newcommand{\pool}[1]{\pscircle[fillstyle=solid,fillcolor=blue,linecolor=blue]#1{1}}
\newcommand{\redpool}[1]{\rput#1{\psframe[fillstyle=solid,fillcolor=red,linecolor=red](0,0)(2,2)}}

\newcommand{\cstart}{C_{\text{start}}}
\newcommand{\cend}{C_{\text{end}}}
\newcommand{\rend}{r_{\text{end}}}

\makeatother

\newcommand{\spiece}{$S$-piece}
\newcommand{\spieces}{$S$-pieces}
\newcommand{\leb}[1]{\textsc{Volume}[#1]}

\newcommand{\range}[2]{\in\{#1,\dots,#2\}}

\usepackage{appendix}
\usepackage{chngcntr}
\usepackage{etoolbox}  
\AtBeginEnvironment{appendices}{%
	\section*{Appendix}
	\addcontentsline{toc}{section}{Appendices}
	\counterwithin{figure}{section}
	\counterwithin{table}{section}
}

\usepackage{natbib}
\NatBibNumeric

\ifdefined\ARXIV

\else

\usepackage[colorlinks=true,breaklinks=true,bookmarks=true,urlcolor=blue,     citecolor=blue,linkcolor=blue,bookmarksopen=false,draft=false]{hyperref}

 \bibpunct[, ]{[}{]}{,}{n}{}{,}%
\TheoremsNumberedBySection
\EquationsNumberedBySection
{\theoremstyle{THkey}\newtheorem{restatetheorem}{XXXXX}}

\fi

\usepackage{xassoccnt}
\DeclareCoupledCountersGroup{informs}
\DeclareCoupledCounters[name=informs]{lemma,corollary,definition,remark}

\ifdefined\ARXIV
\title{\ourtitle{}}
\author{\ourauthors{}}
\fi

\begin{document}

\ifdefined\ARXIV

\maketitle

\else

\TITLE{\ourtitle{}}
\RUNAUTHOR{Segal-Halevi et al.} 
\RUNTITLE{Envy-Free Division of Land}
\ARTICLEAUTHORS{%
\AUTHOR{\ourauthors{}}
\AFF{Ariel University, Bar-Ilan University, Bar-Ilan University, Bar-Ilan University}
\EMAIL{\{erelsgl,shmuelnitzan,avinatanh,yaumann\}@gmail.com}
}

\fi

\ABSTRACT{%
Classic cake-cutting algorithms
enable people with different preferences to divide among them a heterogeneous resource (``cake''), such that the resulting division is fair according to each agent's individual preferences.
However, these algorithms either ignore the geometry of the resource altogether, or assume it is one-dimensional. 
In practice, it is often required to divide multi-dimensional resources, such as land-estates or advertisement spaces in print or electronic media. In such cases, the geometric shape of the allotted piece is of crucial importance. For example, when building houses or designing advertisements, in order to be useful, the allotments should be squares or rectangles with bounded aspect-ratio. We thus introduce the problem of \emph{fair land division} --- 
fair division of a multi-dimensional resource wherein the allocated piece must have a pre-specified geometric shape. We present constructive division algorithms that satisfy the two most prominent fairness criteria, namely \emph{envy-freeness} and \emph{proportionality}. In settings where proportionality cannot be achieved due to the geometric constraints, our algorithms provide a \emph{partially-proportional} division, guaranteeing that the fraction allocated to each agent be at least a certain positive constant. We prove that in many natural settings the envy-freeness requirement is compatible with the best attainable partial-proportionality.
}

\KEYWORDS{Fairness; Land Division; Cake Cutting ; Envy Free ; Two Dimensional ; Geometric constraints }
\HISTORY{
		A preliminary version of this paper appeared in the proceedings of AAAI 2015 \citep{SegalHalevi2015EnvyFree}, where its title was ``Envy-free cake-cutting in two dimensions''. 
		The main additions in the present version are: (a) handling multi-dimensional resources of arbitrary shape rather than just rectangles, (b) handling an arbitrary number $n$ of agents rather than just 2 or 3, (c) rewriting most proofs in a simpler way.
}

\ifdefined\ARXIV
\else
\maketitle
\fi

\section{Introduction}
~\\
Fair division is an active field of research with various applications. A frequently-mentioned potential application is division of land (e.g. \citet{Berliant1988Foundation,Berliant1992Fair,Legut1994Economies,Chambers2005Allocation,DallAglio2009Disputed,Husseinov2011Theory,Nicolo2012Equal}). The basic setting considers a heterogeneous good, such as a land-estate, to be divided among several agents.  The agents may have different preferences over the possible pieces of the good, e.g. one agent prefers the forests while the other prefers the sea shore. The goal is to divide the good among the agents in a way deemed ``fair''. The common fairness criterion in economics is \emph{Envy-freeness}, which means that no agent prefers getting a piece allotted to another agent.

Envy-freeness on its own is trivially satisfied by the empty allocation. The task becomes more interesting when envy-freeness is combined with an \emph{efficiency} criterion. The most common such criterion is Pareto efficiency. Indeed, \citet{Weller1985Fair} has proved that, when the agents' preferences are represented by non-atomic measures over the good, there always exists a Pareto-efficient and envy-free allocation. However, Weller's allocation gives no guarantees about the \emph{geometric shape} of the allotted pieces. A ``piece'' in his allocation might even contain an infinite number of disconnected bits. So Weller's positive result is valid only when agents' preferences ignore the geometry of their allotted pieces. 
While such preferences make sense when dividing a pudding or an ice-cream, they are less sensible when dividing land. 

Many authors have noted the importance of imposing some geometric constraints on the pieces. The most common constraint is \emph{connectivity} --- the good is assumed to be the one-dimensional interval $[0,1]$ and the allotted pieces are sub-intervals (e.g. \citet{Stromquist1980How,Su1999Rental,Nicolo2008Strategic,Azrieli2014Rental}). This is usually justified by the reasoning that higher dimensional settings can always be projected onto one dimension, and hence fairness in one dimension implies fairness in higher dimensions. However, projecting back from the one dimension, the resulting two-dimensional plots are thin rectangular slivers, of little use in most practical applications; it is hard to build a house on a $10\times1,000$ meter plot even though its area is a full hectare, and a thin 0.1-inch wide advertisement space would ill-serve most advertises regardless of its height.

\begin{figure}
	\def\landcake{
		\psframe[fillstyle=none](0,0)(20,20)
		\pool{(3,3)}
		\pool{(17,6)}
		\pool{(17,14)}
	}
	\begin{center}
		\psset{unit=1.5mm,dotsep=1pt,linecolor=brown}
		\begin{pspicture}(25,25)
		\rput(10,22){(a)}
		\psframe[linestyle=dotted,linecolor=green](2,2)(17,17)
		\psframe[linestyle=dotted,linecolor=red](11,6)(19,14)
		\landcake
		\end{pspicture}
		\begin{pspicture}(25,25)
		\rput(10,22){(b)}
		\landcake
		\psframe[linestyle=dotted,linecolor=green](1,1)(5,5)
		\psframe[linestyle=dotted,linecolor=red](15,4)(19,8)
		\end{pspicture}
		\begin{pspicture}(25,25)
		\rput(10,22){(c)}
		\landcake
		\psframe[linestyle=dotted,linecolor=green](1,1)(5,5)
		\psframe[linestyle=dotted,linecolor=red](7,4)(19,16)
		\end{pspicture}
	\end{center}
	\caption[]{
		\label{fig:peef}
	(Impossibility of a Pareto-efficient envy-free land division).
		A square land-estate has to be divided between two people. The land-estate is mostly barren, except for three water-pools (discs). The agents have the same preferences: each agent wants a square land-plot with as much water as possible. The squares must not overlap. Hence:\\
		(a) It is impossible to give both agents more than 1/3 of the water. Hence:\\
		(b) An envy-free division must give each agent at most 1/3 of the water.\\
		(c) But such a division cannot be Pareto-efficient since it is dominated by a division which gives one agent 1/3 and the other 2/3 of the water.
	}
\end{figure}
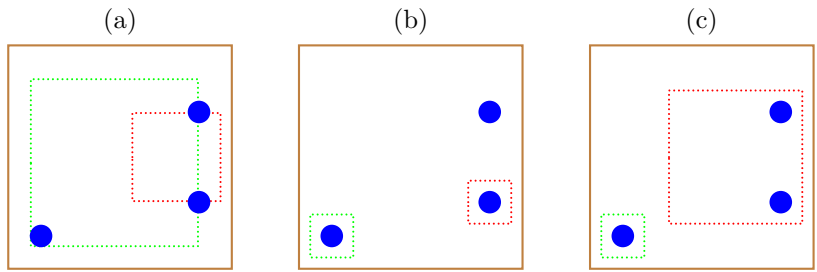

This paper studies the fair division of a multi-dimensional resource with geometric constraints on the pieces. 
We call this problem \emph{fair land division}, 
to differentiate it from the one-dimensional problem often called \emph{fair cake-cutting}
(\citet{Steinhaus1948Problem,Dubins1961How,Brams1996Fair,Robertson1998CakeCutting,Procaccia2015Cake}).
A remarkable feature of the land-division setting is that Pareto-efficient-envy-free allocations might not exist even when there are two agents; see Figure \ref{fig:peef} for a simple example.

Thus, to get an envy-free allocation among agents with geometric preferences, we must replace Pareto-efficiency with a different efficiency criterion. A natural candidate is \emph{proportionality} --- every agent should receive at least $1/n$ of the total resource value. This was the first fairness criterion studied in the context of cake-cutting \citep{Steinhaus1948Problem} and it is still very common in the cake-cutting literature \citep{Robertson1998CakeCutting,Procaccia2015Cake}. With geometric preferences, a proportional division might not exist; see Figure \ref{fig:peef} again for an example. Hence we relax the proportionality requirement and consider \emph{partial proportionality}. Partial proportionality means that each agent receives a piece worth at least a fraction $p$ of the total value, where $p$ is a positive constant in $[0,1]$. Obviously we would like $p$ to be as large as possible. 
In a previous paper \citep{SegalHalevi2017Fair}, we showed that partial-proportionality can be attained in various geometric settings.
For example, in the setting of Figure \ref{fig:peef} (square land and two agents who want square pieces), each agent can be guaranteed at least a fraction 1/4 of the total value, and this is the largest fraction that can be guaranteed. However, these results did not consider envy. 
The present paper (which can be read independently of the previous one)
studies3 the following question:

\begin{quote}
\emph{When each agent wants a plot of land with a given geometric shape, what is the largest fraction of the total value that can be guaranteed to every agent in an envy-free allocation?}
\end{quote}
The following example shows that existing cake-cutting algorithms are insufficient for answering this question.
\begin{example} \label{exm:cutandchoose}
	You and a partner are going to divide a square land-estate. It is 100-by-100 square meters and its western side is adjacent to the sea. Your desire is to build a house near the sea-shore. You decide to use the classic algorithm for envy-free cake-cutting: ``You cut, I choose''. You let your partner divide the land to two plots, knowing that you have the right to choose the plot that is more valuable according to your personal preferences. Your partner makes a cut parallel to the shoreline at a distance of only 1 meter from the sea. %
	\footnote{The reason why your partner decided to cut this way is irrelevant since a fair division algorithm is expected to guarantee that the division is fair for every agent playing by the rules, regardless of what the other
		agents do.%
	} Which of the two plots would you choose? The western plot contains a lot of sea shore, but it is so narrow that it has no room for building anything. On the other hand, the eastern plot is large but does not contain any shore land. Whichever plot you choose, the division is not proportional for you, because your utility is far less than half the utility of the original land estate. 
	
	Of course the land \emph{could} be cut in a more sensible way (e.g. by a line perpendicular to the sea), but existing cake-cutting algorithms say nothing about how exactly to cut in each situation in order to guarantee that the division is fair with respect to the geometric preferences. While the cut-and-choose algorithm still guarantees envy-freeness, it does not guarantee partial-proportionality since it does not guarantee any positive utility to agents who want square pieces.	\Halmos
\end{example}

This paper presents fair division algorithms that guarantee both envy-freeness and partial-proportionality. Our algorithms focus on agents who want \emph{fat pieces} --- pieces with a bounded length/width ratio, such as squares. The rationale is that a fat shape is more convenient to work with, build on, cultivate, etc.

\subsection{Fatness.} \label{sub:Fatness}
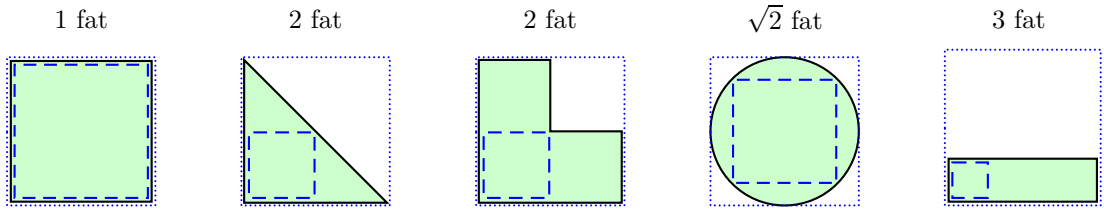
\begin{figure}
	\begin{center}
		\psset{unit=1mm,dotsep=1pt}
		\begin{pspicture}(30,30)
		\psframe[fillstyle=solid,fillcolor=green!20,linestyle=solid,linecolor=black](.5,.5)(19.5,19.5)
		\psframe[linestyle=dotted,linecolor=blue](0,0)(20,20)
		\psframe[linestyle=dashed,linecolor=blue](1,1)(19,19)
		\rput(10,25){1 fat}
		\end{pspicture}
		\begin{pspicture}(30,30)
		\pspolygon[fillstyle=solid,fillcolor=green!20](0.5,0.5)(19.5,0.5)(0.5,19.5)
		\psframe[linestyle=dotted,linecolor=blue](0,0)(20,20)
		\psframe[linestyle=dashed,linecolor=blue](1,1)(10,10)
		\rput(10,25){2 fat}
		\end{pspicture}
		\begin{pspicture}(30,30)
		\pspolygon[fillstyle=solid,fillcolor=green!20](0.5,0.5)(19.5,0.5)(19.5,10)(10,10)(10,19.5)(0.5,19.5)
		\psframe[linestyle=dotted,linecolor=blue](0,0)(20,20)
		\psframe[linestyle=dashed,linecolor=blue](1,1)(10,10)
		\rput(10,25){2 fat}
		\end{pspicture}
		\begin{pspicture}(30,30)
		\pscircle[fillstyle=solid,fillcolor=green!20](10,10){10}
		\psframe[linestyle=dotted,linecolor=blue](0,0)(20,20)
		\psframe[linestyle=dashed,linecolor=blue](3,3)(17,17)
		\rput(10,25){$\sqrt{2}$ fat}
		\end{pspicture}
		\begin{pspicture}(25,30)
		\psframe[fillstyle=solid,fillcolor=green!20](0.5,0.5)(20.5,6.5)
		\psframe[linestyle=dotted,linecolor=blue](0,0)(21,21)
		\psframe[linestyle=dashed,linecolor=blue](1,1)(6,6)
		\rput(10,25){3 fat}
		\end{pspicture}
	\end{center}
	\caption[Fatness of several 2-dimensional geometric shapes]{
		\label{fig:fatness}Fatness of several 2-dimensional geometric shapes. The dashed square is the largest contained cube; the dotted square is the smallest containing parallel cube. The shape is $R$-fat if the ratio of the side-lengths of these squares is at most $R$.
	}
\end{figure}

We use the following formal definition of fatness, which is adapted from the computational geometry literature, e.g. \citet{Agarwal1995Computing,Katz19973D}:
\begin{definition}
	\label{def:rfat}Let $R\geq1$ be a real number. A $d$-dimensional piece is called \emph{$R$-fat}, if it contains a $d$-dimensional cube $B^{-}$ and is contained in a parallel $d$-dimensional cube $B^{+}$, such that the ratio between the side-lengths of the cubes is at most $R$: $\text{{len}(}B^{+})/\text{{len}}(B^{-})\leq R$.
\end{definition}

A 2-dimensional cube is a square. So, for example, the only 2-dimensional 1-fat shape is a square (it is also 2-fat, 3-fat etc.). An $L\times1$ rectangle is $L$-fat; a right-angled isosceles triangle is 2-fat (but not 1-fat) and a circle is $\sqrt{2}$-fat (see Figure \ref{fig:fatness}). 

Note that the fatness requirement is inherently multi-dimensional and cannot be reduced to a 1-dimensional requirement. Hence it cannot be satisfied by methods developed for a 1-dimensional cake.%
\footnote{In contrast, the simpler requirement that the pieces be rectangles with an arbitrary length/width ratio can easily be reduced to a 1-dimensional requirement that the pieces are connected intervals. Such reduction is also possible for the requirements that the pieces be simplices \citep{Ichiishi1999Equitable} or polytopes \citep{DallAglio2009Disputed} with  an unbounded aspect ratio.%
}

\subsection{Results.}
We prove that envy-freeness and partial-proportionality are \emph{compatible} in progressively more general geometric settings. Our proofs are constructive: in every geometric setting (geometric shape of the land and preferred shape of the pieces), we present an algorithm that divides the land with the following guarantees:
\begin{itemize}
	\item \emph{Envy-freeness}: every agent weakly prefers his/her allotted piece over the piece given to any other agent.
	\item \emph{Partial-proportionality}: every agent receives a piece worth for him/her at least a fraction $p$ of the total land value, where $p$ is a positive constant that depends on the geometric requirements.
\end{itemize}
In the following theorems, the partial-proportionality guarantee $p$ is given in parentheses. 

The first theorem handles division between two agents.
\begin{theorem}
\label{thm:2agents}
There is an algorithm for finding an envy-free and partially-proportional allocation of land between two agents in the following cases:
	
(a) The land is square and the usable pieces are squares ($p\geq 1/4$).
	
(b) The land is an R-fat rectangle and the usable pieces are $R$-fat rectangles with $R\geq2$ ($p\geq 1/3$).
	
(c) The land is an arbitrary R-fat object and the usable  pieces are $2R$-fat, where $R\geq1$ ($p\geq 1/2$).
\end{theorem}
\textbf{Value-shape trade-off}: Theorem \ref{thm:2agents} illustrates a multiple-way trade-off between value and shape. Consider two agents who want to divide a square land-estate with no envy. They have the following options: 
\begin{itemize}
	\item By projecting a 1-dimensional division obtained by any classic cake-cutting algorithm, they can achieve a proportional allocation (a value of at least $1/2$) with rectangular pieces but with no bound on the aspect ratio --- the pieces might be arbitrarily thin.
	\item By (a), they can achieve an allocation with square pieces but only partial proportionality --- the proportionality might be as low as $1/4$. 
	\item By (b), they can achieve a proportionality of $1/3$ with 2-fat rectangles, which is a compromise between the previous two options.
	\item By (c), they can achieve an allocation that is \emph{both} proportional \emph{and} with 2-fat pieces, but the pieces might be non rectangular.
\end{itemize}

The proportionality constants in Theorem \ref{thm:2agents} are tight in the following sense: it is not possible to guarantee an allocation with a larger proportionality, even if envy is allowed. 
This means that envy-freeness
is compatible with the largest possible proportionality --- we don't have to compromise on proportionality to prevent envy. 

Moreover, whenever the pieces should be $R$-fat with $R<2$, it might be impossible to guarantee more than $1/4$-proportionality,
and whenever the pieces should be $R$-fat with any finite $R$, it might be impossible to guarantee more than $1/3$-proportionality.
This implies that 2-fat rectangles are a good practical compromise between fatness and fairness: 
if we require fatter pieces ($R<2$) then the proportionality guarantee drops from $1/3$ to $1/4$, while if we allow thinner pieces ($R>2$) the proportionality remains $1/3$ for all $R<\infty$.

The second theorem handles division among any number of agents.
\begin{theorem}
\label{thm:nagents}
There exists an envy-free and partially-proportional allocation 
of land among $n$ agents
in the following cases:
	
	(a) The land is square and the usable pieces are squares ($p> 1/(4 n^{2})$).
	
	(b) The land is an $R$-fat rectangle and the pieces are $R$-fat rectangles, where $R\geq1$ ($p > 1/(4 n^{2})$).
	
	(c) The land is a $d$-dimensional $R$-fat object and the pieces are $\lceil n^{1/d}\rceil R$-fat,%
	\footnote{$\lceil x\rceil$ denotes the \emph{ceiling} of $x$ --- the smallest integer which is larger than $x$. %
	} where $d\geq2$ and $R\geq1$ ($p\geq 1/n$).
\end{theorem}
\textbf{Value-shape trade-off}: Part (a) and part (c) are duals in the following sense:
\begin{itemize}
	\item Part (a) guarantees an envy-free division with perfect pieces (squares) but compromises on the proportionality level; 
	\item Part (c) guarantees an envy-free division with perfect proportionality ($1/n$) but compromises on the fatness of the pieces. 
\end{itemize}

The ``magnitude'' of the first compromise is $4n$, since the proportionality drops from $1/n$ to $1/(4n^2)$. We do not know if this magnitude is tight: we know that it is possible to attain a division with square pieces and a proportionality of $1/O(n)$ which is not necessarily envy-free \cite{SegalHalevi2017Fair}, but we do not know if a proportionality of $1/O(n)$ is compatible with envy-freeness.

The ``magnitude'' of the second compromise is $\lceil n^{1/d}\rceil$. This magnitude is asymptotically tight. We prove that, in order to guarantee a proportional division of an $R$-fat land, with or without envy, we must allow the pieces to be $\Omega(n^{1/d})R$-fat.

\subsection{Challenges and solutions.}
The main challenge in land division is that utility functions depending on geometric shape are not \emph{additive}. For example, consider an agent who wants to build a square house the utility of which is determined by its area. The utility of this agent from a $20\times20$ plot is 400, but if this plot is divided to two $20\times10$ plots, the utility from each plot is 100 and the sum of utilities is only 200. Most existing algorithms for proportional cake-cutting assume that the valuations are additive, so they are not applicable in our case. While there are some previous works on cake-cutting with non-additive utilities, they too cannot handle geometric constraints:
\begin{itemize}
\item \citet{Berliant1992Fair,Maccheroni2003How} focus on sub-additive, or concave, utility functions, in which the sum of the utilities of the parts is \emph{more} than the utility of the whole. These utility functions are inapplicable in our scenario because, as illustrated
in the previous paragraph, utility functions that consider geometry are not necessarily sub-additive --- the sum of the utilities of the parts might be less than the utility of the whole.
	
\item \citet{DallAglio2009Disputed} do not explicitly require sub-additivity, but they require \emph{preference for concentration}: if an agent is indifferent between two pieces $Z_1$ and $Z_2$, then he prefers 100\% of $Z_1$ to 50\% of $Z_1$ plus 50\% of $Z_2$. This axiom may be incompatible with geometric constraints: the agent in the above example is indifferent between the two $20\times 10$ rectangles, but he prefers 50\% of their union (the $20\times 20$ square) to 100\% of a single rectangle.\footnote{We are grateful to Marco Dall'Aglio for his help in clarifying this issue.}
	
\item \citet{Sagara2005Equity,Husseinov2013Existence}
	consider general non-additive utility functions but provide only non-constructive existence proofs. 
	
\item \citet{Su1999Rental,Caragiannis2011Towards,Mirchandani2013Superadditivity} provide practical division algorithms for non-additive utilities, but they crucially assume that the cake is a 1-dimensional interval and cannot handle two-dimensional constraints.

\item \citet{Berliant2004Foundation} study the division of a multi-dimensional good with geometric constraints. Their results are mostly negative: when general value measures are combined with geometric preferences, a competitive-equilibrium might not exist.
This is in contrast to the situation without geometric constraints, in which a competitive equilibrium always exists
\citep{Weller1985Fair,segalhalevi2018monotonicity}.
\end{itemize}

When envy-free division protocols are applied to agents with non-additive utility functions, the division is still envy-free,
but the utility per agent might be arbitrarily small. This is true for cut-and-choose (as shown in Example \ref{exm:cutandchoose} above) and it is also true for all other algorithms for envy-free division that we are aware of (\citet{Stromquist1980How,Brams1995EnvyFree,Reijnierse1998Finding,Su1999Rental,Barbanel2004Cake,Manabe2010MetaEnvyFree,Cohler2011Optimal,Deng2012Algorithmic,Kurokawa2013How,Chen2013Truth,Aziz2016Discrete,SegalHalevi2016Waste}).

Our way to cope with this challenge is to explicitly handle the geometric constraints in the algorithms. The main tool we use is the \emph{geometric knife function}. 

Moving-knife algorithms have been used for envy-free cake-cutting since its earliest years  \citep{Dubins1961How,Stromquist1980How,Brams1997MovingKnife,Saberi2009Cutting}. For example, consider the following simple algorithm for envy-free division among two agents. A referee moves a knife slowly over the cake, from left to right. Whenever an agent feels that the piece to the left of the knife is worth for him exactly half the total cake value, he shouts "stop!". Then, the cake is cut at the current knife location, the shouter receives the piece to its left and the non-shouter receives the piece to its right. 

In this paper we formalize the notion of a knife and add geometric constraints guaranteeing that the final pieces have a sufficiently high utility for agents who care about geometric shape.

\subsection{Other related work.}
The prominent model in the cake-cutting literature assumes that the cake is an interval. Several authors diverge from the interval model by assuming a circular cake (e.g. \citet{Thomson2007Children,Brams2008Proportional,Barbanel2009Cutting}), but they still work in one dimension so the pieces are one-dimensional arcs corresponding to thin wedge-like slivers.

The importance of the multi-dimensional geometric shape of the plots was noted by several authors. 

\citet{Hill1983Determining,Beck1987Constructing,Webb1990Combinatorial,Berliant1992Fair}
study the problem of dividing a disputed territory between several bordering countries, with the constraint that each country should get a piece that is adjacent to its border.

\citet{Ichiishi1999Equitable,DallAglio2009Disputed,segalhalevi2018redividing} require the plots to be convex shapes such as multi-dimensional simplices or rectangles. However, the allocated pieces can be arbitrarily thin --- 
their methods cannot handle  requirements that are inherently two-dimensional, such as squareness.

\citet{Iyer2009Procedure} describe an algorithm for giving each agent a rectangular plot with an aspect ratio determined by the agent.
However, their algorithm is not guaranteed to succeed. If even a single rectangle of Alice intersects two rectangles of George (for example), then the algorithm fails and no agent gets any piece.

In a related paper \cite{SegalHalevi2017Fair}, we considered the problem of partially-proportional division when the pieces must be squares or fat rectangles. We presented an algorithm for dividing a square among $n$ agents such that each agent receives a square piece with a value of at least $1/(4n-4)$. When all agents have the same value function, the proportionality improves to $1/(2n)$. We also proved that the upper bound in both cases is $1/(2n)$. The algorithms in the present paper use very different techniques, in order to guarantee \emph{envy-freeness} in addition to partial-proportionality. Their down-side is that their proportionality guarantee (in the case of square pieces) is only $1/(4n^2)$. Additionally, in the present paper we handle general fat objects rather than just squares and rectangles. Some impossibility results from that paper are replicated here (in Appendix \ref{sec:upper-bounds}) so that each paper can be fully understood without reading the other one.

\subsection{Paper layout.}
The formal definitions and model are provided in Section \ref{sec:The-Model}. Section \ref{sec:geometric} introduces the core geometric concepts and techniques. These geometric techniques are then applied in the construction of the envy-free division algorithms for two agents (Section \ref{sec:2-agents}) and $n$ agents (Section \ref{sec:n-agents}). Some directions for future work are presented in Section \ref{sec:future}.

Appendix \ref{sec:S-good} contains some of the more technical proofs related to continuity of knife-functions. 
Appendix \ref{sec:upper-bounds}
contains negative results --- upper bounds on attainable partial-proportionality in various cases.
Appendix \ref{sec:convex} presents an alternative model in which the pieces should be convex in addition to being fat. 
Appendix \ref{sec:relprop} presents an alternative model in which the proportionality of an allocation is measured relative to the maximum attainable utility rather than the total land value.

\filbreak
\section{Model} \label{sec:The-Model}
\subsection{Land and Pieces.}
The resource to be divided is called a \emph{land-estate} or just \emph{land} for short. It is denoted by $C$. It is assumed to be a Borel subset of a Euclidean space $\mathbb{R}^{d}$. In most of the paper $d=2$. \emph{Pieces} are Borel subsets of $\mathbb{R}^{d}$. \emph{Pieces of $C$} are Borel subsets of $C$.

There is a family $S$ of pieces that are considered usable. An \emph{\spiece} is an element of $S$. 

\subsection{Agents and Utilities.}
There are $n\geq 1$ \emph{agents}. Each agent $i\in\left\{ 1,...,n\right\}$ has a value-density function $v_i$, which is an integrable, non-negative and bounded function on $C$. It represents the quality of each land-spot in the eyes of the agent. It may depend upon factors such as the fertility of soil, the probability of finding oil, the existence of trees, etc. 

The \emph{value} of a piece $Z$ to agent $i$ is denoted by $V_i(Z)$ and it is the integral of the value-density: 
\begin{align*}
V_i(Z)=\int_{z\in Z}v_i(z)dz
\end{align*}
We assume that for all agents $i$, $V_i(\mathbb{R}^d)<\infty$. Hence, each $V_{i}$ is a finite measure which is absolutely-continuous with respect to the Lebesgue measure. In particular, the boundary of a piece has a value of zero to all agents. 

In the standard cake-cutting model  \citep{Weller1985Fair,Chambers2005Allocation,LiCalzi2009Efficient,Chen2013Truth}, the \emph{utility function} of an agent is identical to his/her value measure. The present paper diverges from this model by considering agents whose utility functions depend both on value and on geometric shape. We assume that an agent can derive utility only from an $S$-piece; when his allotted land-plot is not an $S$-piece, he selects the most valuable $S$-piece contained therein and utilizes it. For each agent $i$, we define the \emph{$S$-value function}, which assigns to each piece $Z\subseteq \mathbb{R}^d$ the value of the most valuable usable piece contained therein:
\begin{align*}
V_i^S(Z)=\sup_{Y\in S\,\text{ , }\,Y\subseteq Z}V_i(Y)
\end{align*}
We assume that the utility of agent $i$ is equal to his $S$-value function $V_i^S$. In general, $V_i^S$ is not a measure since it is not additive (it is not even sub-additive). Hence, cake-cutting algorithms that require additivity are not applicable. Note that the two most common cake-cutting models are special cases of our model:
\begin{itemize}
	\item The model in which each agent may receive an arbitrary Borel subset \citep{Weller1985Fair} is a special case in which $S$ is the set of all pieces.
	\item The model in which each agent must receive a connected piece \citep{Stromquist1980How} is a special case in which $C$ is an interval and $S$ is the set of intervals.
\end{itemize}

\subsection{Allocations and Fairness.}
An \emph{allocation} of $C$ 
is a vector of $n$ pieces of $C$, $X = (X_1,...,X_n)$, such that the $X_i$ are pairwise-disjoint%
\footnote{
Throughout the paper, when we talk about ``disjoint pieces'', we allow the pieces to intersect in their boundary. We can ignore the question of which agent receives the boundary, since the value of the boundary is 0 for all agents.
}
 and their union is a subset of $C$. We express the latter two facts succinctly using the ``disjoint union'' operator, $\sqcup$:
\begin{align*}
X_1 \sqcup \cdots \sqcup X_n \subseteq C.
\end{align*}
We assume free disposal --- some of $C$ may remain unallocated. This assumption is natural in division of land: it is common to leave some land unallocated to make it available for public use.

An allocation is called \emph{envy-free} if the utility of an agent from his allotment is at least as large as his utility from every piece allocated to another agent:
\begin{align*}
\forall i,j\in\{1,...,n\}:\,\, V^S_i(X_i)\geq V^S_i(X_j)
\end{align*}
We call an allocation \emph{$p$-proportional}, for some $p\in[0,1]$, if the utility of each agent from his allotment is at least a fraction $p$ of his total land value:
\begin{align*}
\forall i\in\{1,...,n\}:\,\, V^S_i(X_i)\geq p\cdot V_i(C)
\end{align*}
We call an allocation  \emph{$p$-relative-proportional} if the utility of each agent from his allotment is at least a fraction $p$ of his largest attainable \emph{utility}:
\begin{align*}
\forall i\in\{1,...,n\}:\,\, V^S_i(X_i)\geq p\cdot V_i^S(C)
\end{align*}
Since $V_i(C)\leq V_i^S(C)$,
every $p$-proportional allocation is also $p$-relative-proportional. 
The present paper focuses on $p$-proportionality, so all positive results are valid for $p$-relative-proportionality as well.
Moreover, whenever $C$ itself is an element of $S$, $V_i(C)=V_i^S(C)$, so 
$p$-proportionality and $p$-relative-proportionality are equivalent. This is the case in all settings mentioned in our main theorems (\ref{thm:2agents} and \ref{thm:nagents}); therefore they are valid and tight by both criteria.

However, when $C\not\in S$, it may be possible to attain relative-proportionality that is higher than the maximum attainable absolute-proportionality. Exploiting this possibility requires new techniques. Appendix \ref{sec:relprop} presents some results in this direction, leaving a fuller treatment to future work.

\subsection{Fairness Guarantees}
Given the geometric shape of $C$ and the family $S$, we would like to know what proportionality can be guaranteed for any combination of agents, with and without the additional requirement of envy-freeness. Formally:

\begin{definition}\label{def:abs-prop}
Let $C$ be a land-estate, $S$ a family of usable shapes, and $n\geq 1$ an integer.

(a) The \textbf{proportionality guarantee} for $C$, $S$ and $n$, denoted $\prop(C,n,S)$, is the largest fraction $p\in[0,1]$ such that, for every $n$ value measures $(V_1,...,V_n)$, a $p$-proportional allocation exists.

(b) The \textbf{envy-free proportionality guarantee} of $C$, $S$ and $n$, denoted $\prope(C,n,S)$, 
is the largest fraction $p\in[0,1]$ such that, for every $n$ value measures $(V_1,...,V_n)$, an \emph{envy-free}
and $p$-proportional allocation exists.
\end{definition}


For example, classic cake-cutting results can be presented as:
\begin{align*}
\forall C:~~ 
 \prop(C,\, n,\, All Pieces)~~=~~ \prope(C,\, n,\, All Pieces)~~=~~{1/n}
\\
 \prop(Interval,\, n,\, Intervals)~~=~~ \prope(Interval,\, n,\, Intervals)~~=~~{1/n}
\end{align*}
and our results for two agents can be stated as:
\begin{align*}
\prop (Square,2,Squares) ~~=~~ \prope(Square,2,Squares) ~~&=~~ 1/4
\\
\\
\forall R\geq 2: ~~ \prop (R~fat~rectangle,2,R~fat~rectangles) ~~&=~~ 
\\
=~~\prope(R~fat~rectangle,2,R~fat~rectangles) ~~&=~~ 1/3
\\
\\
\forall R\geq 1: ~~ \prop (R~fat~object,2,R~fat~objects) ~~&=~~ 
\\
=~~\prope(R~fat~object,2,2R~fat~objects) ~~&=~~ 1/2
\end{align*}

\subsection{Strategy considerations}
In the present paper we ignore strategic considerations and assume that all agents act according to their true value functions. 
In fact, even without geometric constraints, it is impossible to build a general division protocol that is both fair and strategy-proof \citep{Branzei2015Dictatorship}.
Strategy-proof algorithms exist only in very special cases, for example, when all agents have piecewise-uniform valuations \citep{Chen2013Truth,bei2017cake,bei2018truthful}.

However, the guarantees of our algorithms are valid for any \emph{single} agent who acts according to his own value function. E.g, the algorithm of Theorem \ref{thm:nagents}(c) guarantees that every agent acting according to his true value function receives a piece with a utility of at least $1/n$ and at least as good as the other pieces, regardless of what the other agents do.

\newpage
\section{Geometric Preliminaries} \label{sec:geometric}

\subsection{Covers and choosers.}
The key feature of our geometric setting is that the utility of each agent $i$ is given by $V_i^S$ rather than $V_i$.
Therefore, we would like to bound the ratio between $V_i^S$ and $V_i$.
The key concept we will use is a \emph{cover}.

\begin{figure}
	\psset{unit=0.6mm}
	\begin{pspicture}(60,65)
	\psframe[fillstyle=solid,fillcolor=white!20](0,0)(47,15)
	\psframe[linecolor=black,linestyle=dashed](1,1)(14,14)
	\psframe[linecolor=blue,linestyle=dashed](16,1)(29,14)
	\psframe[linecolor=green,linestyle=dashed](31,1)(44,14)
	\psframe[linecolor=red,linestyle=dashed](33,1)(46,14)
	\rput(30,60){\tiny{$3.1\times 1$ rectangle:}}
	\rput(30,55){\tiny{CoverNum(Z,squares)=4}}
	\end{pspicture}
	\begin{pspicture}(60,65)
	\pspolygon[fillstyle=solid,fillcolor=white!20](0,0)(50,0)(50,30)(30,30)(30,50)(0,50)
	\psframe[linecolor=green,linestyle=dashed](1,49)(29,21)
	\psframe[linecolor=blue,linestyle=dashed](49,1)(21,29)
	\psframe[linecolor=red,linestyle=dashed](1,1)(20,20)
	\rput(30,60){\tiny{L-shape:}}
	\rput(30,55){\tiny{CoverNum(Z,squares)=3}}
	\end{pspicture}
	\begin{pspicture}(60,65)
	\pspolygon[fillstyle=solid,fillcolor=white!20](0,0)(50,0)(50,30)(30,30)(30,50)(0,50)
	\psframe[linecolor=green,linestyle=dashed](1,49)(29,1)
	\psframe[linecolor=blue,linestyle=dashed](49,1)(1,29)
	\rput(30,60){\tiny{L-shape:}}
	\rput(30,55){\tiny{CoverNum(Z,rectangles)=2}}
	\end{pspicture}
	\begin{pspicture}(60,65)
	\pscircle[fillstyle=solid](20,20){20}
	\rput(30,60){\tiny{Disc:}}
	\rput(30,55){\tiny{CoverNum(Z,squares)=$\infty$}}
	\rput(30,50){\tiny{CoverNum(Z,rectangles)=$\infty$}}
	\end{pspicture}
	\caption{
		\label{fig:cover-numbers}
		Cover numbers of several geometric shapes.
		In the first three figures, The dashed squares/rectangles denote minimal covers.
		In the fourth figure, there is no finite cover, so the cover number is defined as $\infty$.
	}
\end{figure}
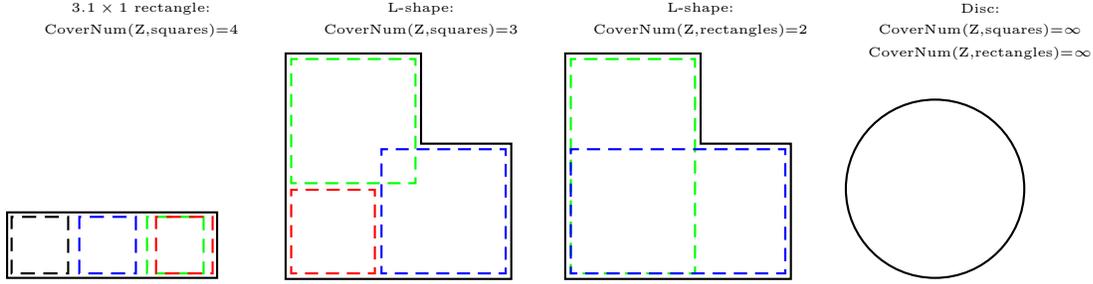

\begin{definition}
(a)
A \emph{cover} of a piece $Z\subseteq \mathbb{R}^d$ is a set of pieces of $Z$ whose union equals $Z$:
\begin{align*}
Z_1\cup\cdots\cup Z_m = Z
\end{align*}

(b)
An \emph{$S$-cover} of $Z$ is a cover in which all pieces are elements of $S$.

(c) The \emph{$S$-cover-number} of $Z$, $\covernum (Z,S)$, is the smallest size of an $S$-cover of $Z$.
\end{definition}
Some examples of cover-numbers are shown in Figure \ref{fig:cover-numbers}.

Consider a piece $Z$ with an $S$-cover of size $m$.
For any value-measure $V$, 
the sum of values of the $m$ \spieces{} is at least $V(Z)$. Therefore, the most valuable \spiece{} in the cover has a value of at least $V(Z)/m$:
\begin{align}
\label{eq:maxvz}
\max_{i\range{1}{m}} V(Z_i) ~\geq~ V(Z)/m.
\end{align}
\ifdefined\CoverLemma
This simple fact has several useful implications.

\begin{lemma}[Cover Lemma]
\label{lem:cover}
For every piece $Z$:
\begin{align*}
V^S(Z) 
~\geq~
 V(Z) / \covernum (Z,S)
\end{align*}
\end{lemma}
\proof{Proof.}
Let $m=\covernum (Z,S)$. Take any $S$-cover of $Z$ of size $m$. By \eqref{eq:maxvz}, for any measure $V$, at least one \spiece{} of that cover has a value of at least $V(Z)/m$. Therefore $V^S(Z)\geq V(Z)/m$.
\Halmos\endproof
\begin{remark}
In many cases, the lower bound on $V^S(Z)/V(Z)$, given by the Cover Lemma, is tight. See Figure \ref{fig:los} and compare to Figure \ref{fig:cover-numbers}.
We know of some cases where it is not tight.
For example, if $Z$ is a square and $S$ is the family of L-shapes, then $\covernum (Z,S)=2$ while for every measure $V$: $V^S(Z) \geq 3 V(Z)/4$.
It is an interesting open question in what cases the Cover Lemma is tight.
\end{remark}

\begin{figure}
	\psset{unit=0.6mm}
	\begin{pspicture}(60,65)
	\psframe[fillstyle=solid,fillcolor=white!20](0,0)(47,15)
	\pool{(1,2)}
	\pool{(16,2)}
	\pool{(31,2)}
	\pool{(46,2)}
	\rput(30,60){\tiny{$3.1\times 1$ rectangle:}}
	\rput(30,55){\tiny{$V(Z)/V^{square}(Z)=4$}}
	\end{pspicture}
	\begin{pspicture}(60,65)
	\pspolygon[fillstyle=solid,fillcolor=white!20](0,0)(50,0)(50,30)(30,30)(30,50)(0,50)
	\pool{(2,2)}	
	\pool{(2,48)}	
	\pool{(48,2)}	
	\rput(30,60){\tiny{L-shape:}}
	\rput(30,55){\tiny{$V(Z)/V^{square}(Z)=3$}}
	\end{pspicture}
	\begin{pspicture}(60,65)
	\pspolygon[fillstyle=solid,fillcolor=white!20](0,0)(50,0)(50,30)(30,30)(30,50)(0,50)
	\pool{(2,48)}	
	\pool{(48,2)}	
	\rput(30,60){\tiny{L-shape:}}
	\rput(30,55){\tiny{$V(Z)/V^{rectangle}(Z)=2$}}
	\end{pspicture}
	\begin{pspicture}(60,65)
	\pscircle[fillstyle=solid](20,20){20}
	\pscircle[fillstyle=solid,linecolor=blue,linewidth=1](20,20){19}
	\rput(30,60){\tiny{Disc:}}
	\rput(30,55){\tiny{$V(Z)/V^{square}(Z)\to\infty$}}
	\rput(30,50){\tiny{$V(Z)/V^{rectangle}(Z)\to\infty$}}
	\end{pspicture}
	\caption{
		\label{fig:los}Worst-case ratios $V(Z)/V^S(Z)$ for several geometric shapes $Z$.
		The value-measure $V$ is 1 in blue regions and 0 outside. 
		In the first three figures, all \spieces{} contain at most 1 blue region so $V^S(Z)=1$ while $V(Z)\geq$ the number of regions.
		In the fourth figure, the blue region is a ring of width $\epsilon>0$ near the perimeter. When $\epsilon\to 0$, the value of any square or rectangle, as a fraction of the total value, approaches 0.
	}
\end{figure}

The next lemma considers general covers. It shows that, any such cover, 
\else
Using this simple fact we now prove that, in any cover of $Z$ (not necessarily an $S$-cover),
\fi
any agent can choose a piece with a utility of at least $1/M$ the total value of $Z$, where $M$ is the sum of cover-numbers of the covering pieces:
\begin{lemma}[Chooser Lemma]
\label{lem:chooser} 
Let $Z$ be a piece covered by $m$ pieces
$Z_1\cup \cdots \cup Z_m = Z$, 
and let $M = \sum_{i=1}^m \covernum (Z_i,C)$.
Then for every value measure $V$:
\begin{align*}
\max_{i\range{1}{m}} V^S(Z_i) ~\geq~ V(Z)/M.
\end{align*}
\end{lemma}
\proof{Proof.}
For every $i\range{1}{m}$, denote $m_i := \covernum (Z_i,S)$.
Suppose that we replace each $Z_i$ in the cover of $Z$ 
by $m_i$ \spieces{} in a minimal $S$-cover of $Z_i$, e.g, $Z_{i,1},\ldots,Z_{i,m_i}$.
We now have an $S$-cover of $Z$ with $M$ \spieces{}.
By \eqref{eq:maxvz}, at least one of these $M$ pieces, say $Z_{i,j}$, has a value of at least $V(Z)/M$.
By definition of $V^S$, $V^S(Z_{i})\geq V(Z_{i,j}) = V(Z)/M$.
\endproof

As a simple corollary of the Chooser Lemma, we get:
\begin{lemma}[Allocation Lemma]
\label{cor:chooser}
Given a land $C$ and some integer $m\geq n$, 
let 
$(C_1,\ldots,C_m)$ be a cover of $C$ 
with a total cover-number of $M$, i.e.:
$M = \sum_{i=1}^m \covernum (Z_i,C)$.
Suppose each agent $i\range{1}{n}$ chooses a piece $C_i$ that gives him a highest utility among the $m$ pieces, and the $n$ choices are pairwise-disjoint.
Then, the resulting allocation of $C$ is envy-free 
and $(1/M)$-proportional.
\end{lemma}

\subsection{Knife functions.} \label{sub:knife-functions}
Moving-knives have been used in fair division procedures ever since the seminal paper of \citet{Dubins1961How}. We generalize the concept of a moving knife to handle geometric shape constraints.

\begin{figure}
\newcommand{\subtitle}[1]{
\rput[lb](0,50){\tiny{\shortstack{#1}}}
}
		\psset{unit=0.5mm,dotsep=1pt,hatchsep=2pt,hatchangle=0}
		\begin{pspicture}(80,75)
		\subtitle{
			(a)
			$C=[0,L]\times[0,1]$			\\
			$K_C(t) = [0, L t]\times[0,1]$  \\
			$\kcovernum(K_C,Squares)=\infty$     \\
			$\kcovernum(K_C,Rectangles)=2$
		}
		\psframe[fillstyle=none](0,0)(50,30)
		\psframe[linestyle=dotted,fillstyle=hlines,hatchcolor=blue](0,0)(7,30) \psframe[linestyle=dotted,fillstyle=none](0,0)(10,30) \psframe[linestyle=dotted,fillstyle=none](0,0)(13,30) \psframe[linestyle=dotted,fillstyle=none](0,0)(16,30)
		\end{pspicture}
		\begin{pspicture}(120,75)
		\subtitle{
			(b)
			$C=[0,1]\times[0,1]$ \\ $K_C(t)=[0,t]\times[0,t]\cup[1-t,1]\times[1-t,1]$ \\
			$\kcovernum(K_C,Rectangles)=4$     \\
			$\kcovernum(K_C,Squares)=4$
		}
		\psframe[fillstyle=none](0,0)(40,40)
		\psframe[linestyle=dotted,fillstyle=hlines,hatchcolor=blue](0,0)(13,13) \psframe[linestyle=dotted,fillstyle=none](0,0)(15,15) \psframe[linestyle=dotted,fillstyle=none](0,0)(17,17) \psframe[linestyle=dotted,fillstyle=none](23,23)(40,40) \psframe[linestyle=dotted,fillstyle=none](25,25)(40,40) \psframe[linestyle=dotted,fillstyle=hlines,hatchcolor=blue](27,27)(40,40)
		\rput(50,0) {
			\psframe[fillstyle=none](0,0)(40,40)
			\psframe[linestyle=dotted,fillstyle=hlines,hatchcolor=blue](0,0)(23,23) \psframe[linestyle=dotted,fillstyle=none](0,0)(25,25) \psframe[linestyle=dotted,fillstyle=none](0,0)(27,27) \psframe[linestyle=dotted,fillstyle=none](13,13)(40,40) \psframe[linestyle=dotted,fillstyle=none](15,15)(40,40) \psframe[linestyle=dotted,fillstyle=hlines,hatchcolor=blue](17,17)(40,40) 
		}
		\end{pspicture}
	\ifdefined\quarterplane
		\begin{pspicture}(80,75)
		\subtitle{
			(c)
			$C=[0,\infty)\times[0,\infty)$ \\ $K_C(t)=[0,{t\over 1-t}]\times[0,{t\over 1-t}]$ \\
			$S$ = rectangles and quarterplanes \\
			$\kcovernum(K_C,S)=3$     \\
		}		
		\psline[linestyle=solid](40,0)(0,0)(0,40)
		\psframe[linestyle=dotted,fillstyle=hlines,hatchcolor=blue](0,0)(13,13)
		\psframe[linestyle=dotted,fillstyle=none](0,0)(15,15)
		\psframe[linestyle=dotted,fillstyle=none](0,0)(17,17)
		\psframe[linestyle=dotted,fillstyle=none](0,0)(19,19)
		\end{pspicture}
	\else
		\begin{pspicture}(85,75)
		\rput(20,5){
			\psellipse[fillstyle=solid,fillstyle=hlines,hatchcolor=blue](16,15)(30,18)
			\psframe[fillstyle=solid,fillcolor=white,linestyle=none](-7,-5)(50,35)
		\psellipse[fillstyle=none,linewidth=1pt](16,15)(30,18)
			\psline[linecolor=black,linestyle=dotted](-6,1)(-6,28)
			\psline[linecolor=black,linestyle=dotted](-4,0)(-4,29)
			\psline[linecolor=black,linestyle=dotted](-2,-1)(-2,30)
		}
		\subtitle{
			(c)
			$C$ = ellipse with major axis 
			\\~~~~~~~~$[0,1]\times \{0\}$\\
			$K_C(t)=C\cap [0,t]\times (-\infty,\infty)$ \\
			$\kcovernum(K_C,convex~objects)=2$
		}
		\end{pspicture}	
	\fi
	
		\begin{pspicture}(80,85)
		\subtitle{
			(d)
			$C=[0,2]\times[0,2]$ \\
			$K_C(t)=[0,t]\times[0,t]$ \\
			$\kcovernum(K_C,Rectangles)=3$     \\
			$\kcovernum(K_C,Squares)=4$
		}				
		\psframe[fillstyle=solid](0,0)(40,40)
		\psline[linestyle=dashed](20,0)(20,40)
		\psline[linestyle=dashed](0,20)(40,20)
		\psframe[linestyle=dotted,fillstyle=hlines,hatchcolor=blue](0,0)(13,13)
		\psframe[linestyle=dotted,fillstyle=none](0,0)(15,15)
		\psframe[linestyle=dotted,fillstyle=none](0,0)(17,17)
		\psframe[linestyle=dotted,fillstyle=none](0,0)(19,19)
		\end{pspicture}
		\begin{pspicture}(100,85)
		\subtitle{
			(e)
			$C=[0,2]\times[0,2]$ \\
			$C_1 := C\setminus [0,1]\times[0,1]$ \\
			$K_C(t)=C_1\cap ([0,2]\times[0,t/2])$\\
			$\kcovernum(K_C,Rectangles)=3$     \\
			$\kcovernum(K_C,Squares)=\infty$
		}		
		\psframe[fillstyle=solid](0,0)(40,40)
		\psline[linestyle=dashed](20,0)(20,20)(0,20)
		\psframe[linestyle=dotted,fillstyle=hlines,hatchcolor=blue](20,0)(40,15)
		\psline[linestyle=dotted](20,17)(40,17)
		\psline[linestyle=dotted](20,19)(40,19)
		\psline[linestyle=dotted](0,21)(40,21)
		\psline[linestyle=dotted](0,23)(40,23)
		\psline[linestyle=dotted](0,25)(40,25)
		\end{pspicture}
		\begin{pspicture}(120,85)
		\subtitle{
			(f)
			$C=[0,1]\times[0,1]$ \\
			$K_C(t) = [0,t]\times[0,1]\cup [1-t,1]\times[0,1]$ \\
			$\kcovernum(K_C,Rectangles)=3$     \\
			$\kcovernum(K_C,Squares)=\infty$
		}
		\def\landcake{\psframe[fillstyle=none](0,0)(40,40)}
		\landcake
		\psframe[linestyle=dotted,fillstyle=hlines,hatchcolor=blue](0,0)(7,40) \psframe[linestyle=dotted,fillstyle=none](0,0)(10,40) \psframe[linestyle=dotted,fillstyle=none](0,0)(13,40) \psframe[linestyle=dotted,fillstyle=none](0,0)(16,40)
		\psframe[linestyle=dotted,fillstyle=hlines,hatchcolor=blue](33,0)(40,40) \psframe[linestyle=dotted,fillstyle=none](30,0)(40,40) \psframe[linestyle=dotted,fillstyle=none](27,0)(40,40) \psframe[linestyle=dotted,fillstyle=none](24,0)(40,40)
		\end{pspicture}
	\caption[Several knife functions.]{
		\label{fig:knives}
		Several knife functions. The area filled with horizontal lines marks $K_C(t)$ in a certain intermediate time $t\in(0,1)$. Dotted lines mark future knife locations. 
	}
\end{figure}
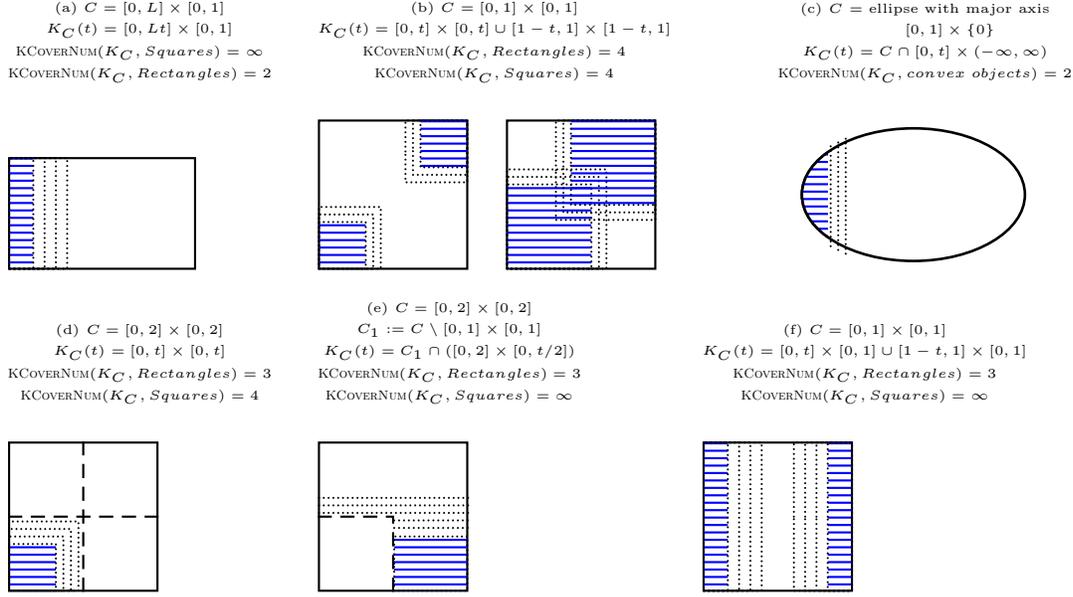

\begin{definition} \label{def:knife}
A \emph{knife function} on a land $C$ is a function $K_C$ from the real interval $[0,1]$ to Borel subsets of $C$, with the following continuity property: 
for every $\epsilon>0$ there is a $\delta>0$ such that  $|t'-t|<\delta$ implies $\leb{K_C(t')\ominus K_C(t)} < \epsilon$.\footnote{
$\leb{Z}$ denotes the Lebesgue measure of a piece $Z$ in $\mathbb{R}^d$.

The symbol $\ominus$ denotes the symmetric set difference: for two sets $A$ and $B$, $A\ominus B := (A\setminus B) \cup (B\setminus A)$.
}
\def\IncDec{}

\ifdefined\IncDec
$K_C$ is called
\emph{increasing} if $t'\geq t \implies K_C(t')\supseteq K_C(t)$,
and \emph{decreasing} if $t'\geq t \implies K_C(t')\subseteq K_C(t)$.
\fi

If $K_C(0)=\cstart$ and $K_C(1)=\cend$, we say that $K_C$ is \emph{a knife function from $\cstart$ to $\cend$}.
\end{definition}
Some examples of knife-functions are shown in Figure \ref{fig:knives}.
In Appendix \ref{sub:growing-ball}
we show a general construction of an increasing knife-function from $\cstart$ to $\cend$ --- the \emph{growing-ball function}.

The \emph{complement} of $K_C$ is denoted $\overline{K_C}$ and defined by $~\overline{K_C}(t) = C\setminus K_C(t)
~~
\forall t\in[0,1]
~$. 
If $K_C$ is a knife-function then $\overline{K_C}$ is a knife-function too.

\subsection{Continuity of utility covered by knife functions.}
The value covered by a knife-function always changes continuously with time. Formally, we prove in Appendix \ref{sub:cont-value} that:
\begin{lemma}
\label{lem:cont-value}
If $K_C$ is a knife-function and $V$ is an absolutely-continuous measure, then $V\circ K_C$ is a uniformly-continuous real functions.
\end{lemma}
However, in our setting the agents care not about $V$ but about $V^S$. In general, the function $V^S\circ K_C$ might be discontinuous. For example, let $K_C$ be the knife-function in Figure \ref{fig:knives}(f), $V$ the area measure, and $S$ the family of rectangles. Then, the function $V^S\circ K_C$ (the largest area of a rectangle covered by the knife)
is discontinuous --- it is less than $1/2$ when $t<1/2$, and jumps to $1$ when $t\geq 1/2$.

To handle this issue we define two different properties of knife functions.

\paragraph{1. $S$-continuity} means (informally) that all \spieces{} in $K_C(t)$ and $\overline{K_C}(t)$ grow or shrink continuously; no \spiece{} with a positive area is created or destroyed abruptly. 
See Appendix \ref{sub:continuity} for a formal definition.
In Figure \ref{fig:knives}, all knives are square-continuous and rectangle-continuous except for knife (f). 
In Appendix \ref{sub:continuity} we prove that:
\begin{lemma}
\label{lem:continuous}
If $K_C$ is an $S$-continuous knife-function
and $V$ is an absolutely-continuous measure,
then both $V^S \circ K_C$ 
and  $V^S\circ \overline{K_C}$ 
are uniformly-continuous real functions.
\end{lemma}
In Appendix \ref{sub:sweeping-plane}
we show a general construction of an increasing $S$-continuous knife-function on $C$ --- the \emph{sweeping-plane function} (Figure \ref{fig:knives}(a,e) are special cases of this function).

\paragraph{2. $S$-smoothness} means (informally) that $K_C(t)$ is a finite union of \spieces{} that grow continuously and $\overline{K_C}(t)$ is a finite union of \spieces{} that shrink continuously. 
To define it formally we need some preliminary definitions:
\begin{itemize}
\item The \emph{union} of two knife-functions, $K_1$ and $K_2$, 
is a knife-function $K_1\cup K_2$ defined by:
\begin{center}
$~(K_1\cup K_2)(t) = K_1(t)\cup K_2(t)~~\forall t\in[0,1]~$
\end{center}
\item  A knife-function $K$ is an \emph{$S$-knife function} if it is into $S$, i.e., $~K(t) \in S~~\forall t\in[0,1]~$.
\item An \emph{$S$-cover} of a knife-function $K_C$ is a set of $S$-knife functions whose union equals $K_C$. 
\end{itemize}

\begin{definition}
\label{def:smooth}
A knife-function $K_C$ is called \emph{$S$-smooth} if:
\begin{itemize}
\item $K_C$ has a finite $S$-cover $K^1\cup\cdots \cup K^m$ for some integer $m\geq 1$.
\item $\overline{K_C}$ has a finite $S$-cover $\overline{K}^1\cup\cdots \cup \overline{K}^{m'}$ for some integer $m'\geq 1$.
\item For every $j\range{1}{m}$, $K^j(1) = K_C(1)$. I.e, the knives covering $K_C$ coincide at $t=1$.%
\footnote{
The latter condition comes to guarantee that there are no jumps in the $S$-value at $t=1$.
It implies that $K_C(1)\in S$.
 }
\end{itemize}

If $K_C$ is $S$-smooth, then we define the \emph{cover number} of $K_C$, $\kcovernum (K_C,S)$, as 
the smallest sum $m+m'$ of integers that satisfy the above definition.
\end{definition}

Lemma \ref{lem:continuous} is not necessarily true for $S$-smooth knife functions --- $V^S\circ K_C$ and $V^S\circ \overline{K_C}$ are not necessarily continuous. Therefore we define the following alternative functions (where $V$ is any value-measure and the $K^j$ are the $S$-knife-functions covering $K_C$):
\begin{align}
\label{eq:vk}
V^{K_C}(t) := \max_{j=1}^m V(K^j(t))
&&
V^{\overline{K_C}}(t) := \max_{j=1}^{m'} V(\overline{K}^j(t)).
\end{align}
Note that for all $t$, $V^{K_C}(t)\leq V^S(K_C(t))$, 
since $V^{K_C}$ considers a most valuable \spiece{} from a finite set of \spieces{} contained in $K_C(t)$,
while $V^S$ considers a supremum over \emph{all} \spieces{} contained in $K_C(t)$,
Similarly, $V^{\overline{K_C}}(t)\leq V^S(\overline{K_C}(t))$.

\begin{lemma}
\label{lem:smooth}
If $K_C$ is an $S$-smooth knife-function 
and $V$ is an absolutely-continuous measure,
then both $V^{K_C}$ and 
$V^{\overline{K_C}}$
are uniformly-continuous real functions.
\end{lemma}
\proof{Proof.}
Each of these functions is a maximum over functions that are uniformly-continuous (by Lemma \ref{lem:cont-value}).
The maximum of uniformly-continuous  functions is uniformly-continuous.
\Halmos\endproof

\paragraph{Examples} (see Figure \ref{fig:knives}):

(a) Both $K_C(t)$ and its complement are
rectangles and their areas are continuous functions of $t$.
Therefore, $K_C$ is rectangle-smooth with $m=m'=1$ and its rectangle-cover-number is $2$.
In contrast, 
$K_C(t)$ cannot be presented as a union of a fixed finite number of squares. Therefore, $K_C$ is not square-smooth.

(b) $K_C(t)$ can be covered by two square-knife-functions: $K^1(t) = [0,t]\times[0,t]$
and $K^2(t) = [1-t,1]\times[1-t,1]$.
Its complement too can be covered by two square-knife-functions: $\overline{K}^1(t) = [0,1-t]\times[t,1]$
and $\overline{K}^2(t) = [t,1]\times[0,1-t]$.
Therefore, $K_C$ is square-smooth with
$\kcovernum (K_C,Squares)=2+2=4$.
$K_C$ is also rectangle-smooth with the same cover-number.

\ifdefined\quarterplane
(c) Here $C$ is the top-right quarter-plane and $S$ the family
of squares and quarter-planes (we consider a quarter-plane to be a
square with infinite side-length). 
$K_C(t)$ is a square that grows continuously and its complement is a union of two quarter-planes that shrink continuously.
$K_C(t)$ is a square and its complement can be covered by two quarter-planes, so the cover number of $K_C$ is at most $1+2=3$.
\else
(c) 
$C$ is an arbitrary convex object and $S$ is the family of convex objects. Both $K_C(t)$ and its complement are convex so $K_C$ is $S$-smooth with a cover-number of 2.
\fi

(d) This is a knife-function from $\emptyset$ to  $[0,1]\times[0,1]$ = the bottom-left quarter of $C$. $K_C(t)$ is a square that grows continuously.
$\overline{K_C}(t)$ is an L-shape, similar to the L-shapes in Figure \ref{fig:cover-numbers}, which can be covered by a union of 3 squares that shrink continuously. Hence $K_C$ is square-smooth with $m=1$ and $m'=3$ and its square-cover-number is 4.

(e) This is a knife-function from $\emptyset$ to $C\setminus [0,1]\times[0,1]$; it sweeps over this L-shape continuously from bottom to top. It is square-continuous (see proof in Appendix \ref{sub:sweeping-plane}), but not square-smooth.

(f) $K_C(t)$ is a union of two rectangles that grow continuously and its complement is a rectangle that shrinks continuously. It is rectangle-smooth with a cover number of 3, but not rectangle-continuous (see proof in Appendix \ref{sub:non-s-cont}).

Examples (e) and (f) show that $S$-continuity and $S$-smoothness are independent properties that do not imply each other.

\section{Envy-Free Division for Two Agents}
\label{sec:2-agents}
~\\
In this section we use the tools developed in Section \ref{sec:geometric} to build complete land division algorithms.

\subsection{Dividing squares.}
Our first algorithm uses a single $S$-smooth knife-function and generalizes the classic cut-and-choose protocol. It is presented in Algorithm \ref{alg:0-parts-smooth}. The following lemma proves that it is correct.

\begin{algorithm}[t]
\caption{Smooth Knife Algorithm}
\label{alg:0-parts-smooth}
INPUT: 

(a) $\cend$ --- an \spiece{} contained in $C$ such that for each agent $i\in\{1,2\}$:
\begin{align*}
V_i^{S}(\cend)\geq V_{i}^S(C\setminus \cend)
\end{align*}
(b) $K_C$ --- an $S$-smooth knife-function from $\emptyset$ to $\cend$ such that, for some $M\geq 2$:
\begin{align*}
\kcovernum (K_C,S)= M
\end{align*}

and the corresponding covers by $S$-knife-functions as in Definition \ref{def:smooth}, where $m+m'=M$:
\begin{align*}
K^1\cup\cdots \cup K^m = K_C
&&
\overline{K}^1\cup\cdots \cup \overline{K}^{m'}
= 
\overline{K_C}
\end{align*}

OUTPUT: 

An envy-free and $(1/M)$-proportional allocation of $C$ between the two agents.

ALGORITHM:

(1)
Agent 1 selects a time $t^*\in [0,1]$ such that (where $V_1^{K_C}$ and $V_1^{\overline{K_C}}$ are as defined in \eqref{eq:vk}):
\begin{align*}
V_1^{K_C}(t^*) = V_1^{\overline{K_C}}(t^*)
\end{align*}

(2)
Agent 2 picks one of the following two options:

(2.1) Agent 2 picks an \spiece{} $K^j(t^*)$
that maximizes $V_2$ for $j\range{1}{m}$,
and agent 1 picks an \spiece{} $\overline{K}^j(t^*)$
that maximizes $V_1$ for $j\range{1}{m'}$.

(2.2) Agent 2 picks an \spiece{} $\overline{K}^j(t^*)$
that maximizes $V_2$ for $j\range{1}{m'}$,
and agent 1 picks an \spiece{} $K^j(t^*)$
that maximizes $V_1$ for $j\range{1}{m}$.
\end{algorithm}
\begin{lemma}[Correctness of Smooth Knife Algorithm]
\label{lem:0-parts}
Let $\cend$ be piece of $C$ such that for each agent $i\in\{1,2\}$:
$V_i^{S}(\cend)\geq V_{i}^S(C\setminus \cend)$.
Let $K_C$ be an $S$-smooth knife-function from $\emptyset$ to $\cend$ such that: $\kcovernum (K_C,\, S)= M$.
Then, Algorithm \ref{alg:0-parts-smooth} produces an envy-free and $(1/M)$-proportional allocation between the two agents.
\end{lemma}
\proof{Proof.}
By Lemma \ref{lem:smooth},
for all $i\in\{1,2\}$,
both $V_i^{K_C}(t)$ and $V_i^{\overline{{K_C}}}(t)$
are continuous functions of $t$.
At $t=0$, 
$V_i^{K_C}(t)\leq V_i^{\overline{{K_C}}}(t)$,%
\footnote{
Since $K_C(0) = \emptyset$, so for all $j\range{1}{m}$, $K^j(0) = \emptyset$,
so $V_i^{K_C}(0) = V_i(\emptyset) = 0$.
}
and at $t=1$, 
$V_i^{K_C}(t)\geq V_i^{\overline{{K_C}}}(t)$,%
\footnote{
Since by definition \ref{def:smooth}, $\forall j\range{1}{m}$, $K^j(1) = K_C(1) = \cend$, 
so $V_i(K^j(1)) = V_i^S(K^j(1)) = V_i^S(\cend)$,
so also $V_i^{K_C}(1) = V_i^S(\cend)$.
On the other hand, $\forall j\range{1}{m}$, $\overline{K_C}^j(1) \subseteq \overline{K}_C(1) = C\setminus \cend$.
Since $\overline{K_C}^j(1)$ is an $S$-piece, 
$V_i(\overline{K_C}^j(1))\leq V_i^S(C\setminus \cend)$.
Since this is true for all $j\range{1}{m}$,
also $V_i^{\overline{K_C}}(1)\leq V_i^S(C\setminus \cend)$.
Now, by assumption, $V_i^S(\cend)\geq V_i^S(C\setminus \cend)$.
}
so by the intermediate value theorem, for some $t^*\in[0,1]$ the two functions are equal.
Therefore, agent 1 can indeed find a $t^*\in[0,1]$ such that $V_1^{K_C}(t^*)=V_1^{\overline{K_C}}(t^*)$, as required in step (1).

Now, in step (2), each agent $i$ receives an \spiece{} that maximizes $V_i$ 
from a covering of size $m+m' = M$.
Hence, by the Allocation Lemma, the allocation is envy-free and $1/M$-proportional.
\Halmos\endproof

Based on this lemma we can now prove our first sub-theorem:

\begin{restatetheorem}[Theorem \ref{thm:2agents}(a)]
	\noindent \emph{$\prope(Square,\,2,\, Squares)\geq1/4$ }
\end{restatetheorem}
\proof{Proof.}
Given a square land $C$,
apply the Single Knife Algorithm with $\cend=C$ and $K_C$ the knife-function shown in Figure \ref{fig:knives}/b. 
As explained in Subsection \ref{sub:knife-functions}, 
this $K_C$ is square-smooth and $\kcovernum(K_C,Squares) = 4$. 
Therefore, by Lemma \ref{lem:0-parts}, the resulting division is envy-free and $(1/4)$-proportional.
\Halmos
\endproof

\ifdefined\quarterplane
Using Algorithm \ref{alg:0-parts-smooth} we can get more results with little additional effort. For example:
\begin{itemize}
	\item Using the knife function of Figure \ref{fig:knives}/b: $\prope(Square,\,2,\: Square\, pairs)\geq1/2$.
	I.e., if each agent has to receive a union of two squares (as is common when dividing land to settlers, e.g. one land-plot for building and another one for agriculture), then a proportional division is possible since the knife function in example (b) has a cover number of 2 relative to the family of square pairs. 
	\item Using the knife function of Figure \ref{fig:knives}/c: $\prope(Quarter\, Plane,\,2,\, Generalized\, Squares)\geq 1/3$ (where the family of generalized squares contains squares and quarter-planes).
\end{itemize}
All lower bounds presented above are tight --- it is not possible to guarantee both agents a larger utility \emph{even if envy is allowed}. This is obvious for the $\geq 1/2$ results, since a proportionality of $1/n$ is the best that can be guaranteed to $n$ agents. For the other results, the matching upper bounds are proved in Appendix \ref{sec:upper-bounds}, Lemmas \ref{lem:negative-square} and \ref{lem:negative-quarter-plane}.
\else
This lower bound is tight --- 
it is not possible to guarantee both agents a larger utility even if envy is allowed. See 
Appendix \ref{sec:upper-bounds}, Lemma \ref{lem:negative-square}.
\fi

\subsection{Dividing cubes and archipelagos.}
\label{sub:archipelagos}
In many cases it is difficult to find a single $S$-smooth knife function that covers the entire land. 
To handle such cases, we first present a subroutine (Algorithm \ref{alg:0-parts-cont}) and then a division algorithm that uses this subroutine (Algorithm \ref{alg:1-parts}).

Below we prove the correctness of the subroutine and then the correctness of the full algorithm.

\begin{algorithm}[t]
\caption{Continuous Knife Subroutine}
\label{alg:0-parts-cont}
INPUT: 

(a) $\cend$ --- a piece of $C$ such that for each agent $i\in\{1,2\}$:
\begin{align*}
V_i^{S}(\cend)\geq V_{i}^S(C\setminus \cend)
\end{align*}
(b) $K_C$ --- an increasing $S$-continuous knife-function from $\emptyset$ to $\cend$.

OUTPUT: 

An envy-free allocation of $C$ in which every agent $i$ gets a utility of at least $V_i^S(C\setminus \cend)$.

ALGORITHM:

(1) Agent 1 selects a time $t^*\in [0,1]$ such that:
\begin{align*}
V_1^S(K_C(t^*)) = V_1^S(\overline{K_C}(t^*)).
\end{align*}
	
(2) Agent 2 picks either $K_C(t^*)$ or $\overline{K_C}(t^*)$; agent 1 receives the remaining piece.
\end{algorithm}

\begin{lemma}[Correctness of Continuous Knife Subroutine]
\label{lem:0-parts-cont}
Let $\cend$ be piece of $C$ such that for each agent $i\in\{1,2\}$:
$V_i^{S}(\cend)\geq V_{i}^S(C\setminus \cend)$.
Let $K_C$ be an increasing $S$-continuous knife-function from $\emptyset$ to $\cend$.
~~~
Then, Algorithm \ref{alg:0-parts-cont} produces an envy-free allocation of $C$ between the two agents such that the utility of agent $i$ is at least $V_i^S(C\setminus \cend)$.
\end{lemma}
\proof{Proof.}
By Lemma \ref{lem:continuous},
for all $i\in\{1,2\}$,
both 
$V_i^S K_C(t))$ and $V_i^S(\overline{K_C}(t))$ are continuous functions.

At $t=0$ the first function is weakly smaller, since  $V_i^S(K_C(0)) = V_i^S(\emptyset) = 0$.

At $t=1$ the first function is weakly larger, since   $V_i^S(K_C(1)) = V_i^S(\cend) \geq V_i^S(C\setminus \cend) = V_i^S(\overline{K_C}(1))$ by assumption. 

Therefore, by the intermediate value theorem, agent 1 can indeed pick a $t^*\in[0,1]$ such that $V_1^{S}(K_C(t^*))=V_1^{S}(\overline{K_C}(t^*))$, as required in step (1).

In step (2), agent 1 is indifferent between the two pieces and agent 2 picks a best piece, so the allocation is envy-free.

Moreover, since $K_C$ is increasing, $\overline{K_C}$ is decreasing, so  $\overline{K_C}(t^*)$ contains $\overline{K_C}(1) = C\setminus \cend$.
Each agent receives either $\overline{K_C}(t^*)$ or a piece with a weakly larger utility. Therefore, the utility of each agent $i$ is at least 
$V_1^S(C\setminus \cend)$.
\Halmos\endproof

\begin{algorithm}[t]
\caption{ּSingle Partition Algorithm}
\label{alg:1-parts}
INPUT: 

(a) a partition of $C$ into $m$ pieces, with a total cover number of at most $M$ (for some integers $M\geq m \geq 2$):
\begin{align*}
\bigsqcup_{j=1}^m C_j  = C
&&
\sum_{j=1}^{m}\covernum(C_{j},S) \leq M
\end{align*}

(b) For every $j\range{1}{m}$, an $S$-smooth knife function $K_{C_j}$ from $\emptyset$ to $C_j$, such that:
\begin{align*}
\forall j\range{1}{m}:\,\kcovernum(K_{C_j},\, S)\leq M
\end{align*}

(c) For every $j\range{1}{m}$, an increasing $S$-continuous knife function $K^{\overline{C_j}}$ from $\emptyset$ to $\overline{C_j} := C\setminus C_j$.

OUTPUT:

An envy-free and $(1/M)$-proportional allocation of $C$ between the two agents.

ALGORITHM:

(1) From the input partition (a), each agent chooses the part $C_j$ that gives him maximum utility. If the choices are different then each agent receives his choice and we are done.

(2) If both agents choose the same part $C_j$, then ask each agent to choose either $C_j$ or its complement  $\overline{C_j}$.
If the choices are different then each agent receives his choice and we are done.

If the choices are identical then there are two cases:
	
(3.1) Both agents chose $C_j$. 
Apply Algorithm \ref{alg:0-parts-smooth} with 
$\cend = C_j$ 
and the $S$-smooth knife-function $K_{C_j}$ of input (b).

(3.2) Both agents chose $\overline{C_j}$. 
Apply Algorithm \ref{alg:0-parts-cont} with 
$\cend = \overline{C_j}$ 
and the $S$-continuous knife-function $K^{\overline{C_j}}$ of input (c).
\end{algorithm}

\begin{lemma}[Correctness of Single Partition Algorithm]
\label{lem:1-parts}
If there exist a partition and knife-functions satisfying the input requirements of Algorithm \ref{alg:1-parts}, then this algorithm produces an envy-free and $(1/M)$-proportional allocation of $C$ between the two agents.
\end{lemma}
\proof{Proof.}
The algorithm may end in step (1), (2), (3.1) or (3.2). We prove that in each of these cases the resulting allocation is envy-free and $(1/M)$-proportional.
	
In steps (1) and (2), if the choices are different, then by the Allocation Lemma (\ref{cor:chooser}) and the condition on input (a), each agent $i$ receives an envy-free share with a utility of at least $V_i(C)/M$.

In step (3.1), we know that both agents prefer $C_j$ over its complement $\overline{C_j}$. Therefore, $\cend=C_j$ 
satisfies requirement (a) of Algorithm \ref{alg:0-parts-smooth}.
The knife-function $K_{C_j}$ is $S$-smooth with a cover number of at most $M$, so it satisfies requirement (b).
Hence, by Lemma \ref{lem:0-parts}, 
Algorithm \ref{alg:0-parts-smooth} gives to each agent $i$ an envy-free share with a utility of at least $V_i(C)/M$.
	
In step (3.2), we know that both agents prefer $\overline{C_j}$ 
over its complement $C_j$. 
Therefore, $\cend=\overline{C_j}$
satisfies requirement (a) of Algorithm \ref{alg:0-parts-cont}.
The knife-function $K^{\overline{C_j}}$ is increasing and $S$-continuous, so it satisfies requirement (b).
Hence, by Lemma \ref{lem:0-parts-cont},
Algorithm \ref{alg:0-parts-cont} gives to each agent $i$ an envy-free share with a utility of at least 
$V_i^S(C\setminus \overline{C_j}) = V_i^S(C_j)$.
Now, in step (1) both agents chose $C_j$ from a partition with a total cover number of at most $M$. Therefore, by the Chooser Lemma (\ref{lem:chooser}), $V_i^S(C_j)\geq V_i(C)/M$.
\Halmos\endproof
\begin{figure}
	\psset{unit=1mm,dotsep=1pt,hatchsep=2pt,hatchangle=0}
	\newcommand{\landcake}{
		\psframe(10,0)(30,10)
		\psframe(20,10)(40,20)
		\psframe(0,20)(35,30)
	}
	\newcommand{\knife}{
		\psframe[fillstyle=hlines,hatchcolor=blue](0,0)(5,10)
		\psline[linestyle=dotted](6,0)(6,10)
		\psline[linestyle=dotted](7,0)(7,10)
		\psline[linestyle=dotted](8,0)(8,10)
	}
	\begin{center}
		\begin{pspicture}(50,30)
		\rput(20,35){(a)}
		\landcake
		\end{pspicture}
		\begin{pspicture}(50,30)
		\rput(20,35){(b)}
		\landcake
		\rput(10,0){\knife}
		\rput(20,10){\knife}
		\rput(0,20){\knife}
		\psline[linestyle=dashed](10,10)(40,10)
		\psline[linestyle=dashed](10,20)(40,20)
		\end{pspicture}
	\end{center}
	\caption[]{
		\label{fig:union-of-rectangles} 
		Knife functions for archipelagos of rectangles.
		\textbf{Left}: A land-estate made of a union of 3 disjoint rectangles. 
		\\
		\textbf{Right}: Three knife functions, each having a cover number of 4, proving that $\prope(C,\,2,\, rectangles)\geq1/4$.
	}
\end{figure}

Several applications of Algorithm \ref{alg:1-parts}  are presented below.

(a) \textbf{Multi-dimensional cubes}: $\prope(d\,dimensional\,cube,2,d\,dimensional\,cubes)\geq1/2^d$. \emph{Proof}: 
$C$ can be partitioned into $2^d$ sub-cubes of equal side-length. 
The total cube-cover-number of this partition is $2^d$, satisfying input condition (a).
For each sub-cube $C_j$, there is an $S$-smooth knife function analogous to Figure \ref{fig:knives}/d --- a cube growing from the corner towards the center of $C$. 
The complement of the cube can always be covered by a union of $2^{d}-1$ cubes (possibly overlapping).
Therefore this knife-function is $S$-smooth with a cover-number of $2^d$, satisfying input condition (b).
For each complement $\overline{C_j}$, 
the sweeping-plane knife-function on $\overline{C_j}$ (see subsection \ref{sub:sweeping-plane} and Figure \ref{fig:knives}/e) is increasing and $S$-continuous, satisfying condition (c). \Halmos

(b) \textbf{Rectangle archipelagos}: Let $C$ be an archipelago which is a union of $m$ disjoint rectangular islands. Then $\prope(C,\,2,\, Rectangles)\geq\frac{1}{m+1}$.
\emph{Proof}: The total rectangle-cover-number of the partition of $C$ into $m$ rectangles is obviously $m<m+1$, satisfying condition (a). For each part $C_j$, define a knife function $K_{C_j}$ based on a line sweeping from one side of the rectangle to the other side, similar to Figure \ref{fig:knives}/a. $K_{C_j}(t)$ is always a rectangle. Its complement is a union of $m$ rectangles: the shrinking rectangle $C_{j}\setminus K_{C_{j}}(t)$, and the remaining $m-1$ fixed rectangular islands%
. 
Hence, $K_{C_j}$ is rectangle-smooth and its cover number is  $1+1+m-1=m+1$, satisfying condition (b) (see Figure \ref{fig:union-of-rectangles}).
For the complements, a sweeping-line knife-function (as in Appendix \ref{sub:sweeping-plane}) satisfies condition (c).

(c) \textbf{Square archipelagos}: Let $C$ be an archipelago which is a union of $m$ disjoint square islands. Then $\prope(C,\,2,\, Squares)\geq\frac{1}{m+3}$.
The proof is the same as in (b), the only difference being that each $S$-smooth knife functions $K_{C_j}$ is a union of two growing squares, similar to Figure \ref{fig:knives}(b).
$K_{C_j}$ is covered by two square-knife-functions and its complement is covered by $2+(m-1)$ square-knife-functions, so $K_{C_j}$ is square-smooth with a cover number of $2+2+m-1=m+3$.

All bounds proved above are tight. The tightness of part (a) can be proved 
analogously to Lemma \ref{lem:negative-square} --- 
a $d$-dimensional cube with a water-pool in each of its $2^d$ corners. 
Part (b) is tight in the following sense: for every integer $m$ there exists some $C$, which is a union of $m$ disjoint rectangles, having
$\prop(C,\,2,\, Rectangles)\leq\frac{1}{m+1}$. See 
Lemma \ref{lem:negative-archipelago}.
Part (c) is tight in a similar sense by a similar proof.

\subsection{Dividing fat rectangles.}
\label{sub:fat-rects}
To prove Theorem \ref{thm:2agents}(b) we add a partition step, as shown in Algorithm \ref{alg:2-parts}.
Note that steps (1) and (2) and (3.2) of that algorithm are the same as in Algorithm \ref{alg:1-parts}; they are repeated for completeness.
Step (3.1) is refined.

\begin{algorithm}
\caption{Multiple Partition Algorithm}
\label{alg:2-parts}
INPUT: 

(a) a partition of $C$ into $m$ pieces, with a total cover number of at most $M$:
\begin{align*}
\bigsqcup_{j=1}^m C_j  = C
&&
\sum_{j=1}^{m}\covernum(C_{j},S) \leq M
\end{align*}
	
(b) For every part $C_j$, a partition such that, if $C_j$ is replaced with its partition, then the total cover number of the resulting partition of $C$ is at most $M$, i.e. for every $j$ there exist $C_j^{1},\dots,C_j^{m_j}$ with:
\begin{align*}
\bigsqcup_{k=1}^{m_j}C_j^{k} = C_j
&&
\sum_{j'\neq j}\covernum(C_{j'},S)+\sum_{k=1}^{m_{j}}\covernum(C_{j}^{k},S)\leq M
\end{align*}

(c) For every $j\range{1}{m}$ and $k\range{1}{m_j}$, an $S$-smooth knife function $K_{C_{j}^{k}}$ from $\emptyset$ to $C_{j}^{k}$, with a cover number of at most $M$:
\begin{align*}
\forall j:\,\kcovernum(K_{C_{j}^{k}},\, S)\leq M
\end{align*}

(d) For every $j\range{1}{m}$ and $k\range{1}{m_j}$, an increasing $S$-continuous knife function from $\emptyset$ to $C_j$ and to $\overline{C_j}$ and to $\overline{C_{j}^{k}}$.

OUTPUT:

An envy-free and $(1/M)$-proportional allocation of $C$ between the two agents.

ALGORITHM:

\textbf{(1)} From the input partition (a), each agent chooses the part $C_j$ that gives him maximum utility. If the choices are different then each agent receives his choice and we are done.

\textbf{(2)} If both agents choose the same part $C_j$, then ask each agent to choose either $C_j$ or its complement  $\overline{C_j}$.
If the choices are different then each agent receives his choice and we are done.

If the choices are identical then there are two cases, denoted below by \textbf{(3.1)} and \textbf{(3.2)}:
~\\

\textbf{(3.1)} Both agents chose $C_j$. Refine the partition of $C$ by replacing $C_j$ with its sub-partition:
\begin{align*}
	(\bigsqcup_{j'\neq j}C_{j'})\sqcup (\bigsqcup_{k=1}^{m_j}C_j^{k}) = C
\end{align*}
Let each agent choose a best part from this refined partition.

\begin{itemize}
\item (3.11) If the choices are different then each agent receives his choice and we are done.

\item (3.12) If both agents chose the same part from the main partition, e.g. $C_{j'}$ for some $j'\neq j$, then 
apply Algorithm \ref{alg:0-parts-cont}
with $\cend = C_j$ [where $C_j$ is the piece chosen by both agents in step (3.1)]
and the $S$-continuous knife-function from $\emptyset$ to $C_j$ of input (d).

\item (3.13) If both agents chose the same part from the sub-partition, e.g. $C_j^k$ for some $k$, then 
ask each agent to choose either $C_j^k$ or $\overline{C_j^k}$ (where $\overline{C_j^k}:=C\setminus C_j^k$). If the choices are different, then each agent receives his choice and we are done.

If the choices are identical then there are two sub-cases:
	
\item (3.141) Both agents chose $C_j^k$. 
Apply Algorithm \ref{alg:0-parts-smooth}
with $\cend = C_j^k$ 
and the $S$-smooth knife-function of input (c).

\item (3.142) Both agents chose $\overline{C_j^k}$. 
Apply Algorithm \ref{alg:0-parts-cont}
with $\cend = \overline{C_j^k}$ 
and the $S$-continuous knife-function from $\emptyset$ to $\overline{C_j^k}$ of input (d).
\end{itemize} 

~\\

\textbf{(3.2)} Both agents chose $\overline{C_j}$. 
Apply Algorithm \ref{alg:0-parts-cont} with 
$\cend = \overline{C_j}$ 
and the $S$-continuous knife-function from $\emptyset$ to $\overline{C_j}$ of input (d).

\end{algorithm}

\begin{lemma}[Correctness of Multiple Partition Algorithm]
\label{lem:2-parts}
If there exist partitions and knife-functions satisfying the input requirements of Algorithm \ref{alg:2-parts}, then this algorithm produces an envy-free and $(1/M)$-proportional allocation of $C$ between the two agents.
\end{lemma}
\proof{Proof.}
Steps (1), (2) and (3.2) are the same as in Algorithm \ref{alg:1-parts} and their correctness proof is the same too. 
It remains to prove that, if the algorithm ends in one of the sub-steps of (3.1), then the resulting allocation is envy-free and $(1/M)$-proportional.

In step (3.11), if the choices are different, then by the Allocation Lemma and the condition on input (b), each agent $i$ receives an envy-free share with a utility of at least $V_i(C)/M$. 

In step (3.12), we know that both agents prefer $C_j$ to its complement $\overline{C_j}$ (from the choice of step (3.1)). Therefore, $\cend=C_j$ and the $S$-continuous knife-function from input (d) satisfy the input requirements of Algorithm \ref{alg:0-parts-cont}. 
The algorithm gives each agent an envy-free share with utility at least
$V_i^S(\overline{C_{j}})$. 
This $\overline{C_{j}}$ contains all other parts of the main partition, including $C_{j'}$. The fact that both agents chose $C_{j'}$ in the refined partition proves, by the Chooser Lemma, that $V_i^S(C_{j'})\geq V_i(C)/M$.  Hence also $V_i^S(\overline{C_{j}})\geq V_i(C)/M$.
	
In step (3.13), if the choices are different, then by the Allocation Lemma and the condition on input (b), each agent $i$ receives an envy-free share worth at least $V_i(C)/M$.

In step (3.141), we know that both agents prefer $C_j^k$ to its complement $\overline{C_j^k}$. 
Therefore, $\cend=C_j^k$ and the $S$-smooth knife-function from input (c)
satisfy the input requirements of Algorithm \ref{alg:0-parts-smooth}.
The cover-number of this knife-function 
is at most $M$. 
Hence, Algorithm \ref{alg:0-parts-smooth} gives to each agent $i$ an envy-free share with a utility of at least $V_i(C)/M$.

In step (3.142), we know that both agents prefer $\overline{C_j^k}$ to its complement $C_j^k$. 
Therefore, $\cend=\overline{C_j^k}$ and the $S$-continuous knife-function from input (d)
satisfy the input requirements of Algorithm \ref{alg:0-parts-cont}.
Each agent receives an envy-free share with utility at least
$V_i^S(C\setminus \overline{C_j^k}) = V_i^S(C_j^k)$.
Now, in step (3.13) both agents chose $C_j^k$ from a partition with a total cover number of at most $M$. Therefore, by the Chooser Lemma, $V_i^S(C_j^k)\geq V_i(C)/M$.
\Halmos\endproof

\begin{figure}
	\begin{centering}
		\psset{unit=1mm,dotsep=1pt,hatchsep=2pt,hatchangle=0}
		\newcommand{\landcake}{
			\psframe[fillstyle=none](0,0)(50,30)
		}
		\begin{pspicture}(60,30)
		\landcake
		\rput(25,35){(a)}
		\psline[linestyle=dashed](25,0)(25,30)
		\psline[linestyle=dashed](0,15)(25,15)
		\psframe[linestyle=dotted,fillstyle=hlines,hatchcolor=blue](0,0)(15,9)
		\psframe[linestyle=dotted,fillstyle=none](0,0)(17.5,10.5)
		\psframe[linestyle=dotted,fillstyle=none](0,0)(20,12)
		\psframe[linestyle=dotted,fillstyle=none](0,0)(22.5,13.5)
		\end{pspicture}
		\caption[Knife-function for a fat rectangle]{
			\label{fig:2-fat-rectangles} 
			A 2-fat rectangle is partitioned into two 2-fat rectangles. One of them is further partitioned to two smaller 2-fat rectangles.
			On the bottom-left one, there is a knife-function with a cover number (relative to the family of 2-fat rectangles) of at most 3.
			The partitions are knife-function are used in the proof of Theorem \ref{thm:2agents}(b).
		}
	\end{centering}
\end{figure}
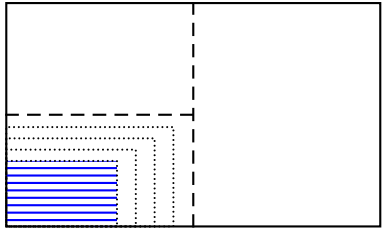

Based on this lemma we can now prove the second part of our first theorem:

\begin{restatetheorem}[Theorem \ref{thm:2agents}(b)]
	For every $R\geq2$: 
	\begin{align*}
	\prope(R\, fat\, rectangle,\,2,\, R\, fat\, rectangles)\geq1/3
	\end{align*}
\end{restatetheorem}
\proof{Proof.}
Let $C$ be an $R$-fat rectangle.
We divide $C$ using Algorithm \ref{alg:2-parts} with $M=3$, in the following way (see Figure \ref{fig:2-fat-rectangles}).

The main partition (input (a)) is a partition of $C$ into two halves in the middle of its longer side.
Since $R\geq 2$, the two halves are $R$-fat too so the total cover number of the partition is $1+1<3$.

The sub-partitions (input (b)) are again partitions of each half to two quarters in the middle of its longer side. 
When a part is replaced by its sub-partition,
the resulting partition contains three $R$-fat rectangles so its total cover number is $3$.

The $S$-smooth knife-functions on the quarters (input (c)) are rectangles growing from the corner towards the center, as in Figure \ref{fig:2-fat-rectangles}.
$K_C$ is a single $R$-fat rectangle and its complement can always be covered by two $R$-fat rectangles, so 
the cover number of these knife-functions is at most $3$.

The increasing $S$-continuous knife-functions (input (d)) are sweeping-lines (see subsection \ref{sub:sweeping-plane} and Figure \ref{fig:knives}/e). 

All input conditions are satisfied so the resulting division is envy-free  and $1/3$-proportional.
\Halmos\endproof
The fraction $1/3$ is tight; see Lemma \ref{lem:negative-Rfat}.

Algorithm \ref{alg:2-parts} can be further refined by adding more sub-partition steps. For example, by adding a third sub-partition step we can prove that if $C$ is an archipelago of $m$ disjoint R-fat rectangles (with $R\geq2$) then:
\begin{align*}
\prope(C,\,2,\, R\textrm{-fat rectangles}) \geq \frac{1}{m+2}
\end{align*}
and this bound is tight. We omit the proof details as the proof is analogous to examples (b) and (c) after Lemma \ref{lem:1-parts}.

\subsection{Dividing fat lands of arbitrary shape.} \label{sub:General-fat-shapes}
Our most general result involves land-estates that are arbitrary Borel sets. The result is proved for any number of dimensions; Figure \ref{fig:prop2-fatshapes} illustrates the proof for $d=2$ dimensions.
\begin{figure}
	\begin{center}
		\psset{unit=0.6mm,dotsep=1pt,hatchsep=2pt,hatchangle=0}
		\newcommand\landcake{
			\pspolygon[fillstyle=none](-5,40)(0,0)(30,0)(35,40)(20,30)
		}
		\begin{pspicture}(50,55)\rput(5,0){
			\landcake
			\rput(15,50){\scriptsize (a)}
			\rput(20,35){C}
			\psframe[linestyle=dashed,linecolor=blue](0,0)(30,30)
			\rput(1,32){$B^{-}$}
			\pscircle[fillstyle=solid,fillcolor=blue](15,15){0.5}
		}\end{pspicture}
		\begin{pspicture}(50,55)\rput(5,0){
			\landcake
			\rput(15,50){\scriptsize (b)}
			\rput(6,40){$C_1$}
			\rput(28,40){$C_2$}
			\psline[linecolor=black,linestyle=solid](15,0)(15,32)
			\psframe[linestyle=dashed,linecolor=blue](1,16)(14,29)
			\psframe[linestyle=dashed,linecolor=blue](16,1)(29,14)
			\psframe[linestyle=dotted](0,0)(14,30)
			\psframe[linestyle=dotted](16,0)(30,30)
		}\end{pspicture}
		\hskip 1.2cm
		\begin{pspicture}(50,55)\rput(5,0){
			\landcake
			\rput(15,50){\scriptsize (c)}
			\psline[linecolor=black,linestyle=solid](15,0)(15,32)
			\psframe[fillstyle=hlines,hatchcolor=blue](0,0)(9,18)
			\psframe[linestyle=dotted](0,0)(11,22)
			\psframe[linestyle=dotted](0,0)(13,26)
			\psframe[linestyle=dotted](0,0)(15,30)
		}\end{pspicture}
		\begin{pspicture}(50,55)\rput(5,0){
			\rput(15,50){\scriptsize (d)}
			\landcake
			\psline[linecolor=black,linestyle=solid](15,0)(15,32)
			\psframe[fillstyle=hlines,hatchcolor=blue](0,0)(15,30)
			\psline[linestyle=dotted](-2,16)(-2,30)(0,30)
			\psline[linestyle=dotted](-4,32)(15,32)
			\psline[linestyle=dotted](-4.25,34)(11,34)
			\psline[linestyle=dotted](-4.5,36)(7,36)
		}\end{pspicture}
	\end{center}
	\caption[]{
		\label{fig:prop2-fatshapes} Dividing a general $R$-fat land-estate between two people.
		\\
		(a) The $R$-fat land $C$ and its largest contained square $B^-$ (the smallest containing square $B^+$ is not shown).
		\\
		(b) The parts $C_{1}$ and $C_{2}$ (solid), the two rectangles $B_1$ and $B_2$ (dotted) and their largest contained squares (dashed).
		\\
		(c) The knife function on $C_1$ in $t\in[0,\frac{1}{2}]$.
		\\
		(d) The knife function on $C_1$ in $t\in[\frac{1}{2},1]$.
	}
\end{figure}

\begin{restatetheorem}[Theorem \ref{thm:2agents}(c)]
	For every $R\geq1$, If $C$ is $R$-fat and $S$ is the family of $2R$-fat pieces then: 
	\begin{align*}
	\prope(C,2,S)=\prop(C,2,S)={1/2}
	\end{align*}
\end{restatetheorem}
\proof{Proof.}
The proof uses Algorithm \ref{alg:1-parts} (the Single Partition Algorithm). 
We show a partition of $C$ into two parts and a knife-function on each part.

Scale, rotate and translate $C$ such that the largest cube contained in $C$ is $B^{-}=[-1,1]^d$ (Figure \ref{fig:prop2-fatshapes}/a). By definition of fatness (see Subsection \ref{sub:Fatness}), $C$ is now contained in a cube $B^{+}$ of side-length at most $2 R$.
	Using the hyperplane $x=0$, bisect the cube $B^{-}$ into two boxes $B_{1} = [-1,0]\times[-1,1]^{d-1}$ and $B_{2} = [0,1]\times[-1,1]^{d-1}$. This hyperplane also bisects $C$ into two parts, $C_1$ and $C_2$ (Figure \ref{fig:prop2-fatshapes}/b). Every $C_j$ contains $B_j$ which contains a cube with a side-length of $1$. Every $C_j$ is of course still contained in $B^{+}$ which is cube with a  side-length of $2R$. Hence every $C_j$ is $2R$-fat. Hence the total cover number of the partition $C=C_1\sqcup C_2$, w.r.t. the family of $2R$-fat objects, is 2, satisfying input condition (a).
	
For every $j\in\{1,2\}$, define the following knife function $K_{C_j}$ on $C_j$ (see Figure \ref{fig:prop2-fatshapes}/c,d):
\begin{itemize}
	\item For $t\in[0,\frac{1}{2}]$, $K_{C_j}(t)=(B_j)^{2t}$, i.e., the box $B_j$ dilated by a factor of $2t$. Hence $K_{C_j}(0)=\emptyset$ and $K_{C_j}(\frac{1}{2})=B_j$.
	\item For $t\in[\frac{1}{2},1]$, $K_{C_j}(t)$ is any knife-function from $B_j$ to $C_j$, for example, the growing-ball function of Appendix \ref{sub:growing-ball}.
\end{itemize}

$K_{C_j}(t)$ is always $2R$-fat, since in $[0,\frac{1}{2}]$ it is a scaled-down version of the box $B_j$ (which is 2-fat) and in $[\frac{1}{2},1]$ it contains $B_j$ and is contained in the cube $B^{+}$. Also $\overline{K_{C_j}(t)}$ is $2R$-fat, since it contains the box $B_{3-j}$ and is contained in $B^{+}$. 
Hence, $K_{C_j}$ is an $S$-smooth knife-function with a cover number of $1+1=2$, satisfying input condition (b).

For the complements we can use, for example, the sweeping-plane knife-function of Appendix \ref{sub:sweeping-plane},
satisfying input condition (c).

By Lemma \ref{lem:1-parts},
Algorithm \ref{alg:1-parts} finds an envy-free and $(1/2)$-proportional division.
\Halmos\endproof

Theorem \ref{thm:2agents}(c) implies that we can satisfy the two main fairness requirements: proportionality and envy-freeness, while keeping the allocated pieces sufficiently fat. The fatness guarantee means that each allotted piece: 
(a) contains a sufficiently \emph{large} square, (b) is contained in a sufficiently \emph{small} square. In the context of land division, these guarantees can be interpreted as follows: (a) Each land-plot has sufficient room for building a large house having a convenient shape (square); (b) The parts of the land that are valuable to the agent are close together, since they are bounded in a sufficiently small square.

Finally we note that a different technique leads to a version of Theorem \ref{thm:2agents}(c) which guarantees that the pieces are not only $2R$-fat but also \emph{convex} (if the original land is convex); hence an agent can walk in a straight line from his square house to his valuable spots without
having to enter or circumvent the neighbor's fields. See Appendix \ref{sec:convex} for details.

\subsection{Between envy-freeness end proportionality.}
For all lands $C$ and families of usable pieces $S$ studied in this section, we proved that there exists a positive constant $p$ such that $\prope(C,2,S)\geq p$ and $\prop (C,2,S)\leq p$. Since $\prope(C,2,S)\leq \prope(C,2,S)$ always, we get that for all settings studied here:
\begin{align*}
\prope(C,2,S)=\prop(C,2,S)
\end{align*}
In other words, in these cases, envy-freeness is compatible with the best possible proportionality guarantee.
It is an open question whether this equality holds for \emph{every} combination of lands $C$ and families $S$.

What \emph{can} we say about the relation between proportionality and envy-freeness for arbitrary $C$ and $S$? In addition to the trivial upper bound $\prope(C,2,S)\leq\prop(C,2,S)$, we have the
following lower bound:
\begin{lemma}
For every $C$ and $S$:
	\begin{align*}
	\prope(C,2,S)\geq p_S\cdot \prop(C,2,S)
	\end{align*}
where $p_S := \inf_{Z\in S}\prope(Z,2,S)$.
\end{lemma}
\proof{Proof.}
Let $p_C=\prop(C,2,S)$. The following meta-algorithm yields an envy-free allocation of $C$ in which the utility of each agent $i$ is at least $p_S\cdot p_C\cdot V_i(C)$.

By the definition of $\prop(C,2,S)$, there exists an allocation of $C$, say $X=(X_1,X_2)$, with a partial-proportionality of at least $p_C$, i.e, for each agent $i$, $V_i^S(X_i) \geq p_C\cdot V_i(C)$.
This means that, for each $i$, the piece $X_i$ contains an \spiece{} $Z_i$ with $V_i(Z_i) \geq p_C \cdot V_i(C)$.

Ask each agent whether he prefers $Z_1$ or $Z_2$ and proceed accordingly.

(a) If each agent $i$ prefers $Z_i$, then the allocation $(Z_1,Z_2)$ is envy-free. Both the value and the utility of each agent $i$ are at least $p_C\cdot V_i(C)$, which is at least $p_S \cdot p_C\cdot V_i(C)$ (since $p_S\leq 1$).
	
(b) If each agent $i$ prefers $Z_{3-i}$, then the allocation $(Z_2,Z_1)$ is envy-free. Both the value and the utility of each agent $i$ are now even more than $p_C\cdot V_i(C)$.
	
(c) The remaining case is that both agents prefer the same piece, say $Z_2$.
So for each agent $i$, $V_i(Z_2)\geq p_C\cdot V_i(C)$.
By the assumptions of the lemma, since $Z_2\in S$, $\prope(Z_2,2,S)\geq p_S$. Therefore, there exists an envy-free allocation of $Z_2$ in which the utility of each agent $i$ is at least $p_S\cdot V_i(Z_2) \geq p_S\cdot p_C\cdot V_i(C)$.
\Halmos\endproof
So by previous results we have the following bounds for \emph{every} $C$:
\begin{itemize}
	\item $\prop(C,\,2,\, Squares)\geq\prope(C,\,2,\, Squares)\geq\frac{1}{4}\prop(C,\,2,\, Squares)$
	\item $\prop(C,\,2,\, R\, fat\, rects)\geq\prope(C,\,2,\, R\, fat\, rects)\geq\frac{1}{3}\prop(C,\,2,\, R\, fat\, rects)$
	(for $R\geq2$)
	\item $\prop(C,\,2,\, Rectangles)\geq\prope(C,\,2,\, Rectangles)\geq\frac{1}{2}\prop(C,\,2,\, Rectangles)$
\end{itemize}

\section{Envy-Free Division for Many Agents}
\label{sec:n-agents}

\subsection{The one-dimensional algorithm.}
\label{sec:n-agents-1d}

\begin{figure}
\begin{center}
\psset{unit=0.6mm,dotsep=1pt,hatchsep=2pt,hatchangle=0}
\psset{xunit=0.7mm,yunit=1mm}
\def\subsimplex{
\pspolygon[fillstyle=none,linecolor=green!60](0,0)(18,0)(9,9)
}
\def\assignment(#1,#2,#3){
\rput(0,0){#1}
\rput(9,9){#2}
\rput(18,0){#3}
}
\def\simplex{
\rput(0,0){\subsimplex}
\rput(18,0){\subsimplex}
\rput(36,0){\subsimplex}
\rput(9,9){\subsimplex}
\rput(27,9){\subsimplex}
\rput(18,18){\subsimplex}
}
\begin{pspicture}(-5,0)(65,40){\rput(0,5){
\simplex
\rput(0,27){(1)}
\rput(0,0){\assignment(A,B,C)}
\rput(18,0){\assignment(C,A,B)}
\rput(36,0){\assignment(B,C,A)}
\rput(9,9){\assignment(B,C,A)}
\rput(27,9){\assignment(A,B,C)}
\rput(18,18){\assignment(C,A,B)}
}}\end{pspicture}
\begin{pspicture}(-5,0)(65,40){\rput(0,5){
\simplex
\rput(0,27){(2,3)}
\rput(0,0){\assignment(A:1,,)}
\rput(18,0){\assignment(C:3,,B:1)}
\rput(36,0){\assignment(,,A:3)}
\rput(9,9){\assignment(B:1,C:2,A:1)}
\rput(27,9){\assignment(A:1,B:2,C:3)}
\rput(18,18){\assignment(,A:2,)}
\rput(36,12){*}
}}\end{pspicture}
\begin{pspicture}(-5,0)(65,40){\rput(0,5){
\rput(0,27){(4)}
\scalebox{.5}{
	\simplex
	\rput(54,0)\simplex
	\rput(27,27)\simplex
	\rput{180}(80,27)\simplex
}
}}\end{pspicture}
\end{center}
\caption[An illustration of the Simmons-Su algorithm.]{
\label{fig:simmons-su} An illustration of the Simmons-Su algorithm for $n=3$ agents, A B and C.
\\
(1) A triangulation of the simplex of partitions in which each vertex is assigned to an agent.\protect 
\\
(2,3) Each vertex is labeled with the index of the piece preferred by its assigned agent. The fully-labeled triangle is starred.
\\
(4) The process is repeated with a finer triangulation of the original simplex.
}
\end{figure}
Existence of envy-free allocations in one dimension was first proved by \citet{Stromquist1980How}. Later, \citet{Su1999Rental} presented an algorithm, attributed to Simmons, for generating an infinite sequence of allocations that converges to an envy-free allocation. 
In this section we generalize their algorithm to handle geometric constraints. We briefly describe the 1-dimensional algorithm below. 

$C$ is a 1-dimensional interval $[0,1]$ and $S$ is the family of intervals. A partition of $C$ to $n$ intervals can be described by a vector of length $n$ whose coordinates are the lengths of the intervals. The sum of all lengths in a partition is 1, so the set of all partitions is equivalent to 
$\Delta^{n-1}$ ---
the standard $(n-1)$-dimensional simplex in $\mathbb{R}^n$. The algorithm proceeds as follows (see Figure \ref{fig:simmons-su}): 

(1) \emph{Preparation}. Triangulate the simplex of partitions to a collection of $(n-1)$-dimensional sub-simplices.  Assign each vertex of the triangulation to one of the $n$ agents, such that in each sub-simplex, all $n$ agents are represented. Su shows that there always exists such a triangulation.

(2) \emph{Evaluation}. For each vertex $v$ of the triangulation, ask its assigned agent: ``if $C$ is partitioned as prescribed by the coordinates of $v$, which piece would you prefer?''. The answer is an integer $j\range{1}{n}$; label that vertex with $j$.

(3) \emph{Selection}. The labeling created in step (2) 
satisfies \emph{Sperner's boundary condition}:
in every face of the simplex, 
every vertex is labeled with one of the labels on the endpoints of that face, since the other labels correspond to empty pieces (See Figure \ref{fig:simmons-su}/b, where the three vertices of the large triangle are labeled by 1, 2 and 3). 
By Sperner's lemma, there exists a \emph{fully-labeled sub-simplex} --- a sub-simplex in which all vertices are labeled differently.

(4) \emph{Refinement}. Steps (1), (2), (3) can be repeated again and again, each time with a finer triangulation. This yields an infinite sequence of fully-labeled simplices. By compactness of the simplex, there is a subsequence that converges to a single point. By the continuity of the agents' valuations, this point corresponds to a partition in which each of the $n$ agents prefers a different piece. By definition, this partition is envy-free.

Note that the above algorithm is infinite --- the envy-free partition is found only at the limit of an infinite sequence. In fact, \citet{Stromquist2008Envyfree} proved that when $n\geq3$, an envy-free partition among $n$ agents with connected pieces cannot be found by a finite algorithm. Therefore, Simmons' infinite algorithm is the best that can be hoped for. 


\subsection{Knife tuples.}
Both Stromquist's existence proof and the Simmons--Su 
algorithm
do not work directly on $C$ --- they work on the unit simplex, each point of which represents a partition of $C$.
Therefore, we can extend these results to multiple dimensions if we find an appropriate way to map each point of the unit simplex to a partition of a multi-dimensional land.

Our main tool is a \emph{knife-tuple} --- a generalization of the  knife-function defined in Definition \ref{def:knife}.
\begin{definition} \label{def:knife-tuple}
An \emph{$n$-knife-tuple on $C$} is a vector of $n$ functions $(K_1,\dots,K_n)$ from $\Delta^{n-1}$ to pieces of $C$, such that:
\begin{itemize}
\item To every $t\in \Delta^{n-1}$, the tuple assigns a partition of $C$, so that $K_1(t)\sqcup \cdots \sqcup K_n(t) = C$.
\item For every $\ell\range{1}{n}$ and every $t\in \Delta^{n-1}$ such that $t_\ell=0$, $K_\ell(t) = \emptyset$.
\item \emph{Continuity}: for every $\epsilon>0$ there is a $\delta>0$ such that  $|t'-t|<\delta$ implies:\\$~~~~~~~~~~\forall \ell\range{1}{n}: \leb{K_\ell(t')\ominus K_\ell(t)} < \epsilon$,
\\
where $|t-t'|$ denotes the distance $\sum_{\ell=1}^n |t_\ell-t'_\ell$.
\end{itemize}
\end{definition}

The definitions of $S$-continuity and $S$-smoothness can be easily generalized from knife-functions to knife-tuples. For the present paper it is sufficient to generalize $S$-smoothness (definition \ref{def:smooth}):
\begin{definition}
\label{def:smooth-tuple}
A knife-tuple $(K_1,\ldots,K_n)$ is $S$-smooth if, for every $\ell\range{1}{n}$, 
the function $K_\ell$ has a finite $S$-cover $K_\ell^1\cup\cdots \cup K_\ell^{m_\ell}$ for some integer $m_\ell\geq 1$.

If a knife-tuple $K$ is $S$-smooth, then we define its \emph{cover number}, $\kcovernum (K,S)$, as 
the smallest sum $\sum_{\ell=1}^n{m_\ell}$ of integers that satisfy the above definition.
\end{definition}

Given an $S$-smooth knife-tuple $(K_1,\ldots,K_n)$
and an absolutely-continuous measure $V$,
define the function $V^{K_\ell}$ (analogously to \eqref{eq:vk}) as the maximum value of an \spiece{} in the given cover of $K_\ell(t)$:
\begin{align*}
\forall \ell\range{1}{m}:&&
V^{K_\ell}(t) := \max_{j=1}^{m_\ell} V(K_\ell^j(t)).
\end{align*}

Lemma \ref{lem:smooth} can be generalized as follows:
\begin{lemma}
\label{lem:smooth-n}
If $(K_1,\ldots,K_n)$ is an $S$-smooth knife-tuple
and $V$ is an absolutely-continuous measure,
then for every $\ell \range{1}{n}$,
$V^{K_\ell}$ is a uniformly-continuous real function.
\end{lemma}
The proof is the same --- the maximum of uniformly-continuous functions is uniformly-continuous.

\subsection{Constructing knife-tuples.}
Knife-tuples can be constructed from knife-functions.
\begin{lemma}
\label{lem:2-tuple}
Let $K$ be a knife-function from $\emptyset$ to $C$, and let $K_1,K_2$ be the following functions from $\Delta^1$ to pieces of $C$:
\begin{align*}
K_1(t_1,t_2) &:= K(t_1)
\\
K_2(t_1,t_2) &:= C\setminus K(1-t_2)
\end{align*}
Then, $(K_1,K_2)$ is a 2-knife-tuple on $C$.

Moreover, if the knife-function $K_C$ is $S$-smooth and its cover-number is $M$,
then the knife-tuple $(K_1,K_2)$ is $S$-smooth and its cover-number is $M$ too.
\end{lemma}
\proof{Proof.} 
We prove that $(K_1,K_2)$ satisfies the conditions of Definition \ref{def:knife-tuple}.
\begin{itemize}
\item For every $t\in \Delta^1$, $1 - t_2 = t_1$. Hence, 
$K_2(t) = C \setminus K(1 - t_2) = C \setminus K(t_1) = C\setminus K_1(t)$.
Hence $K_1(t),K_2(t)$ is indeed a partition of $C$.
\item If $t_1=0$ then $K_1(t) = K(0) = \emptyset$;
if $t_2=0$ then $K_2(t) = C\setminus K(1) = C\setminus C = \emptyset$.
\item For every $\epsilon>0$, 
since $K$ is a knife-function,
there exists a $\delta>0$ such that $|t_1-t'_1|<\delta$ implies $\leb{K(t_1)\ominus K(t'_1)}<\epsilon$.
For every $t, t'\in \Delta^1$, $|t-t'| = |t_1-t'_1| + |t_2-t'_2| = |t_1-t'_1| + |(1-t_1)-(1-t'_1)| = 2|t_1-t'_1|$.
If $|t-t'|<\delta$ then $|t_1-t'_1|<\delta/2 < \delta$, so 
$\leb{K_1(t)\ominus K_1(t')}<\epsilon$ and $\leb{K_2(t)\ominus K_2(t')}<\epsilon$.
\end{itemize}

Now, if $K$ is $S$-smooth, then $K$ can be covered by $m_1$ $S$-knife-functions and $\overline{K}$ can be covered by $m_2$ $S$-knife-functions (for some integers $m_1,m_2\geq 1$), so $K_1$ and $K_2$ can be covered by the same $S$-knife-functions, so $(K_1,K_2)$ is $S$-smooth with the same cover-number.
\Halmos\endproof

For example, from the knife-function in Figure \ref{fig:knives}/b --- a growing pair of squares --- we get the following 2-knife-tuple:
\begin{align}
\label{eq:k1k2}
K_1(t) &= [0,t_1]\times[0,t_1] ~\cup~ [1-t_1,1]\times[1-t_1,1]
\\
\notag
K_2(t) &= [0,t_2]\times[1-t_2,1] ~\cup~ [1-t_2,1]\times[0,t_2]
\end{align}
It is square-smooth with a square-cover-number of 4.

Longer knife-tuples can be constructed by splitting existing knife-tuples.
Let $(K_1,\dots,K_{n})$ be an $n$-knife-tuple on $C$. 
Suppose that for some $\ell\range{1}{n}$, for every $t\in\Delta^{n-1}$ for which $t_\ell>0$, we have a knife-function $K^t$ from $\emptyset$ to $K_\ell(t)$. 

We create a new tuple by replacing the index $\ell$ with two indices $\ell1$ and $\ell 2$ and replacing the function 
$K_\ell$ with two complementary functions $K_{\ell1}'$ and $K_{\ell2}'$, split by the knife-function $K^t$. This gives a new vector of $n+1$ functions  $(K_1',\dots,K_{\ell1}',K_{\ell2}',\dots,K_n')$ from $\Delta^n$ to partitions of $C$:
\begin{align*}
K'^{\ell1}(t_1,\dots,t_{\ell1},t_{\ell2},\dots,t_n) & := 
\begin{cases}
K^{t_1,\dots,t_{\ell1}+t_{\ell2},\dots,t_{n}}
\bigg(\frac{t_{\ell1}}{t_{\ell1}+t_{\ell2}}\bigg) & [t_{\ell1}+t_{\ell2}>0]
\\
\emptyset & [t_{\ell1}=t_{\ell2}=0]
\end{cases}
\\
K'^{\ell2}(t_1,\dots,t_{\ell1},t_{\ell2},\dots,t_n) & := 
\begin{cases}
K_\ell(t_1,\dots,t_{\ell1}+t_{\ell2},\dots,t_{n})\setminus &
\\
~~~~~~K^{t_1,\dots,t_{\ell1}+t_{\ell2},\dots,t_{n}}
\bigg(\frac{t_{\ell1}}{t_{\ell1}+t_{\ell2}}\bigg) & [t_{\ell1}+t_{\ell2}>0]
\\
\emptyset & [t_{\ell1}=t_{\ell2}=0]
\end{cases}
\\
\forall j\neq \ell:
~~~
K'^j(t_1,\dots,t_{\ell1},t_{\ell2},\dots,t_n) & := 
K_j(t_1,\dots,t_{\ell1}+t_{\ell2},\dots,t_n)
\end{align*}
It is easy to see that, 
to every $t\in \Delta^n$, 
the new tuple indeed assigns a partition of $C$, and that 
whenever one of the new $n+1$ time variables is $0$ the corresponding part of the knife-tuple returns an empty set. The continuity or smoothness of the new knife-tuple does not follow automatically from the construction --- it has to be verified separately.

As an example we apply the above construction with $n=2$, the 2-knife-tuple $(K_1,K_2)$ of \eqref{eq:k1k2}, and $\ell=2$.

For every $(t_1,t_2)\in\Delta^{1}$, we have to define a knife-function $K^{t_1,t_2}$ from $\emptyset$ to $K_2(t_1,t_2)$. 
Recall that $K_2(t_1,t_2)$ is a union of two squares.
For each such square, we create a square-pair knife-function
analogous to Figure \ref{fig:knives}(b) --- two squares growing from opposite corners. 
We define $K^{t_1,t_2}$ as the union of these two square-pairs; see Figure \ref{fig:Knife-square-squares}. 
The above construction yields the following three functions:
\begin{align*}
K_1'(t_1,t_2,t_3) &= K_1(t_1,t_2+t_3) = [0,t_1]\times[0,t_1] ~\cup~ [1-t_1,1]\times[1-t_1,1]
\\
K_2'(t_1,t_2,t_3) &= 
[0,t_2]\times [1-t_2-t_3,1-t_3]
~\cup~
[t_3,t_2+t_3]\times[1-t_2,1]
\\
&~\cup~
[1-t_2-t_3,1-t_3]\times [0,t_2]
~\cup~
[1-t_2,1]\times[t_3,t_2+t_3]
\\
K_3'(t_1,t_2,t_3) &= 
[0,t_3]\times [1-t_3,1]
~\cup~
[t_2,t_2+t_3]\times[1-t_2-t_3,1-t_2]
\\
&~\cup~
[1-t_2-t_3,1-t_2]\times [t_2,t_2+t_3]
~\cup~
[1-t_3,1]\times[0,t_3]
\end{align*}
It can be verified that $(K_1',K_2',K_3')$ is
indeed a 3-knife-tuple: for every $t\in\Delta^2$ it returns a partition of $C$,
whenever $t_\ell=0$ the corresponding $K'_\ell$ is empty, and for every $\ell\in\{1,2,3\}$, the function $K_\ell(t)$ satisfies the continuity requirement of Definition \ref{def:knife-tuple} (since it is a union of squares whose boundaries are continuous functions of $t$).

Moreover, $K_1'$ is covered by two square-knife-functions while each of $K_2',K_3'$ is covered by four square-knife-functions, so $(K_1',K_2',K_3')$ is square-smooth and its cover number is $2+4+4=10$.

In exactly the same manner, we can replace $K'_1$ --- the growing square-pair --- with two growing square-quadruplets. This yields a new 4-knife-tuple that is square-smooth with a cover-number of $4+4+4+4=16$.

\subsection{Land division using knife-tuples}
Using knife-tuples, Lemma \ref{lem:0-parts} can be generalized as follows:
\begin{lemma}
\label{lem:nagents} Let $C$ be a land and $S$ a family of pieces. If there is an $S$-smooth $n$-knife-tuple on $C$ with a cover number of at most $M$, then there exists an envy-free and $1/M$-proportional division of $C$ among the $n$ agents.
\end{lemma}

\proof{Proof.}
The proof generalizes the infinite algorithm of Simmons-Su (Subsection \ref{sec:n-agents-1d}).\footnote{
When $n=3$, the three-knives algorithm of \citet{Stromquist1980How} can be used instead of Simmons' algorithm. See the conference version \citep{SegalHalevi2015EnvyFree} for details.%
}

(1) The \emph{preparation} step is entirely the same: triangulate the standard simplex $\Delta^{n-1}$ and assign each triangulation vertex to an agent such that in each sub-simplex, all agents are represented.

(2) The \emph{evaluation} step is different:
for each vertex $t = (t_1,\dots,t_n)$ of the triangulation, a partition of $C$ is defined by the given $n$-knife-tuple: $K_1(t),\dots,K_n(t)$.
Each part $K_\ell(t)$ in this partition is covered by $m_\ell$ \spieces{}.
Ask the owner of vertex $t$ (e.g. agent $i$) to calculate, for each $\ell\range{1}{n}$, the value of $V_i^{K_\ell}(t)$, i.e, the most valuable \spiece{} 
from the set of $m_\ell$ \spieces{} covering $K_\ell(t)$.
Then, find the maximum of these $n$ maxima:
\begin{align*}
\arg\max_{\ell\range{1}{n}} V_j^{K_\ell}(t),
\end{align*}
and label the vertex $t$ with the result.

(3) By the definition of a knife-tuple, whenever $t_\ell=0$, $K_\ell(t)=\emptyset$, so for every agent $i$, $V_i^{K_\ell}(t)=0$. Hence, any agent asked to label vertex $t$, will never label it with $\ell$. 
Hence, the resulting labeling satisfies Sperner's labeling condition, so in the \emph{selection} step, a fully-labeled sub-simplex exists.

(4) By repeating steps (1), (2), (3) infinitely many times with finer and finer triangulations, we get a 
sequence of smaller and smaller fully-labeled simplices.
This sequence has a subsequence that converges to a single point $t^*$. 
Because the knife-tuple is $S$-smooth, 
by Lemma \ref{lem:smooth-n},
all agents' utilities are continuous functions of $t$. Therefore, 
in the partition corresponding to the limit point, $K_1(t^*),\ldots,K_n(t^*)$, 
each agent is allocated a different \spiece{} with a maximum value.

The cover number of the knife-tuple is at most $M$. Therefore, 
by the Allocation Lemma (\ref{cor:chooser}), 
this allocation is envy-free and $1/M$-proportional.
\Halmos\endproof

\subsection{Dividing squares and rectangles.}
We apply Lemma \ref{lem:nagents} to prove our second theorem.

\begin{figure}
\psset{unit=0.65mm,dotsep=1pt,hatchsep=2pt,hatchangle=0} \def\landcake{\psframe[fillstyle=none](0,0)(40,40)}
\begin{center}
\begin{pspicture}(-10,0)(50,50) \landcake
\rput(20,45){\small $t_1=0.3, t_2=0.1, t_3=0.6$} \psframe[linestyle=dotted,fillstyle=hlines,hatchcolor=blue](0,0)(13,13) \psframe[linestyle=dotted,fillstyle=vlines,hatchcolor=green](0,13)(5,18) \psframe[linestyle=dotted,fillstyle=vlines,hatchcolor=green](13,0)(18,5) \psframe[linestyle=dotted,fillstyle=hlines,hatchcolor=blue](27,27)(40,40) \psframe[linestyle=dotted,fillstyle=vlines,hatchcolor=green](27,40)(22,35) \psframe[linestyle=dotted,fillstyle=vlines,hatchcolor=green](40,27)(35,22) 
\end{pspicture}
\begin{pspicture}(-10,0)(50,50) \landcake
\rput(20,45){\small $t_1=0.3, t_2=0.6, t_3=0.1$} \psframe[linestyle=dotted,fillstyle=hlines,hatchcolor=blue](0,0)(13,13) \psframe[linestyle=dotted,fillstyle=vlines,hatchcolor=green](0,13)(20,33) \psframe[linestyle=dotted,fillstyle=vlines,hatchcolor=green](13,0)(33,20) \psframe[linestyle=dotted,fillstyle=hlines,hatchcolor=blue](27,27)(40,40) \psframe[linestyle=dotted,fillstyle=vlines,hatchcolor=green](27,40)(7,20) \psframe[linestyle=dotted,fillstyle=vlines,hatchcolor=green](40,27)(20,7) 
\end{pspicture}
\begin{pspicture}(-10,0)(50,50) \landcake
\rput(20,45){\small $t_1=0.6, t_2=0.1, t_3=0.3$} \psframe[linestyle=dotted,fillstyle=hlines,hatchcolor=blue](0,0)(23,23) \psframe[linestyle=dotted,fillstyle=vlines,hatchcolor=green](0,23)(5,28) \psframe[linestyle=dotted,fillstyle=vlines,hatchcolor=green](23,0)(28,5) \psframe[linestyle=dotted,fillstyle=hlines,hatchcolor=blue](17,17)(40,40) \psframe[linestyle=dotted,fillstyle=vlines,hatchcolor=green](17,40)(12,35) \psframe[linestyle=dotted,fillstyle=vlines,hatchcolor=green](40,17)(35,12) 
\end{pspicture}
\begin{pspicture}(-10,0)(50,50) \landcake
\rput(20,45){\small $t_1=0.6, t_2=0.3, t_3=0.1$} \psframe[linestyle=dotted,fillstyle=hlines,hatchcolor=blue](0,0)(23,23) \psframe[linestyle=dotted,fillstyle=vlines,hatchcolor=green](0,23)(10,33) \psframe[linestyle=dotted,fillstyle=vlines,hatchcolor=green](23,0)(33,10) \psframe[linestyle=dotted,fillstyle=hlines,hatchcolor=blue](17,17)(40,40) \psframe[linestyle=dotted,fillstyle=vlines,hatchcolor=green](17,40)(7,30) \psframe[linestyle=dotted,fillstyle=vlines,hatchcolor=green](40,17)(30,7) 
\end{pspicture}
\end{center}
\caption[]{
\label{fig:Knife-square-squares}
A 3-knife-tuple on a square cake.
Four partitions induced by the 3-knife-tuple $(K_1',K_2',K_3')$ in different points $(t_1,t_2,t_3)$ of the unit-simplex. $K_1'(\cdot)$ is filled with horizontal blue lines, $K_2'(\cdot)$ is filled with vertical green lines and $K_3'(\cdot)$ is blank. 
}
\end{figure}
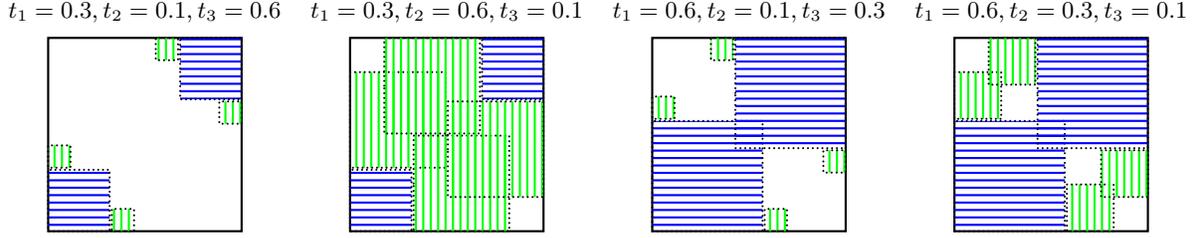

\begin{restatetheorem}[Theorem \ref{thm:nagents}(a)]
For every $n\geq1$:
\begin{align*}
\prope(Square,n,Squares)\geq\frac{1}{2^{2\lceil log_{2}n\rceil}} > \frac{1}{4 n^{2}}
\end{align*}
\end{restatetheorem}
\proof{Proof.}
We first consider the case in which $n$ is a power of $2$.
We construct an $n$-knife-tuple $(K_1,\dots,K_n)$, in which for every $t\in \Delta^{n-1}$, and for every $\ell \range{1}{n}$, $K_\ell(t)$ is a union of at most $n$ squares. Hence, the cover number of the $n$-knife-tuple $(K_1,\dots,K_n)$ is at most $n\cdot n = n^2$.

The construction is recursive. The base is $n=2$. Take the knife-function in Figure \ref{fig:knives}/b --- a union of two corner-squares growing towards the center.
By Lemma \ref{lem:2-tuple}, this knife-function defines a 2-knife-tuple which we denote by $(K_1,K_2)$. For each $t_1$ and $t_2$, $K_1(t_1,t_2)$ and $K_2(t_1,t_2)$ are square-pairs --- unions of 2 squares.

Consider next the case $n=4$. 
In every square-pair in the above 2-knife-tuple, define a knife-function as shown in Figure \ref{fig:Knife-square-squares} --- a union of four squares growing from opposite corners towards the center. This yields a 4-knife-tuple $(K_1',K_2',K_3',K_4')$.  For each $\ell \in\{1,2,3,4\}$ and for each $t\in \Delta^3$,  $K'_\ell(t)$ is a union of four squares.

After $k$ steps, we have a $2^k$-knife-tuple in which each component is a union of $2^k$ squares. We split each component using a knife-function made of a union of $2^{k}$ square-pairs --- a square-pair for each of the $2^k$ squares in the cover.\footnote{
In each square-pair, the two squares should grow from opposite corners of the square in the cover. It does not matter which pair of opposite corners is used.
} This gives a new, $2^{k+1}$-knife-tuple in which each component is a union of $2^{k+1}$ squares. After $\log_2{n}$ steps, we get an $n$-knife-tuple.
It is square-smooth since each element of the tuple can be covered by $n$ square-knife-functions --- $n$ squares changing continuously with $t$. 
Its square-cover-number is therefore at most $n^2$.

Applying Lemma \ref{lem:nagents} with this knife-tuple implies that there exists an envy-free and $(1/n^2)$-proportional division of $C$ among the $n$ agents.

Now, suppose $n$ is not a power of two.
Define $n' = 2^{\lceil log_{2}n\rceil} = $the smallest power of two larger than $n$. 
Add $n'-n$ dummy agents and apply the proof of the first case.
There exists an envy-free and $(1/n'^2)$-proportional division of $C$ among the $n'$ agents.
Since free disposal is assumed, the pieces allocated to the $n'-n$ dummy agents can be discarded. We remain with an envy-free division with a proportionality of at least 
$1/2^{2\lceil log_{2}n\rceil} > 1 / 4n^2$.
\Halmos\endproof

The second part of the second theorem is a simple corollary of the first part:

\begin{restatetheorem}[Theorem \ref{thm:nagents}(b)]
If $C$ is an $R$-fat rectangle and $S$ the family of $R$-fat rectangles then:
\begin{align*}
\prope(C,n,S)\geq\frac{1}{2^{2\lceil log_{2}n\rceil}} > \frac{1}{4 n^{2}}
\end{align*}
\end{restatetheorem}
\proof{Proof.}
Rescale the axes such that $C$ becomes a square. 
By Theorem \ref{thm:nagents}(a), there exists an allocation in which each agent $i$ receives a piece that contains a square $Z_i$ with a value of at least $V_i(C)/2^{2\lceil log_{2}n\rceil}$.
Rescale the axes back. Now, each $Z_i$ is an $R$-fat rectangle.
\Halmos\endproof

We do not know if the $1/(4n^{2})$ lower bound is asymptotically tight. 
The best upper bound currently known \citep{SegalHalevi2017Fair} is  $\prop(Square,\allowbreak n,\allowbreak squares)\leq1/(2n)$. Moreover, there is an algorithm for non-envy-free division that proves $\prop(Square,\allowbreak n,\allowbreak squares)\geq1/(4n-4)$. We do not know if it is possible to attain an envy-free division with a proportionality of $1/O(n)$. 

\subsection{Dividing fat objects of arbitrary shape.}
In this subsection we show that it is possible to attain an envy-free \emph{and proportional} division for every $n$, in return to a compromise on the fatness of the pieces.
\begin{figure}
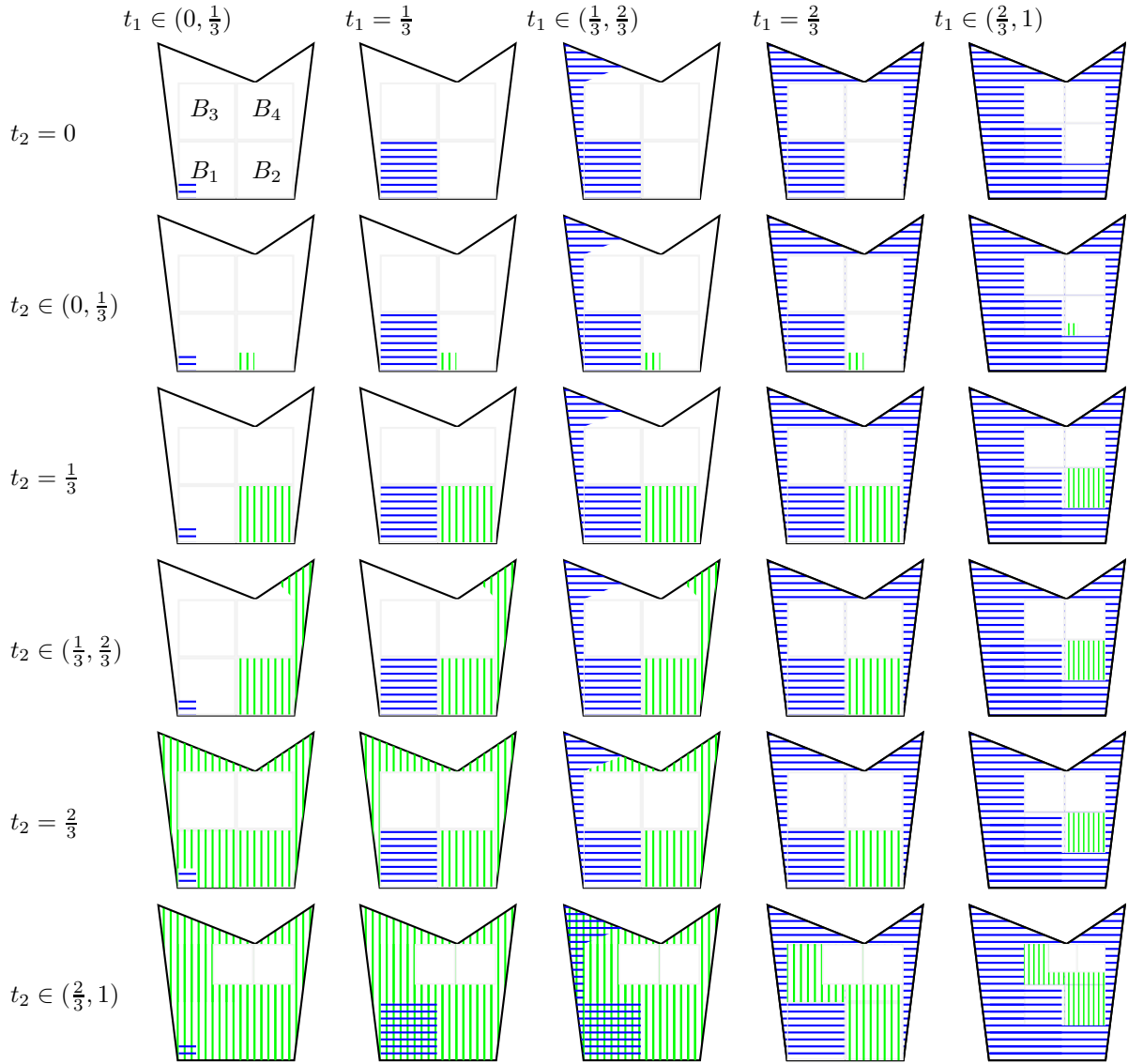

\psset{unit=0.55mm,dotsep=1pt,hatchsep=2pt,hatchangle=0,linestyle=none}
\newcommand\landcake[1]{\pspolygon[#1](-5,40)(0,0)(30,0)(35,40)(20,30)}
\newcommand\landcakeempty{\landcake{linestyle=solid,fillstyle=none}}
\newcommand\landcakefilled{\landcake{fillstyle=solid,fillcolor=white}\landcake{linestyle=solid,fillstyle=hlines,hatchcolor=blue}}
\newcommand\landcakefilledb{\landcake{fillstyle=solid,fillcolor=white}\landcake{linestyle=solid,fillstyle=vlines,hatchcolor=green}}
\newcommand\landcakesquares{
\psframe[fillstyle=solid,fillcolor=white,linestyle=solid,linecolor=gray!10](0,0)(15,15)
\psframe[fillstyle=solid,fillcolor=white,linestyle=solid,linecolor=gray!10](0,15)(15,30)
\psframe[fillstyle=solid,fillcolor=white,linestyle=solid,linecolor=gray!10](15,0)(30,15)
\psframe[fillstyle=solid,fillcolor=white,linestyle=solid,linecolor=gray!10](15,15)(30,30)
}
\newcommand\landcakesquarestop{
\psframe[fillstyle=solid,fillcolor=white,linestyle=solid,linecolor=gray!10](0,15)(15,30)
\psframe[fillstyle=solid,fillcolor=white,linestyle=solid,linecolor=gray!10](15,15)(30,30)
}
\newcommand\stagea{\psframe[fillstyle=hlines,hatchcolor=blue](0,0)(5,5)}
\newcommand\stageb{\psframe[fillstyle=hlines,hatchcolor=blue](0,0)(15,15)}
\newcommand\stagec{\psframe[fillstyle=hlines,hatchcolor=blue](0,0)(15,15)\pspolygon[fillstyle=hlines,hatchcolor=blue](0,0)(-5,40)(10,34)(0,30)}
\newcommand\staged{\landcakefilled \landcakesquares \psframe[fillstyle=hlines,hatchcolor=blue](0,0)(15,15)}
\newcommand\stagee{\landcakefilled\rput(9,9){\scalebox{0.7}{\landcakesquares}}\psframe[fillstyle=hlines,hatchcolor=blue](0,0)(19,19)}

\newcommand\stepa{\psframe[fillstyle=vlines,hatchcolor=green](15,0)(20,5)}
\newcommand\stepb{\psframe[fillstyle=vlines,hatchcolor=green](15,0)(30,15)}
\newcommand\stepc{\psframe[fillstyle=vlines,hatchcolor=green](15,0)(30,15)\pspolygon[fillstyle=vlines,hatchcolor=green](30,0)(35,40)(26,34)(30,30)}
\newcommand\stepe{\pspolygon[fillstyle=vlines,hatchcolor=green](15,0)(30,0)(30,30)(0,30)(0,15)(15,15) \rput(9,9){\scalebox{0.7}{\landcakesquarestop}}}

\newcommand\rowheading[1]{\rput[Bl](0,15){#1}}

\begin{psmatrix}[colsep=0.8cm,rowsep=0.8cm]
&&
$t_1 \in (0,\frac{1}{3})$ && 
$t_1 = \frac{1}{3}$ && 
$t_1 \in(\frac{1}{3},\frac{2}{3})$ && 
$t_1 = \frac{2}{3}$ && 
$t_1 \in (\frac{2}{3},1)$ 
\\ \\

\rowheading{$t_2 = 0$} &&
\landcakeempty\landcakesquares\stagea 
\rput(7,7){$B_1$} 
\rput(23,7){$B_2$} 
\rput(7,23){$B_3$} 
\rput(23,23){$B_4$} 
&& 
\landcakeempty\landcakesquares\stageb && 
\landcakeempty\landcakesquares\stagec && 
\landcakeempty\landcakesquares\staged && 
\landcakeempty\landcakesquares\stagee &&
\\ \\

\rowheading{$t_2 \in (0,\frac{1}{3})$} &&
\landcakeempty\landcakesquares\stagea \stepa && 
\landcakeempty\landcakesquares\stageb \stepa && 
\landcakeempty\landcakesquares\stagec \stepa && 
\landcakeempty\landcakesquares\staged \stepa && 
\landcakeempty\landcakesquares\stagee \rput(9,9){\scalebox{0.7}{\stepa}}
\\ \\

\rowheading{$t_2 = \frac{1}{3}$} &&
\landcakeempty\landcakesquares\stagea \stepb && 
\landcakeempty\landcakesquares\stageb \stepb && 
\landcakeempty\landcakesquares\stagec \stepb && 
\landcakeempty\landcakesquares\staged \stepb && 
\landcakeempty\landcakesquares\stagee \rput(9,9){\scalebox{0.7}{\stepb}}
\\ \\

\rowheading{$t_2 \in(\frac{1}{3},\frac{2}{3})$} &&
\landcakeempty\landcakesquares\stagea \stepc && 
\landcakeempty\landcakesquares\stageb \stepc && 
\landcakeempty\landcakesquares\stagec \stepc && 
\landcakeempty\landcakesquares\staged \stepb && 
\landcakeempty\landcakesquares\stagee \rput(9,9){\scalebox{0.7}{\stepb}}
\\ \\

\rowheading{$t_2 = \frac{2}{3}$} &&
\landcakeempty\landcakesquares\stagea \psframe[fillstyle=vlines,hatchcolor=green](15,0)(30,15)\pspolygon[fillstyle=vlines,hatchcolor=green](30,0)(35,40)(20,30)(-5,40)(0,0)(0,5)(5,5)(5,0)(15,0)(15,15)(0,15)(0,30)(30,30) && 
\landcakeempty\landcakesquares\stageb \psframe[fillstyle=vlines,hatchcolor=green](15,0)(30,15)\pspolygon[fillstyle=vlines,hatchcolor=green](30,0)(35,40)(20,30)(-5,40)(0,0)(0,30)(30,30) && 
\landcakeempty\landcakesquares\stagec \psframe[fillstyle=vlines,hatchcolor=green](15,0)(30,15)\pspolygon[fillstyle=vlines,hatchcolor=green](30,0)(35,40)(20,30)(10,34)(0,30)(30,30)&& 
\landcakeempty\landcakesquares\staged \stepb && 
\landcakeempty\landcakesquares\stagee \rput(9,9){\scalebox{0.7}{\stepb}}
\\ \\

\rowheading{$t_2 \in(\frac{2}{3},1)$} &&
\landcakefilledb \stagea \stepe&& 
\landcakefilledb \stageb \stepe&& 
\landcakefilledb \stagec \stepe&& 
\landcakeempty\landcakesquares\staged \stepe && 
\landcakefilledb \stagee \rput(9,9){\scalebox{0.7}{\stepe}}
\\ \\

\end{psmatrix}

\caption[Dividing a general $R$-fat cake to 3 people.]{
\label{fig:propn-fatshapes}
Dividing a general $R$-fat cake to $n=3$ people. 
$K_1$ is filled with horizontal
lines, $K_2$ is filled with vertical lines and $K_3$ is white. Note that each of these three pieces is $2R$-fat, where $R$ is the fatness of the original cake.
}
\end{figure}

\begin{restatetheorem}[Theorem \ref{thm:nagents}(c)]
\label{cor:fat-fat}
Let $C$ be a $d$-dimensional $R$-fat land and $n\geq 2$ an integer. Let $S$ be the family of $mR$-fat pieces, where $m$ be the smallest integer such that $n\leq m^{d}$ (i.e. $m=\lceil n^{1/d}\rceil$).
Then:
\begin{align*}
\prope(C,n,S)={1 / n}
\end{align*}
\end{restatetheorem}

\proof{Proof.}
The proof is illustrated in Figure \ref{fig:propn-fatshapes} for the case of $d=2$ dimensions. Let $C$ be an $R$-fat $d$-dimensional land. By definition of fatness it contains a cube $B^{-}$ of side-length $x$ and it is contained in a parallel cube $B^{+}$ of side-length $R\cdot x$, for some $x>0$. 

Partition the cube $B^{-}$ to a grid of $m^d$ sub-cubes, $B_{1},...,B_{m^{d}}$, each of side-length $\frac{x}{m}$. For every $i\range{1}{n-1}$, denote by $B_{>i}$ the union of the $m^{d}-i$ squares with indices larger than $i$, i.e:
\begin{align*}
B_{>i} :=  \bigcup_{j > i} B_j
\end{align*}
Denote by $\overline{B^{-}}$ the land outside the enclosed cube, i.e:
\begin{align*}
\overline{B^{-}} := C\setminus B^{-}
\end{align*}
Define the following knife function $K$ from $\emptyset$ to $C$:
\begin{itemize}
\item For $t\in[0,\frac{1}{3}]$: $K(t)=(B_1)^{3t}$, i.e., the cube $B_1$ dilated by a factor of $3t$. Hence $K(0)=\emptyset$ and $K(\frac{1}{3})=B_1$.
\item For $t\in[\frac{1}{3},\frac{2}{3}]$: $K(t)$ 
is 
any knife-function from $B_1$ to $C\setminus B_{>1}$, such as the growing-ball function of Appendix \ref{sub:growing-ball}.
\begin{align*}
\end{align*}
\item For $t\in[\frac{2}{3},1]$: $K(t)$ is $C\setminus[(B_{>1}){}^{3(1-t)}]$,
i.e., the land not yet covered by the knife is $B_{>1}$ dilated by a factor proportional to the remaining time. Hence $K(1)=C$.
\end{itemize}

By Lemma \ref{lem:2-tuple}, $K$ induces a 2-knife-tuple $(K_1,K_2)$. 
For every $t_1,t_2$ with $t_1+t_2=1$, $K_1(t_1,t_2)$ is $mR$-fat:
\begin{itemize}
\item When $t_1\in [0,\frac{1}{3}]$, $K_1$ it is a cube, which is 1-fat. 
\item When $t_1\in [\frac{1}{3},1]$, $K_1$ contains the cube $B_1$, whose side-length is $x/m$, and is contained in the cube $B^{+}$, whose side-length is $x\cdot R$.
\end{itemize}
$K_2(t_1,t_2)$ is $mR$-fat too:
\begin{itemize}
\item When $t_1\in [0,\frac{2}{3}]$, $K_2$ contains e.g. the cube $B_n$, whose side-length is $x/m$, and is contained in the larger cube $B^{+}$, whose side-length is $x\cdot R$.
\item When $t_1\in [\frac{2}{3},1]$, $K_2$ contains a dilated $B_n$ and it is contained in a dilated $B^-$; since they are dilated by the same factor, the ratio between their side-lengths remains $m$, so $K_2(t)$ is $m$-fat.
\end{itemize}
Therefore, $(K_1,K_2)$ is an $S$-smooth 2-knife-tuple with a cover number of 2.

For every $t_1,t_2$ with $t_1+t_2=1$, we now define a knife-function from $\emptyset$ to $K_2(t_1,t_2)$. $K^{t_1,t_2}$ is analogous to $K$ but uses the sub-cube $B_2$. This is possible because:
\begin{itemize}
\item When $t_1\in [0,\frac{2}{3}]$, $K_2$ contains the cube $B_2$ itself;
\item When $t_1\in (\frac{2}{3},1]$, $K_2$ contains a dilated $B_2$, which is contained in a dilated $B^{-}$.
\end{itemize}
The function $K^{t_1,t_2}$ is defined as follows:
\begin{itemize}
\item For $t\in[0,\frac{1}{3}]$: $K^{t_1,t_2}(t)=(B_2)^{3t}$.
\item For $t\in[\frac{1}{3},\frac{2}{3}]$: $K^{t_1,t_2}(t)$ is any knife-function from $B_2$ to $K_2(t_1,t_2)\setminus B_{>2}$ (e.g. the growing-ball function of Subsection \ref{sub:growing-ball}).
\item For $t\in[\frac{2}{3},1]$: $K^{t_1,t_2}(t)$ is $K_2(t_1,t_2)\setminus[(B_{>2}){}^{3(1-t)}]$.
\end{itemize}
This process induces a 3-knife-tuple $(K_1',K_2',K_3')$; see Figure \ref{fig:propn-fatshapes}.

To define an $n$-knife-tuple, proceed in a similar way for the pieces $B_3,\ldots,B_n$. All components in the knife-tuple are $mR$-fat. Therefore, the knife-tuple is $S$-smooth
with a cover-number of $n$.
By Lemma \ref{lem:nagents},
there is an envy-free and $1/n$-proportional division of $C$ with $mR$-fat pieces.
\Halmos\endproof
Figure \ref{fig:propn-fatshapes} shows an example of the construction for $d=2$ dimensions and $n=3$ agents. Here $m=\lceil\sqrt{3}\rceil=2$ so each agent receives an envy-free $2R$-fat land-plot with a utility of at least
$1/3$.

Theorem \ref{thm:nagents}(c) implies that we can guarantee the highest possible level of proportionality ($1/n$) by compromising on the fatness of the pieces --- allowing the pieces to be thinner than the original land by a factor of $\lceil n^{1/d}\rceil$. This factor is asymptotically optimal even when envy is allowed:

\begin{figure}
\psset{unit=3cm}
\begin{pspicture}(0,-.5)(5,1.5)
\pspolygon[fillstyle=solid,fillcolor=black!20](0,1)(0,0)(5,0)(5,.2)(1,.2)(1,1)(0,1)
\psframe[fillstyle=solid,fillcolor=blue](.02,.02)(.98,.98)
\pscircle[fillstyle=solid,fillcolor=blue](4.9,.1){.1}

\rput(.5,1.1){Length = 1}
\rput(3,  .3){Length = $R$}

\rput[l](0,-.1){\color{blue} $V=n-1+\epsilon$}
\rput[r](5, -.1){\color{blue} {$V=1-\epsilon$}}
\end{pspicture}
\caption[]{
\label{fig:negative-non-convex}
A fat land-estate in which every proportional division must use slim pieces.
See Lemma \ref{lem:negative-example-non-convex}.
}
\end{figure}
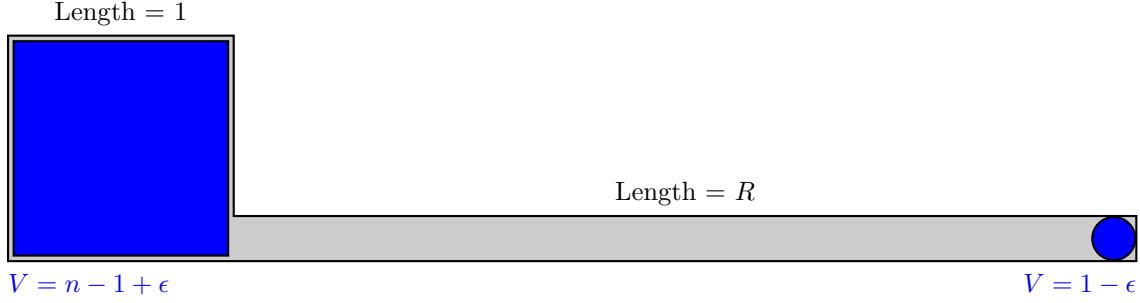

\begin{lemma}
\label{lem:negative-example-non-convex}For every $R\geq1$, there is an $(R+1)$-fat land $C$ for which, for every $m'\leq(n-1)^{1/d}$:
\begin{align*}
\prope(C,\, n,\, m'R\, fat\, objects)\leq
\prop(C,\, n,\, m'R\, fat\, objects)<
{1/n}
\end{align*}
\end{lemma}
\proof{Proof.}
Let $\delta, \epsilon$ be small positive constants.
Let $C$ be a land with the following two components:
\begin{itemize}
\item The \West component is a cube with all sides of length 1;
\item The \East component is a box with one side of length $R$ and the other sides of length $\delta$.
\end{itemize}
See Figure \ref{fig:negative-non-convex} for an illustration for $d=2$. $C$ is contained in a cube of side-length $R+1$ and it contains a cube of side-length 1, so it is $(R+1)$-fat.

$C$ represents a desert with the following water sources: 
\begin{itemize}
\item The \West cube contains $n-1+\epsilon$ water units; 
\item A small disc at the end of the \East box contains $1-\epsilon$ water units. 
\end{itemize}
$C$ has to be divided among $n$ agents whose value measures are equal to the amount of water. To get a proportional division, each agent must receive exactly 1 unit of water. This means that at least one piece, e.g. $X_i$, must overlap both the \East pool and the \West pool. 

The smallest cube containing $X_i$ has a side-length of at least $R$. For the largest cube contained in $X_i$, there are two options:
\begin{itemize}
\item If the largest contained cube is in the \West side, then its side-length must be at most $\bigg(\frac{1}{n-1+\epsilon}\bigg)^{1/d}$, since it must contain at most 1 unit of water.
\item If the largest contained cube is in the \East side, then its side-length must be at most $\delta$.
\end{itemize}
If $\delta$ is sufficiently small (in particular, $\delta<\bigg(\frac{1}{n-1}\bigg)^{1/d}$), then the piece $X_i$ is not $m'R$-fat for every $m'\leq(n-1)^{1/d}$. This means that, if all pieces must be $m'R$-fat, a proportional division is impossible.
\Halmos\endproof

\section{Conclusion and future work.}
\label{sec:future}
Fair division algorithms
hold a great promise for resolving material disputes between people.
But to realize this promise, these algorithms must consider practical requirements such as the geometry of the pieces. 
The present paper contributed to this objective by presenting several algorithms for envy-free division
considering geometric constraints.
For two agents, the algorithms have the best possible partial-proportionality guarantees in various geometric settings. For $n$ agents, the algorithms guarantee a positive partial-proportionality, and it is an open question whether this guarantee can be improved.

The tools developed in this paper are generic and can work for various geometric shapes. In fact, these tools reduce the envy-free division problem to a geometric problem --- the problem of finding appropriate knife functions.
~~
Some topics not covered in the present paper are: 
\begin{itemize}
\item Utility functions that take into account both the value contained in the best usable piece and the total value of the piece, e.g.: $U(Z)=w\cdot V^{S}(Z)+(1-w)\cdot V(Z)$,
where $w$ is some constant.
\item Absolute size constraints on the usable pieces instead of the relative fatness constraints studied here, e.g. let $S$ be the family of all rectangles with length and width of at least 10 meters. 
\item Personal geometric preferences --- letting each agent $i$ specify a different family $S_{i}$ of pieces.
\end{itemize}

\ifdefined\ARXIV
\appendix
\else
\begin{APPENDICES}
\fi

\section{Knife-functions: existence and continuity}\label{sec:S-good}
~\\
This appendix fills in some technical details related to Subsection \ref{sub:knife-functions}.

\subsection{Existence of increasing knife-functions.}
\label{sub:growing-ball}

Given two bounded Borel subsets of $\mathbb{R}^{d}$, $\cstart$ and $\cend$, does there always exist an increasing knife function $K$ from $\cstart$ to $\cend$?

Since $K$ should be increasing, a necessary condition is that $\cstart\subset \cend$. Below we show that this condition is also sufficient.%
\footnote{
Based on an answer by Christopher Fish here: http://math.stackexchange.com/a/1015267/29780
}
We denote by $D(r)$ be the open $d$-dimensional ball of radius $r$ around the origin.
\begin{definition}
\label{def:growing-ball}
Let $\cstart$ and $\cend$ be two bounded Borel subsets of $\mathbb{R}^{d}$, with $\cstart\subsetneq \cend$. 
Let $\rend$ be a radius of a ball that contains $\cend$ (it exists because $\cend$ is bounded).
Let $D^*(t) := D(t\cdot \rend)$.
The \emph{growing ball from $\cstart$ to $\cend$} is the following function from $[0,1]$ to Borel subsets of $\cend$:
\begin{align*}
K(t):=[\cstart\cup D^{*}(t)]\cap \cend
\end{align*}
\end{definition}

\begin{lemma}
\label{lem:growing-ball}
If $\cstart$ and $\cend$ are two bounded Borel subsets of $\mathbb{R}^{d}$ with $\cstart\subsetneq \cend$,
then the growing ball from $\cstart$ to $\cend$
is an increasing knife-function from $\cstart$ to $\cend$.
\end{lemma}
\proof{Proof.}
Clearly, $K(0)=\cstart$, $K(1)=\cend$ and $K$ is increasing. It remains to show that $K$ satisfies the continuity property of a knife-function.
For every two times $t,t'$:
\begin{align*}
K(t)\ominus K(t') = [D^*(t) \ominus D^*(t')]\cap [\cend\setminus \cstart]
\end{align*}
So:
\begin{align*}
\leb{K(t)\ominus K(t')} \leq  \leb{D^*(t) \ominus D^*(t')}
=
|\leb{D^*(t)} - \leb{D^*(t')}|
\end{align*}
Since $D^*(t)$ is a ball whose radius is a uniformly-continuous function of $t$, for every $\epsilon>0$ there exists a $\delta>0$ such that $|t-t'|<\delta$ implies that the right-hand side is smaller than $\epsilon$.
\Halmos\endproof
\begin{remark}
The growing-ball function is not necessarily a ``nice'' knife-function. For example, it may return disconnected pieces. 
It is useful mainly as a proof of existence.
\end{remark}

\subsection{Continuity of value covered by knife}
\label{sub:cont-value}
\begin{restatetheorem}[Lemma \ref{lem:cont-value}]
If $K$ is a knife-function and $V$ is an absolutely-continuous measure, then $V\circ K$ is a uniformly-continuous real functions.
\end{restatetheorem}
\proof{Proof.}
We have to prove that for every $\epsilon'>0$
there exists $\delta>0$ such that $|t-t'|<\delta$ implies $|V(K(t))-V(K(t'))|<\epsilon'$.

Indeed, for every $\epsilon'>0$, 
by the absolute-continuity of $V$, 
there is an $\epsilon>0$ such that
$\leb{Z}<\epsilon$ implies $V(Z)<\epsilon'$.
~~
Given that $\epsilon$, 
by the definition of a knife-function, 
there is a $\delta>0$ such that 
$|t'-t|<\delta$ implies $\leb{K(t')\ominus K(t)} < \epsilon$,
which implies $V(K(t')\ominus K(t))<\epsilon'$.

By the additivity of $V$, for every two Borel sets $A,B$, $V(A)-V(B)\leq V(A\setminus B)$. Therefore:
\begin{align*}
|V(K(t')) - V(K(t))| &= 
\max~\big(~V(K(t')) - V(K(t))~,~ V(K(t)) - V(K(t'))~\big)
\\
&\leq~\max~\big(~V(K(t') \setminus K(t))~,~ V(K(t) \setminus  K(t'))~\big)
\\
&\leq V(K(t') \setminus K(t)) ~+~ V(K(t) \setminus  K(t'))
\\
&= V( (K(t') \setminus K(t)) ~\cup~ (K(t) \setminus  K(t')))
\\
&= V(K(t') \ominus K(t))
\\
&<\epsilon'.
\end{align*}
So $V\circ K$ is a uniformly-continuous real function.
\endproof

\subsection{S-continuity of knife-functions.}
\label{sub:continuity}
In Subsection \ref{sub:knife-functions} we informally defined a knife-function as $S$-continuous if ``all \spieces{} in $K_C(t)$ and $\overline{K_C}(t)$ grow or shrink continuously; no \spiece{} with a positive area is created or destroyed abruptly.''
Below we define this property formally. 

\begin{definition}
A piece $Z$ is called an \emph{$\epsilon$-predecessor} of a piece $Z'$ if  $Z\subseteq Z'$ and $\leb{Z'\setminus Z}<\epsilon$.
\end{definition}

\begin{definition} \label{def:S-cont}
	Let $S$ be a family of pieces. A knife function $K(t)$ is called \emph{$S$-continuous} if for every $\epsilon>0$ there exists $\delta>0$ such that, for all $t$ and $t'$ having $|t'-t|<\delta$:
	
	(a) Every \spiece{} $Z_{t'}\subseteq K(t')$ has an $\epsilon$-predecessor \spiece{} $Z_t\subseteq K(t)$.
	
	(b) Every \spiece{} $Z_{t'}\subseteq \overline{K}(t')$ has an $\epsilon$-predecessor \spiece{} $Z_t\subseteq \overline{K}(t)$.
\end{definition}

We now prove that $S$-continuity implies continuity of utility: 

\begin{restatetheorem}[Lemma \ref{lem:continuous}]
If $K_C$ is an $S$-continuous knife-function
and $V$ is an absolutely-continuous measure,
then both $V^S \circ K_C$ 
and  $V^S\circ \overline{K_C}$ 
are uniformly-continuous real functions.
\end{restatetheorem}
\proof{Proof.}
Given $\epsilon'>0$, we show the
existence of $\delta>0$ such that, for every $t,t'$, if $|t'-t|<\delta$ then $|V^S(K(t'))-V^S(K(t))|<\epsilon'$.

Given $\epsilon'$, by the absolute-continuity of $V$, there is an $\epsilon>0$ such that:
\begin{align}
\label{eq:epsilon'}
\leb{Z}<\epsilon && \implies && V(Z)<\epsilon'
\end{align}

Given that $\epsilon$, by the $S$-continuity of $K$ there is a $\delta>0$ such that, if $|t'-t|<\delta$, then every \spiece{} $Z_{t'}\subseteq K(t')$ has an $\epsilon$-predecessor \spiece{} $Z_t\subseteq K(t)$. This means that $Z_t\subseteq Z_{t'}$ and:
\begin{align*}
\leb{Z_{t'}\setminus Z_t}<\epsilon
\end{align*}
which by (\ref{eq:epsilon'}) implies
\begin{align*}
V(Z_{t'}\setminus Z_t)<\epsilon'
\end{align*}
which by additivity of $V$ implies
\begin{align*}
V(Z_t)>V(Z_{t'})-\epsilon'  
\end{align*}
The latter inequality is true for every \spiece{} $Z_{t'}\subseteq K(t')$, so it is also true for the supremum:
\begin{align*}
V(Z_t) > \sup_{Z_{t'}\in S, Z_{t'}\subseteq K(t')}V(Z_{t'})-\epsilon'  
\end{align*}
By definition, the $S$-value is the supremum, so:
\begin{align*}
V(Z_t)  > V^S(K(t'))-\epsilon'  
\end{align*}
Since $Z_t$ is an \spiece{} in $K(t)$, By definition of $V^S$, 
$V^S(K(t)\geq V(Z_t)$. Therefore:
\begin{align*}
V^S(K(t))  > V^S(K(t'))-\epsilon'  
\end{align*}
By symmetric arguments (replacing the roles of $t$ and $t'$), $V^S(K(t'))>V^S(K(t))-\epsilon'$. Hence $|V^S(K(t'))-V^S(K(t))|<\epsilon'$ as we wanted to prove.

An analogous proof applies to the function $V^S \circ \overline{K}$.
\Halmos\endproof

\subsection{Existence of $S$-continuous knife-functions.}
\label{sub:sweeping-plane}

We show below how $S$-continuous knife-functions can be constructed.
We denote by $H(r)$ the open half-space of $\mathbb{R}^d$ defined by: $x < r$.
\begin{definition}
\label{def:sweeping-plane}
Let $C$ a bounded $d$-dimensional polytope in $\mathbb{R}^d$.
Let $x_0 < x_1$ be real numbers such that $C$ lies entirely to the right of the hyperplane $x=x_0$ and to the left of $x=x_1$ (they exist since $C$ is bounded).
Define $H^*(t) = H(x_0 + t\cdot (x_1-x_0))$.

The \emph{sweeping-plane function on $C$} is the following function from $[0,1]$ to Borel subsets of $C$:
\begin{align*}
K(t) = C\cap H^*(t)
\end{align*}
\end{definition}
\begin{lemma}
\label{lem:sweeping-plane}
Let $C$ be a bounded $d$-dimensional object in $\mathbb{R}^d$.
Let $S$ be the family of $d$-dimensional cubes.
The sweeping-plane function on $C$ is an $S$-continuous increasing knife-function from $\emptyset$ to $C$.
\end{lemma}
\proof{Proof.}
Clearly, $K(0)=\emptyset$, $K(1)=C$ and $K$ is increasing.
The continuity of $K$ can be proved exactly as in Lemma \ref{lem:growing-ball}.
It remains to prove that $K$ is $S$-continuous.
To simplify the proof, we scale and translate $C$ such that it is contained in the unit cube $[0,1]^d$. In this case we can take $x_0=0$ and $x_1=1$ so that $H^*(t) \equiv H(t)$.
	\begin{figure}
		\begin{center}
			\psset{unit=0.7mm,dotsep=1pt,hatchsep=2pt,hatchangle=0}
			\begin{pspicture}(-5,0)(30,60)
			\rput[l](10,55){$t$}
			\psline[linestyle=solid,linecolor=black,linewidth=2pt](10,0)(10,50)
			\rput[l](15,55){$t+\delta$}
			\psline[linestyle=dotted,linecolor=black,linewidth=2pt](15,0)(15,50)
			
			\rput{0}(0,5){	
				\psframe[linecolor=blue,linestyle=dotted](0,0)(14,14)
				\psframe[linecolor=blue,linestyle=solid](0,0)(9,9)
			}
			
			\rput{-45}(-5,35){	
				\psframe[linecolor=blue,linestyle=dotted](0,0)(14,14)
				\psframe[linecolor=blue,linestyle=solid](0,0)(10,10)
			}
			\end{pspicture}
		\end{center}
		\caption[Square-continuity of a knife-function.]{
			\label{fig:sweeping-plane} 
			Square-continuity of the knife-function defined in Lemma \ref{lem:sweeping-plane}.
			\\
			The solid line describes the knife location at time $t$; the dotted line describes its location at time $t+\delta$. 
			\\
			The dotted squares are squares contained in $H(t+\delta)$; the solid squares are their predecessors in $H(t)$.
			\\
			At the bottom, the side-length of the solid square $Z_t$ is smaller than the dotted square $Z_{t+\delta}$ by exactly $\delta$. 
			\\ 
			At the top, the side-length of the solid square $Z_t$ is smaller than the dotted square $Z_{t+\delta}$ by less than $\delta$. 
		}
	\end{figure}
	
The proof of $S$-continuity is based on the following geometric fact: 
for every cube $Z_{t'}$ contained in the half-space $H(t+\delta)$, there exists a cube $Z_t\subseteq Z_{t'}$ contained in the half-space $H(t)$, such that the side-length of $Z_t$ is smaller than that of $Z_{t'}$ by at most $\delta$ (it is smaller by exactly $\delta$ when $Z_{t'}$ is adjacent to the rightmost side of $H(t+\delta)$ and parallel to the axes; see Figure \ref{fig:sweeping-plane} for an illustration of the two-dimensional case).
Suppose $Z_{t'}$ is also contained in $C$. Since $C$ is contained in the unit cube, the side-length of $Z_{t'}$ is at most 1. Therefore, the volume of $Z_t$ is smaller than that of $Z_{t'}$ by at most $1-(1-\delta)^d \leq d\cdot \delta$.

Consider now the definition of $S$-continuity. For every $\epsilon>0$, take $\delta:= \epsilon/d$, let $t'=t+\delta$ and let $Z_{t'}$ be an $S$-piece contained in $K(t')$. By definition of $K$, $Z_{t'}$ is contained in both $C$ and $H(t')$. By the geometric fact, $Z_{t'}$ has an $\epsilon$-predecessor $Z_t$ that is contained in $H(t)$. Since $Z_t\subseteq Z_{t'}$, it is also contained in $C$. Hence, it is contained in $K(t)$.

The $S$-continuity of $\overline{K}$ can be proved analogously.
\Halmos\endproof
Using similar arguments, it is possible to prove that the sweeping-plane function on $C$ is $S$-continuous also when $S$ is the family of $d$-dimensional boxes or fat boxes. 

\subsection{Existence of non-$S$-continuous knife-functions.}
\label{sub:non-s-cont}
We show below how to prove that a knife-function is \emph{not} $S$-continuous.
\begin{lemma}
Let $C := [0,1]\times[0,1]$ and $K_C$ be the following knife-function on $C$ (Figure \ref{fig:knives}(f)):
\begin{align*}
K_C(t) = [0,t]\times[0,1]\cup [1-t,1]\times[0,1]
\end{align*}
Then $K_C$ is not square-continuous.
\end{lemma}
\proof{Proof.}
Intuitively, a square of side-length 1 is created at time $t=0.5$, when the two components of $K_C(t)$ meet. Formally, let $\epsilon=0.75$ and let's prove that there does not exist any $\delta$ satisfying the requirements of square-continuity.

For every $\delta>0$, select $t=0.5-\frac{\delta}{3}$ and $t'=0.5+\frac{\delta}{3}$. Then $K_C(t')$ contains the square $Z'=[0,1]\times[0,1]$, but
all squares $Z\subseteq K_C(t)$ have a side-length of less than $0.5$, hence $\leb{Z'\setminus Z}>0.75=\epsilon$.
\Halmos\endproof

\section{Upper bounds on proportionality.} \label{sec:upper-bounds}
To complement the positive results presented in Section \ref{sec:2-agents}, we present here some negative results --- upper bounds on the attainable proportionality. 

The proof technique is explained in more detail in \cite{SegalHalevi2017Fair}.
Assume that $C$ is a desert with $M$ water-pools.  Consider two agents whose value measure is proportional to the amount of water in their land-plot. Suppose it is possible to give each agent an \spiece{} containing a single water pool, but impossible to give both agents more than one pool since there is room for at most a single \spiece{} touching two pools. Therefore, at least one agent has at most one pool and a utility of at most $1/M$. The arrangements of pools and  corresponding upper bounds are presented below.

\filbreak
\begin{lemma}
\label{lem:negative-square}
$\prop(Square,2,Squares)\leq 1/4$. ~~~ \emph{Proof:}
\end{lemma}
	\psset{unit=0.8mm,dotsep=1pt,hatchsep=2pt,hatchangle=0}
	\begin{center}
		\begin{pspicture}(60,40)
		\psframe[fillstyle=none,linewidth=2pt](0,0)(40,40)
		\pool{(2,2)}
		\pool{(38,2)}
		\pool{(2,38)}
		\pool{(38,38)}
		\end{pspicture}
	\end{center}
The same impossibility holds if we replace ``Square'' with ``$R$-fat rectangle'' for any $R<2$.

\ifdefined\QuarterPlane
\filbreak
\begin{lemma}
\label{lem:negative-quarter-plane}
$\prop(Quarter Plane,2,Squares)\leq 1/3$. ~~~ \emph{Proof:}
\end{lemma}
	\begin{center}
		\begin{pspicture}(60,40)
		\psline[linestyle=solid,linewidth=2pt](40,0)(0,0)(0,40)
		\pool{(2,2)}
		\pool{(38,2)}
		\pool{(2,38)}
		\end{pspicture}
	\end{center}
\fi

\filbreak
\begin{lemma}
\label{lem:negative-Rfat}
For any real $R$,
$\prop(R~fat~rectangle, 2,R\, fat\, rectangles)\leq 1/3$. ~~~ \emph{Proof:}
\end{lemma}
	\begin{center}
		\begin{pspicture}(50,30)
		\newcommand{\landcake}{
			\psframe[fillstyle=none,linewidth=2pt](0,0)(50,30)
		}
		\landcake
		\pool{(2,2)}
		\pool{(48,2)}
		\pool{(2,28)}
		\end{pspicture}
	\end{center}

\filbreak
\begin{lemma}
\label{lem:negative-archipelago}
$\prop(3~rectangles,2,Rectangles)\leq 1/4$. ~~~ \emph{Proof:}
\end{lemma}
	\psset{unit=1mm,dotsep=1pt,hatchsep=2pt,hatchangle=0}
	\begin{center}
		\begin{pspicture}(50,30)
	\newcommand{\landcake}{
		\psframe[linewidth=2pt](10,0)(30,10)
		\psframe[linewidth=2pt](20,10)(40,20)
		\psframe[linewidth=2pt](0,20)(35,30)
	}

		\landcake
		\pool{(28,2)}
		\pool{(12,8)}
		\pool{(38,12)}
		\pool{(2,22)}
		\end{pspicture}
	\end{center}

\filbreak
\section{Convex Lands and Rotating Knives}
\label{sec:convex}
~\\
\newcommand{\half}{{1\over 2}}
Theorem \ref{thm:2agents}(c) 
allows the pieces to be arbitrary fat objects.
Often, land-plots are required to be \emph{convex} in addition to being fat.
To see why the convexity requirement is challenging,
suppose that $C$ is an ellipse
and we want to divide it using a knife-function from $\emptyset$ to $C$, like the knife-functions we used in Section \ref{sec:2-agents}.
If both pieces have to be convex, then they must be separated by a straight line. 
But then, when $t$ approaches 0, 
$K_C(t)$ approaches $\emptyset$, so $K_C(t)$ might be unboundedly slim (see Figure \ref{fig:prop2-fatshapes-convex}/Left).

To handle this challenge we define 
a new property of a knife-function:\footnote{
The rotating-knife technique was introduced by \citet{Robertson1998CakeCutting}[pages 77-78]
as a tool for envy-free division among 3 agents without geometric constraints.
}
\begin{definition} \label{def:rotating-knife}
A knife-function $K_C$ on a land $C$ is called \emph{rotating}
if for every $t\in[0,\half]$, $K_C(t+\half) = \overline{K_C}(t)$ (which is defined as $C\setminus K_C(t)$).
\end{definition}
The definition implies that $K_C(1) = K_C(0)$. At time 0, the knife divides $C$ to two pieces, e.g, ``top'' and ``bottom''; at time $1/2$, the knife has made half a round, so the piece that was previously ``top'' is now ``bottom'' and vice versa; at time $1$ the knife has completed a full round so the ``top'' and ``bottom'' pieces are as in time 0.

The Smooth Knife Algorithm (Algorithm \ref{alg:0-parts-smooth})
can be used with a rotating-knife-function.
We replace the input requirements (a) and (b) of Algorithm \ref{alg:0-parts-smooth} with the following input requirement:

(a') $K_C$ --- an $S$-smooth \emph{rotating}-knife-function on $C$ such that, for some $M\geq 2$:
\begin{align*}
\kcovernum (K_C,S)= M
\end{align*}

We prove the following analogue of Lemma \ref{lem:0-parts}.
\begin{lemma}[Rotating Knife Algorithm]
\label{lem:0-parts-rotated}
Let $K_C$ be rotating-knife function on $C$ with $\kcovernum(K_C,\, S)= M$.
Then, Algorithm \ref{alg:0-parts-smooth} produces an envy-free and $(1/M)$-proportional allocation between the two agents.
\end{lemma}
\proof{Proof.}
We have to prove that, for every absolutely-continuous value-measure $V_1$, it is possible to select a time $t^*\in[0,1]$ such that: $V_1^{K_C}(t^*) = V_1^{\overline{K_C}}(t^*)$.

Suppose that $V_1^{K_C}(0)\leq V_1^{\overline{K_C}}(0)$
By definition of rotating-knife, $K_C(\half)=\overline{K_C}(0)$ and $K_C(0)=\overline{K_C}(\half)$,
so 
$V_1^{K_C}(\half)=V_1^{\overline{K_C}}(0)$
and 
$V_1^{\overline{K_C}}(\half)=V_1^{K_C}(0)$,
so 
$V_1^{K_C}(\half)\geq V_1^{\overline{K_C}}(\half)$.

By Lemma \ref{lem:smooth}, 
both $V^{K_C}$ and 
$V^{\overline{K_C}}$
are continuous real functions.
Therefore, by the intermediate value theorem,
there exists some $t^* \in[0,\half]$ such that 
$V_1^{K_C}(t^*) = V_1^{\overline{K_C}}(t^*)$.

The case  $V_1^{K_C}(0)\geq V_1^{\overline{K_C}}(0)$ is analogous.

From here, the proof is exactly the same as Lemma \ref{lem:0-parts}.
\Halmos\endproof

\begin{figure}
\begin{centering}
\psset{unit=0.7mm}
\newcommand{\landcake}{	\psellipse[fillstyle=solid,linewidth=1pt](16,15)(30,18)
}
\begin{pspicture}(75,40)
\rput(20,5){
	\landcake{}
	\psline[linecolor=black,linestyle=dotted](-13,-2)(-13,32)
}
\end{pspicture}
\begin{pspicture}(75,40)
\rput(20,5){
	\landcake{}
	\psline[linecolor=black,linestyle=dotted](5,-5)(25,35)
	\psframe[linestyle=dashed](1,16)(14,29)
	\psframe[linestyle=dashed](16,1)(29,14)
	\psframe[linestyle=dashed,linecolor=blue](0,0)(30,30)
}
\end{pspicture}
\par
\end{centering}
\caption[Dividing a convex fat land into convex fat pieces.]{
\label{fig:prop2-fatshapes-convex}
Dividing a convex fat land into two convex fat pieces.
\\
\textbf{Left:}
a knife-function from $\emptyset$ to $C$ 
where both sides are convex necessarily makes at least one piece arbitrarily slim.
\\
\textbf{Right:}
A rotating-knife function 
yields two pieces that are both convex and fat.
This is a convex variant of Figure \ref{fig:prop2-fatshapes}.
}
\end{figure}

Now we can prove the convex analogue of Theorem \ref{thm:2agents}(c).
\begin{theorem}
	\noindent For every $R\geq1$, If $C$ is an $R$-fat 2-dimensional convex figure and
	$S$ is the family of convex $2R$-fat pieces, then: 
	\begin{align*}
	\prope(C,2,S)={1 / 2}
	\end{align*}
\end{theorem}
\proof{Proof.}
Scale, rotate and translate $C$ such that the largest square
contained in $C$ is $B^{-}=[-1,1]\times[-1,1]$. By definition of
fatness, $C$ is now contained in a square $B^{+}$ of side-length
at most $2R$. 
For every $t\in[0,1]$,
consider the line passing through the origin at angle $2\pi t$
from the x axis, and let $K_C(t)$ be the part of $C$ at the right-hand side of this line
(see Figure \ref{fig:prop2-fatshapes-convex}/Right). 
This is a rotating-knife function, since for every $t\in[0,\half]$, $K_C(t+\half)$ is the part of $C$ to the left-hand side of the same line. 

This line cuts the contained square $B^{-}$ into two quadrangles, each
of which contains a square with side-length 1. Because $C$ is convex,
this line also cuts the boundary of $C$ at exactly two points, splitting
$C$ into two convex pieces. Each of these two pieces is $2 R$-fat since
it contains a square with side-length 1 and it is contained in $B^{+}$
whose side-length is $2R$.
Therefore, $K_C$ is $S$-smooth with $\kcovernum(K_C,S)=2$.

By Lemma \ref{lem:0-parts-rotated}, Algorithm \ref{alg:0-parts-smooth} yields an envy-free and 1/2-proportional allocation.
\Halmos\endproof
Generalizing the rotating-knife technique to $n$ agents is an interesting topic for future work.

\section{Relative Proportionality}
\label{sec:relprop}
Throughout the paper our goal was to attain a $p$-proportional allocation --- in which the utility of each agent $i$ is at least $p\cdot V_i(C)$. 
In this appendix our goal is to attain a \emph{$p$-relative-proportional allocation} --- in which the utility of each agent $i$ is at least $p\cdot V_i^S(C)$. 

Since $V_i(C)\leq V_i^S(C)$, a $p$-proportional allocation is always $p$-relative-proportional, so all our positive results for $p$-proportionality are valid here too. However, in many cases $p$-proportional allocations might not exist for any positive $p$. As an example, suppose $C$ is a disc and $S$ is the family of squares. The value-measure of an agent might be concentrated in a arbitrarily thin ring around the perimeter of $C$%
\ifdefined\CoverLemma
(see e.g. Figure \ref{fig:los}/Right)
\fi
;
then every square contains only an arbitrarily small fraction of $V_i(C)$.
In such cases, relative-proportionality may be a more appropriate goal.

Analogously to Definition \ref{def:abs-prop}, we define:
\begin{definition}\label{def:rel-prop}
Let $C$ be a land-estate, $S$ a family of usable shapes, and $n\geq 1$ an integer.

(a) The \textbf{relative proportionality guarantee} for $C$, $S$ and $n$, denoted $\proprel(C,n,S)$, is the largest fraction $p\in[0,1]$ such that, for every $n$ value measures $(V_1,...,V_n)$, a $p$-relative-proportional allocation exists (where for each agent $i$, $V_i^S(X_i)\geq p\cdot V_i^S(C)$).

(b) The \textbf{envy-free relative proportionality guarantee} of $C$, $S$ and $n$, denoted $\properel(C,n,S)$, 
is the largest fraction $p\in[0,1]$ such that, for every $n$ value measures $(V_1,...,V_n)$, an \emph{envy-free}
and $p$-relative-proportional allocation exists.
\end{definition}

This appendix provides some initial results regarding relative-proportionality, focusing on the case of $n=2$ agents and rectangular or square pieces.

\subsection{Dividing archipelagos}
As a warm-up, we consider the case that $C$ is an ``archipelago'' --- a union of $m$ pairwise-disjoint ``islands'' of a simple shape. Recall the following proportionality guarantees from Subsections \ref{sub:archipelagos} and \ref{sub:fat-rects}:
\begin{align*}
\prope (m~disjoint~rectangles,2,Rectangles) &= 1/(m+1)
\\
\prope (m~disjoint~R~fat~rectangles,2,R~fat~Rectangles) &= 1/(m+2) \text{~~(for $R\geq 2$)}
\\
\prope (m~disjoint~squares,2,Squares) &= 1/(m+3)
\end{align*}
In contrast, the relative-proportionality guarantees are higher and independent of $m$:
\begin{theorem}
\label{thm:relprop-archipelago}
For every $m\geq 1$:
\begin{align*}
\properel (m~disjoint~rectangles,2,Rectangles) &= 1/2
\\
\properel (m~disjoint~R~fat~rectangles,2,R~fat~Rectangles) &= 1/3 \text{~~~(for $R\geq 2$)}
\\
\properel (m~disjoint~squares,2,Squares) &= 1/4
\end{align*}
\end{theorem}
\proof{Proof.}
The lower bound can be attained using 
Algorithm \ref{alg:relprop-archipelago}.
It asks each agent to specify his most valuable island. If they specify different islands then each agent gets his favorite island and we are done; if they specify the same island then it is divided between them using existing algorithms for dividing \spieces{} to \spieces{}.
By the results of Section \ref{sec:2-agents}, 
this gives each agent a utility of at least $V_i^S(C)/2$
for rectangles, $V_i^S(C)/3$ for $R$-fat rectangles with $R\geq 2$, and $V_i^S(C)/4$ for squares.

The matching upper bound is proved by taking $m=1$ and using the corresponding upper bound for a rectangle or fat-rectangle or square land
(see Appendix \ref{sec:upper-bounds}).
\Halmos
\endproof

\begin{algorithm}[t]
\caption{ּRelative-proportional division of an archipelago between  2 agents}
\label{alg:relprop-archipelago}
INPUT: 

An algorithm for finding an envy-free and $p$-proportional allocation of an \spiece{} between two agents who want \spieces{}.

OUTPUT: 

An envy-free and $p$-relative-proportional allocation of $C$ --- an archipelago of $m$ \spieces{}.

ALGORITHM:

For each agent $i$, let $Z_i$ be the most valuable \spiece{} contained in $C$ --- the \spiece{} for which $V_i^S(C) = V_i(Z_i)$.
Since the $m$ islands comprising $C$ are disjoint, $Z_i$ must be one of these islands. There are two cases.
\begin{itemize}
\item Case A: $Z_1 \neq Z_2$. Give each $Z_i$ to agent $i$. The allocation is envy-free and 1-relative-proportional since the utility of each agent is $1\cdot V_i(Z_i)$.
\item Case B: $Z_1 = Z_2$. Divide this island between the agents using the algorithm of input (b). The resulting allocation is envy-free and $p$-relative-proportional since the utility of each agent is $p\cdot V_i(Z_i) = p\cdot V_i^S(C)$.
\end{itemize}
\end{algorithm}

\subsection{Dividing general lands}
In general lands, in addition to the two cases in Algorithm \ref{alg:relprop-archipelago}, there is a more challenging case in which the agents' best \spieces{}, $Z_1$ and $Z_2$, are different but overlapping. Coping with this challenge requires a different technique that is presented as Algorithm \ref{alg:relprop-general}.

\begin{algorithm}[t]
\caption{ּRelative-proportional division of a general land between  2 agents}
\label{alg:relprop-general}
INPUT: 

For each agent $i$, a set of $m\geq 2$ pairwise-disjoint \spieces{}, each with utility exactly $p\cdot V_i^S(C)$.

PRECONDITION: 

There must be at least one \spiece{} of agent 1 and one \spieces{} of agent 2 that are disjoint.

OUTPUT: 

An envy-free and $p$-relative-proportional allocation of $C$ between the two agents.

ALGORITHM:

Ask each agent to specify a most valuable \spiece{}
of the $2 m $ \spieces{} provided as input.

\begin{itemize}
\item Case A: agent 1 prefers one of the $m$ \spieces{} of agent 2. Give agent 1 his preferred \spiece{} and give agent 2 one of his $m-1$ remaining \spieces{}. Agent 1 does not envy since he got his best \spiece{}. His utility is larger than the utility of each of his own $m$ \spieces{}, which is $p\cdot V_1^S(C)$. Agent 2 does not envy since he is indifferent between all his $m$ \spieces{}. His utility is exactly $p\cdot V_2^S(C)$.
\item Case B: agent 2 prefers one of the $m$ \spieces{} of agent 1. This is analogous to Case A.
\item Case C: each agent prefers his own $m$ \spieces{} to the $m$ \spieces{} of the other agent. By 
the precondition, there is a pair of disjoint \spieces{}, one per agent. Give each agent a disjoint \spiece{}. Both agents feel no envy and their utility is at least $p\cdot V_i^S(C)$.
\end{itemize}
\end{algorithm}

The algorithm requires each agent $i$ to specify a set of pairwise-disjoint \spieces{} with the same value --- a constant fraction $p$ of the value of his best piece $V_i(Z_i)$.
In general, some \spieces{} of agent 1 might overlap some \spieces{} of agent 2. 
A crucial precondition of the algorithm is that \emph{from any two sets specified by the agents, it is possible to select one \spiece{} per agent such that the selected \spieces{} are disjoint}.

In the following subsections we present several cases in which this precondition is satisfied.

\subsection{Dividing general lands to axis-parallel rectangles}
When dividing urban lands, it is natural to require that the pieces be not only rectangular but also parallel to some pre-specified axes. 
With this assumption, a relative-proportional division always exists.

\begin{theorem}
\label{thm:relprop-parallel-rectangles}
For any land $C$:
\begin{align*}
\properel (C,2,axes~parallel~rectangles) = 1/2
\end{align*}
\end{theorem}
\proof{Proof.}
Let $Z_i$ be an axis-parallel rectangle in $C$ that agent $i$ finds the most valuable.%
\footnote{
A compactness assumption is needed to ensure that the supremum defining $V_i^S(C)$ is indeed attained by some square $Z_i\subseteq C$. 
This requires to define a metric space of pieces. Appendix C of \citet{SegalHalevi2017Fair} presents a detailed way to do this.
} 
Ask each agent $i$ to divide $Z_i$ into two axis-parallel rectangles with a value of $V_i(Z_i)/2$, using a horizontal line (See Figure \ref{fig:proprel-parallel-rectangles}). 

Since the division lines are parallel, one of them is above the other: if agent 1's line is above agent 2's line, then the top part of $Z_1$ and the bottom part of $Z_2$ are disjoint; otherwise, the top part of $Z_2$ and the bottom part of $Z_1$ are disjoint.

Therefore, the precondition to Algorithm \ref{alg:relprop-general} is satisfied and we can apply it with $p=1/2$ and get an envy-free and $1/2$-relative-proportional division.
\Halmos\endproof

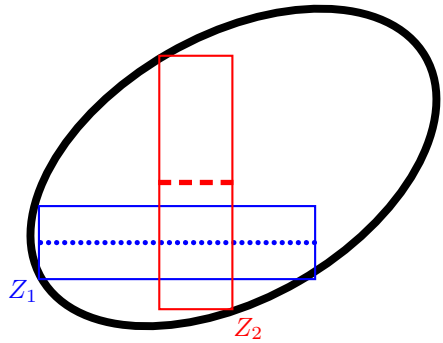
\begin{figure}

\begin{centering}
	\psset{unit=1mm}
	\begin{pspicture}(-10,-10)(80,60)
	\rput(20,5){
		\psellipse[fillstyle=solid,rot=30,linewidth=3pt](16,15)(30,18)

		\psframe[linestyle=solid,linecolor=blue](-10,0)(27,10)
		\rput[tr](-10,0){\textcolor{blue}{$Z_1$}}
		\psline[linestyle=dotted,linecolor=blue,linewidth=2pt](-10,5)(27,5)

		\psframe[linestyle=solid,linecolor=red](6,-4)(16,30)
		\rput[tl](16,-5){\textcolor{red}{$Z_2$}}
		\psline[linestyle=dashed,linecolor=red,linewidth=2pt](6,13)(16,13)

	}
	\end{pspicture}
	\par
\end{centering}

\caption[Dividing general land to axis-parallel rectangles.]{
\label{fig:proprel-parallel-rectangles}
Dividing a general land-estate (for example, an ellipse) where the pieces have to be axis-parallel rectangles.
Each agent $i$ divides his best rectangle $Z_i$ to two parts of equal value by a horizontal line.
Here, agent 1 (blue-dashed) receives the bottom part of $Z_1$ and agent 2 (red-dotted) receives the top part of $Z_2$.
}
\end{figure}

\subsection{Dividing general lands to rotated rectangles}
When the $Z_i$ are not parallel to the axes, 
we must work harder in order to satisfy the preconditions of Algorithm \ref{alg:relprop-general}.
We need a geometric definition and a lemma.

Define a \emph{2x2-partition} of a rectangle as a partition to four sub-rectangles using 
one guillotine cut parallel to its longer side and two cuts parallel to its shorter side, as in Figure \ref{fig:relprop-rectangles}/Left.

\begin{figure}
\begin{center}
\psset{unit=1.5mm,dotsep=1pt,hatchsep=2pt,hatchangle=0}
\begin{pspicture}(30,30)
	\pspolygon[linecolor=black,fillstyle=none,linewidth=3pt](0,5)(5,0)(30,25)(25,30)		
	\psline[linecolor=green](2.5,2.5)(27.5,27.5)
	\psline[linecolor=green](5,10)(7.5,7.5)
	\psline[linecolor=green](20,15)(17.5,17.5)	
\end{pspicture}
~~~
\begin{pspicture}(-20,-20)(30,30)
	\pscircle[fillcolor=black,fillstyle=solid](0,0){.7}  
	
	\psframe[linecolor=red,fillstyle=none](-10,-20)(10,30)
	\psline[linecolor=red,linestyle=dashed,linewidth=2pt](0,-20)(0,30)
	\psline[linecolor=red,linestyle=dashed,linewidth=2pt](-10,-10)(0,-10)	
	\psline[linecolor=red,linestyle=dashed,linewidth=2pt](10,10)(0,10)

	\pspolygon[linecolor=blue,fillstyle=none](0,-20)(-15,-5)(15,25)(30,10)
	\psline[linecolor=blue,linestyle=dotted,linewidth=2pt](-7.5,-12.5)(22.5,17.5)
	\psline[linecolor=blue,linestyle=dotted,linewidth=2pt](-1.5,-6.5)(6,-14)
	\psline[linecolor=blue,linestyle=dotted,linewidth=2pt](1.5,-3.5)(-6,4)

\end{pspicture}

	\end{center}
\caption{
\label{fig:relprop-rectangles}
\textbf{Left:} a 2x2-partition of a rectangle.
\textbf{Right:} 2x2-partitions of two rectangles, red-dashed and blue-dotted. They are rotated and scaled as in the proof of Lemma \ref{lem:2x2-partition}.
The black dot denotes the origin.
}
\end{figure}
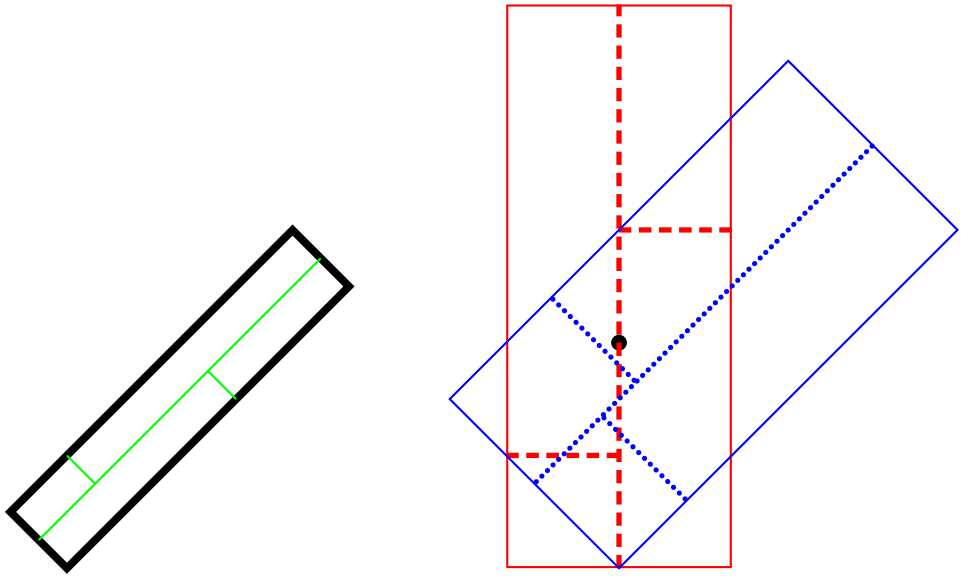

\begin{lemma}
\label{lem:2x2-partition}
For every two rectangles --- red and blue, 
and for every 2x2-partitions of these rectangles, 
there exists a pair of disjoint sub-rectangles --- one red and one blue.
\end{lemma}
\proof{Proof.}
\footnote{
The proof is based on an answer by alcana in
https://math.stackexchange.com/a/2884696/29780.}
Rotate and scale the coordinate system such that the long cut of the red rectangle is at $x=0$, and its two short cuts are at $y=\pm 1$ (as in Figure \ref{fig:relprop-rectangles}/Right).

Suppose by contradiction that all blue sub-rectangles overlap all red sub-rectangles. Consider one such blue sub-rectangle $B$. This $B$ must have a point $(x,y)$ with $x<0,y<-1$ and another point with $x<0,y>-1$, therefore by convexity it must have a point with $x<0,y=-1$.
Similarly it must have a point with $x>0,y=1$.
Therefore by convexity it must have a point with $x=0, y\in (-1,1)$.
In other words, $B$ must overlap a point of the red \emph{central segment} --- the segment $x=0,y\in[-1,1]$ connecting the two red short cuts.
The same is true for all blue sub-rectangles. This means that the red central segment must pass through all blue sub-rectangles.

By analogous arguments, the blue central segment must pass through all red sub-rectangles. So this segment must have a point $(x,y)$ with $x>0,y>1$ and a point with $x<0, y<-1$. Hence, the projection of the blue central segment on the $y$ axis must strictly contain the red central segment $x=0,y\in[-1,1]$. In particular, this projection must be longer than the red central segment.

By analogous arguments, the projection of the red central segment on the blue central segment must be longer than the blue central segment. But the two latter facts are in contradiction.
\Halmos\endproof

Using this lemma, we can prove:

\begin{theorem}
\label{thm:relprop-rotated-rectangles}
For any land $C$:
\begin{align*}
\properel (C,2,rectangles) \geq 1/4
\end{align*}
\end{theorem}
\proof{Proof.}
Ask each agent $i$ to make a 2x2-partition of his best rectangle $Z_i$ such that the value of each sub-rectangle is exactly $V_i(Z_i)/4$.

By Lemma \ref{lem:2x2-partition}, there is a pair of disjoint sub-rectangles --- one per agent. Therefore, we can apply Algorithm \ref{alg:relprop-general} with $p=1/4$.
\Halmos\endproof

To complement this positive result, we show that, in contrast to the axis-parallel case, with rotated rectangles
it may be impossible to guarantee a relative-proportionality of $1/2$.
\begin{theorem}
\label{thm:relprop-rectangles-negative}
There exist lands $C$ for which 
\begin{align*}
\proprel (C,2,rectangles) < 1/2
\end{align*}
\end{theorem}
\proof{Proof.}
Consider the scenario illustrated in Figure \ref{fig:relprop-rectangles-negative}.
The land $C$ (denoted by a thick black line) is a polygon with 16 sides --- a union of two overlapping squares. 
The value-measure of agent 1 is uniform inside the axis-parallel square $Z_1$ (blue) and the value-measure of agent 2 is uniform inside the rotated square $Z_2$ (red).

In order to give an agent a utility of $V_i(Z_i)/2$, we must give a rectangle that touches the center of $C$ (the black dot).
However, each such rectangle of agent 1 overlaps all such rectangles of agent 2 (see Figure \ref{fig:relprop-rectangles-negative}/Middle).
Therefore, in any division, the utility of at least one agent $i$ will be less than 
$V_i(Z_i)/2$.
\Halmos\endproof

\begin{remark}
In the above scenario, the largest utility that we managed to give to both agents simultaneously is $0.46$ (see Figure \ref{fig:relprop-rectangles-negative}/Right).
Therefore we conjecture that $\proprel (C,2,rectangles) \leq 0.46$.
However, proving this formally requires geometric analysis that is beyond the scope of the present appendix.
Our current results thus leave a gap between the lower bound of $1/4$ and the upper bound of less-than-$1/2$.
\Halmos
\end{remark}

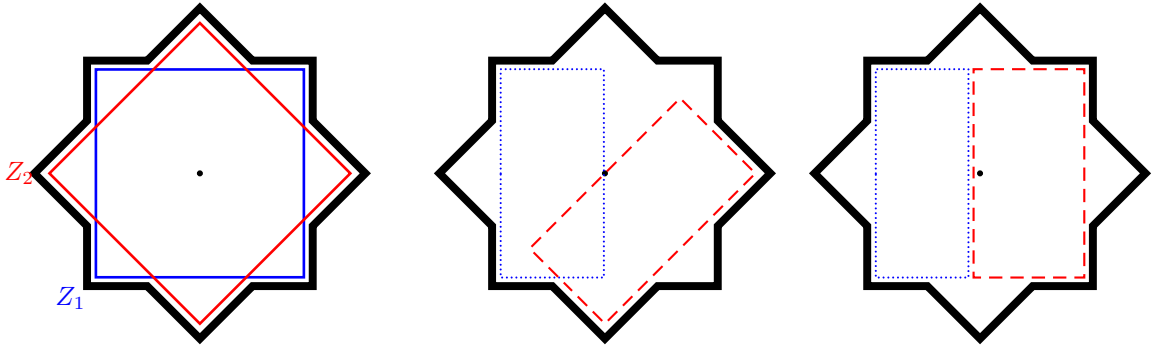
\begin{figure}
\psset{unit=2mm,dotsep=1pt,hatchsep=2pt,hatchangle=0}
\begin{center}
\newcommand{\landcake}{		
	\pspolygon[linestyle=solid,linecolor=black,linewidth=3pt]
	(0,-11)(3.5,-7.5)(7.5,-7.5)(7.5,-3.5)(11,0) (7.5,3.5)(7.5,7.5)(3.5,7.5)
	(0, 11)(-3.5,7.5)(-7.5,7.5)(-7.5,3.5)(-11,0)(-7.5,-3.5)(-7.5,-7.5)(-3.5,-7.5)
	\pscircle[fillcolor=black,fillstyle=solid](0,0){.2}
	
}
\newcommand{\BlueZ}{
		\psframe[linecolor=blue,linestyle=solid,linewidth=1pt,fillstyle=none](-7,-7)(7,7)
		\rput[tr](-7.6,-7.6){\textcolor{blue}{$Z_1$}}
}
\newcommand{\RedZ}{
		\pspolygon[linecolor=red,linestyle=solid,linewidth=1pt,fillstyle=none](0,-10)(10,0)(0,10)(-10,0)
		\rput[mr](-11,0){\textcolor{red}{$Z_2$}}
}
		\begin{pspicture}(-13,-11)(13,11)
		\landcake
		\BlueZ
		\RedZ
		\end{pspicture}
		~~~
		\begin{pspicture}(-11,-11)(11,11)
		\landcake
		\psframe[linestyle=dotted,linecolor=blue,fillstyle=none](-7,-7)(0,7)
		\pspolygon[linestyle=dashed,linecolor=red,fillstyle=none](0,-10)(10,0)(5,5)(-5,-5)
		\end{pspicture}
		~~~
		\begin{pspicture}(-11,-11)(11,11)
		\landcake
		\psframe[linestyle=dotted,linecolor=blue,fillstyle=none](-7,-7)(-0.7,7)
		\psframe[linestyle=dashed,linecolor=red,fillstyle=none](-0.5,-7)(7,7)
		\end{pspicture}
	\end{center}
\caption{
\label{fig:relprop-rectangles-negative}
A land in which each agent $i$ can get less than $1/2$ of the value of his best rectangle $Z_i$.
\textbf{Left:} the land $C$ (thick black) and the two best rectangles $Z_1,Z_2$ (thin  blue/red).
\textbf{Middle:} for each agent $i$, a representative rectangle with a value of $V_i^S(C)/2$. 
The rectangle of agent 1 is dotted-blue and the rectangle of agent 2 is dashed-red.
Each such rectangle must touch the center of $C$ (the black dot). Therefore, each such rectangle of agent 1 overlaps all such rectangles of agent 2.
\textbf{Right:} for each agent $i$, a rectangle with a value of $0.46 V_i^S(C)$. 
Note that agent 2's rectangle is larger, but the area of its intersection with $Z_2$ is the same as agent 1's rectangle intersection with $Z_1$.
}
\end{figure}

\subsection{Dividing general lands to axis-parallel squares}
Suppose the pieces have to be axis-parallel squares.
We need two lemmas in order to guarantee that the preconditions of Algorithm \ref{alg:relprop-general} can be satisfied.

\begin{lemma}
\label{lem:3-red-3-blue}
Given three red pairwise-disjoint axis-parallel squares and three blue pairwise-disjoint axis-parallel squares, 
there always exists a pair of disjoint squares, one red and one blue.
\end{lemma}
\proof{Proof.}
\footnote{
The proof is based on an answer by Abel in
https://math.stackexchange.com/a/412881/29780.}
Since the red squares are pairwise-disjoint and convex, we can rotate the plane such that Red$_1$ is entirely to the left of Red$_2$.

Suppose by contradiction that every red square overlaps all blue squares (and vice-versa). Then:

(a) Since all blue squares overlap both Red$_1$ and Red$_2$, the projections of all blue squares on the x-axis must contain the interval between the rightmost x-coordinate of Red$_1$ and the leftmost x-coordinate of Red$_2$. 

(b) Since the blue squares are pairwise-disjoint, convex and axis-parallel, they must by (a) lie one above the other. 

(c) Since all red squares overlap each blue square, the projection of all red squares on the y axis must properly contain the projection of the middle blue square. 

(d) Since the red squares are pairwise-disjoint, convex and axis-parallel, they must by (c) lie one to the left of the other. 

(e) Since all blue squares overlap each red square, the projection of all blue squares on the x axis must properly contain the projection of the middle red square. 

By (c), the middle red square must be larger than the middle blue square. By (e), the middle blue square is larger. This is a contradiction.
\Halmos\endproof

\begin{lemma}
\label{lem:algorithm-2k}
There is an algorithm that, 
for every integer $k\geq 2$ 
and every value-measure $V$ on an axis-parallel square $Z$, 
returns $k$ pairwise-disjoint axis-parallel squares inside $Z$, such that the value of each square is at least $V(Z)/(2 k)$.
The factor $1/(2 k)$ is the maximum factor that can be guaranteed.
\end{lemma}
\proof{Proof.}
The algorithm is presented in subsection 5.5 of  \citet{SegalHalevi2017Fair}.
\Halmos\endproof

Using these lemmas we can prove:
\begin{theorem}
\label{thm:relprop-squares}
For any land $C$: 
\begin{align*}
\properel (C,2,axes~parallel~squares) \geq 1/6
\end{align*}
\end{theorem}
\proof{Proof of Theorem \ref{thm:relprop-squares}.}
Ask each agent $i$ to 
apply Lemma \ref{lem:algorithm-2k}
to 
his most valuable square $Z_i$, with  $k=3$.
So each agent $i$ finds inside $Z_i$ 
three axis-parallel squares with a value of at least 
$V_i(Z_i)/6$. 
Shrink these squares such that their value is exactly $V_i(Z_i)/6$.

By Lemma \ref{lem:3-red-3-blue}, there are at least two disjoint squares, one per agent. 
Therefore, we can apply Algorithm \ref{alg:relprop-general} with $p=1/6$ and get an envy-free and $1/6$-relative-proportional division.
\Halmos\endproof

The upper bound of $1/4$ from Lemma \ref{lem:negative-square} is of course valid here too, since $C$ may be a square. But if $C$ is even slightly more complicated than a square (e.g, a convex hexagon), the upper bound is tighter.
\begin{theorem}
\label{thm:relprop-squares-negative}
There exist lands $C$ for which 
\begin{align*}
\proprel (C,2,axes~parallel~squares) \leq 1/5
\end{align*}
\end{theorem}

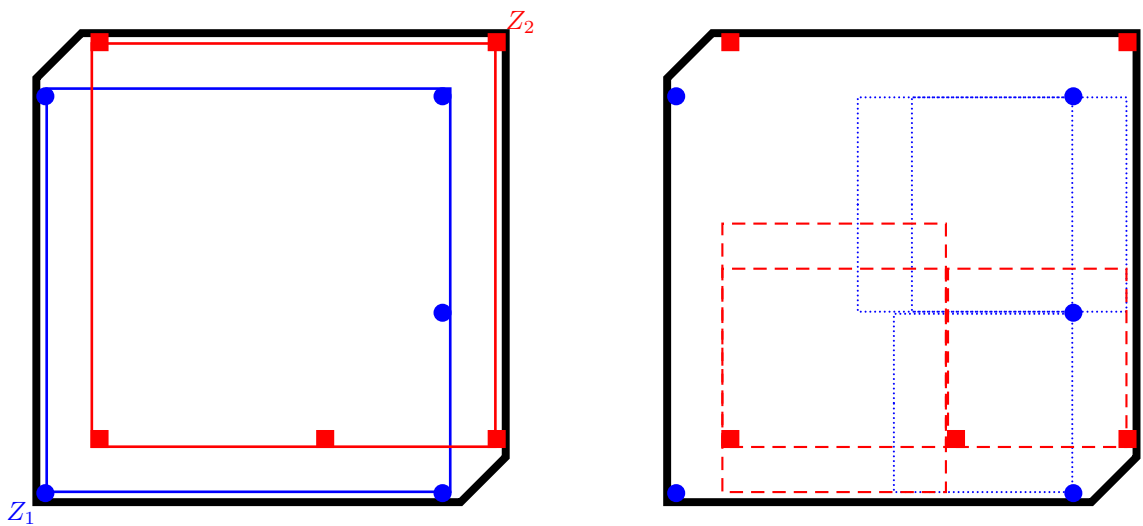
\begin{figure}
\psset{unit=1.2mm,dotsep=1pt,hatchsep=2pt,hatchangle=0}
\begin{center}
\newcommand{\landcake}{		
	\pspolygon[linestyle=solid,linecolor=black,linewidth=3pt](-1,-1)(46,-1)(51,4)(51,51)(4,51)(-1,46)(-1,-1)
}
\newcommand{\BlueZ}{
		\psframe[linecolor=blue,linestyle=solid,linewidth=1pt,fillstyle=none](0,0)(45,45)
		\rput[tr](-1,-1){\textcolor{blue}{$Z_1$}}
}
\newcommand{\BluePools}{
		\pool{(0,0)}
		\pool{(0,44)}
		\pool{(44,0)}
		\pool{(44,44)}
		\pool{(44,20)}
}
\newcommand{\RedZ}{
		\psframe[linecolor=red,linestyle=solid,linewidth=1pt,fillstyle=none](5,5)(50,50)
		\rput[bl](51,51){\textcolor{red}{$Z_2$}}
}
\newcommand{\RedPools}{
		\redpool{(5,5)}
		\redpool{(30,5)}
		\redpool{(49,5)}
		\redpool{(49,49)}
		\redpool{(5,49)}
}

		\begin{pspicture}(-5,-5)(60,60)
		\landcake
		\BlueZ
		\BluePools
		\RedZ
		\RedPools
		\end{pspicture}
		~~~
		\begin{pspicture}(-5,-5)(60,60)
		\landcake
		\BluePools
		\psframe[linestyle=dotted,linecolor=blue,fillstyle=none](24,0)(44,20)
		\psframe[linestyle=dotted,linecolor=blue,fillstyle=none](20,20)(44,44)
		\psframe[linestyle=dotted,linecolor=blue,fillstyle=none](26,20)(50,44)
		\RedPools
		\psframe[linestyle=dashed,linecolor=red,fillstyle=none](30,5)(50,25)
		\psframe[linestyle=dashed,linecolor=red,fillstyle=none](5,5)(30,30)
		\psframe[linestyle=dashed,linecolor=red,fillstyle=none](5,0)(30,25)
		\end{pspicture}
		
	\end{center}
\caption{
\label{fig:relprop-squares-negative}
A land in which each agent $i$ can get at most $1/5$ of the value of his best axis-parallel square $Z_i$.
\textbf{Left:} the land $C$ (thick black), the two best squares $Z_1,Z_2$ (thin  blue/red), and the value-measures (blue discs / red boxes).
\textbf{Right:} for each agent, a representative sample of 3 squares with a value of more than 1. 
Agent 1's squares are dotted-blue and agent 2's squares are dashed-red.
Each square of one agent overlaps all 3 squares of the other agent.
}
\end{figure}

\proof{Proof.}
Consider the scenario illustrated in Figure \ref{fig:relprop-squares-negative}.
The land $C$ is a hexagon (denoted by a thick black line).

The value-measure of agent 1 is 1 in each of the 5 small blue discs and 0 elsewhere. All these discs are contained in the square $Z_1$. Therefore $V_1^S(C) = V_1(Z_1) = 5$.

The value-measure of agent 2 is 1 in each of the 5 small red boxes and 0 elsewhere. All these boxes are contained in the square $Z_2$. Therefore $V_2^S(C) = V_2(Z_2) = 5$.

It is easy to give each agent a utility of 1. 
To give each agent a utility of more than 1, we must give 
agent 1 a square touching two blue discs 
and agent 2 a square touching two red boxes.
By checking all possibilities (see Figure \ref{fig:relprop-squares-negative}/Right), we see that each such square of agent 1 overlaps all such squares of agent 2, so it is impossible to give both agents a utility of more than 1.
\Halmos\endproof

Our current results leave a gap between the lower bound of $1/6$ and the upper bound of $1/5$.

Moreover, we currently do not have results for general (rotated) squares and for $R$-fat rectangles. We believe such results can be obtained by proving variants of the geometric lemmas 
\ref{lem:2x2-partition} and
\ref{lem:3-red-3-blue}.

\ifdefined\ARXIV
\else
\end{APPENDICES}
\fi

\section*{Acknowledgments.}
This research was funded in part by the Doctoral Fellowships of Excellence Program at Bar-Ilan University, the Mordecai and Monique Katz Fellowship, the ISF grants 1083/13 and 1241/12 and the BSF grant 2012344.
	
We are grateful to the anonymous referees of the AAAI 2015 conference, the participants of that conference, the Dagstuhl seminar on fair division, and Nerly Biton for their helpful comments on a preliminary version of this manuscript.
We are grateful to the editor and anonymous referees of Mathematics of Operations Research for their helpful comments in this manuscript.
	
Many members of the math.stackexchange.com, mathoverflow.com and cs.stackexchange.com communities kindly helped in coping with specific issues arising in this paper. In particular, we are grateful to: Abel, alcana (Jack Young), Julián Aguirre, Christian Blatter, Ilya Bogdanov, Beni Bogosel, Boris Bukh, Michael Burr, Anthony Carapetis, Scott Carnahan, Christopher Culter, Gabriel C. Drummond-Cole, David Eppstein, Yuval Filmus, Christopher Fish, Peter Franek, Nick Gill, John Gowers, Michael Greinecker, Dafin Guzman, W\l odzimierz Holszty\'{n}ski, Marcus Hum, Robert Israel, Barry Johnson, Joonas Ilmavirta, V. Kurchatkin, Jonathan Lubin, Raymond Manzoni, Ross Millikan, Lee Mosher, Rahul Narain, Sebastian Negraszus, Mariusz Nowak, Boris Novikov, Joseph O'Rourke, Emanuele Paolini, Raphael Reitzig, David Richerby, András Salamon, Realz Slaw, B. Stoney, Eric Stucky, Steven Taschuk, Marc van Leeuwen, Martin van der Linden, Hagen von Eitzen, Martin von Gagern, Jared Warner, Frank W., Ittay Weiss, Willie Wong, Phoemue X and Amitai Yuval.


\ifdefined\ARXIV
\bibliographystyle{apalike} 
\else
\bibliographystyle{informs2014} 
\fi
\bibliography{../erelsegal-halevi}

\begin{thebibliography}{}

\bibitem[Agarwal et~al., 1995]{Agarwal1995Computing}
Agarwal, P.~K., Katz, M.~J., and Sharir, M. (1995).
\newblock {Computing depth orders for fat objects and related problems}.
\newblock {\em Computational Geometry}, 5(4):187--206.

\bibitem[Aziz and Mackenzie, 2016]{Aziz2016Discrete}
Aziz, H. and Mackenzie, S. (2016).
\newblock {A Discrete and Bounded Envy-Free Cake Cutting Protocol for Any
  Number of Agents}.
\newblock In {\em FOCS 2016}, pages 416--427.

\bibitem[Azrieli and Shmaya, 2014]{Azrieli2014Rental}
Azrieli, Y. and Shmaya, E. (2014).
\newblock {Rental harmony with roommates}.
\newblock {\em Journal of Economic Theory}, 153:128--137.

\bibitem[Barbanel and Brams, 2004]{Barbanel2004Cake}
Barbanel, J.~B. and Brams, S.~J. (2004).
\newblock {Cake division with minimal cuts: envy-free procedures for three
  persons, four persons, and beyond}.
\newblock {\em Mathematical Social Sciences}, 48(3):251--269.

\bibitem[Barbanel et~al., 2009]{Barbanel2009Cutting}
Barbanel, J.~B., Brams, S.~J., and Stromquist, W. (2009).
\newblock {Cutting a pie is not a piece of cake}.
\newblock {\em American Mathematical Monthly}, 116(6):496--514.

\bibitem[Beck, 1987]{Beck1987Constructing}
Beck, A. (1987).
\newblock {Constructing a Fair Border}.
\newblock {\em The American Mathematical Monthly}, 94(2):157--162.

\bibitem[Bei et~al., 2017]{bei2017cake}
Bei, X., Chen, N., Huzhang, G., Tao, B., and Wu, J. (2017).
\newblock Cake cutting: envy and truth.
\newblock In {\em International Joint Conference on Artificial Intelligence
  (IJCAI)}.

\bibitem[Bei et~al., 2018]{bei2018truthful}
Bei, X., Huzhang, G., and Suksompong, W. (2018).
\newblock {Truthful Fair Division without Free Disposal}.
\newblock In {\em Proceedings of IJCAI-18}.

\bibitem[Berliant and Dunz, 2004]{Berliant2004Foundation}
Berliant, M. and Dunz, K. (2004).
\newblock {A foundation of location theory: existence of equilibrium, the
  welfare theorems, and core}.
\newblock {\em Journal of Mathematical Economics}, 40(5):593--618.

\bibitem[Berliant and Raa, 1988]{Berliant1988Foundation}
Berliant, M. and Raa, T. (1988).
\newblock {A foundation of location theory: Consumer preferences and demand}.
\newblock {\em Journal of Economic Theory}, 44(2):336--353.

\bibitem[Berliant et~al., 1992]{Berliant1992Fair}
Berliant, M., Thomson, W., and Dunz, K. (1992).
\newblock {On the fair division of a heterogeneous commodity}.
\newblock {\em Journal of Mathematical Economics}, 21(3):201--216.

\bibitem[Brams et~al., 2008]{Brams2008Proportional}
Brams, S.~J., Jones, M., and Klamler, C. (2008).
\newblock {Proportional pie-cutting}.
\newblock {\em International Journal of Game Theory}, 36(3-4):353--367.

\bibitem[Brams and Taylor, 1995]{Brams1995EnvyFree}
Brams, S.~J. and Taylor, A.~D. (1995).
\newblock {An Envy-Free Cake Division Protocol}.
\newblock {\em The American Mathematical Monthly}, 102(1):9--18.

\bibitem[Brams and Taylor, 1996]{Brams1996Fair}
Brams, S.~J. and Taylor, A.~D. (1996).
\newblock {\em {Fair Division: From Cake Cutting to Dispute Resolution}}.
\newblock Cambridge University Press, Cambridge UK.

\bibitem[Brams et~al., 1997]{Brams1997MovingKnife}
Brams, S.~J., Taylor, A.~D., and Zwicker, W.~S. (1997).
\newblock {A Moving-Knife Solution to the Four-Person Envy-Free Cake Division
  Problem}.
\newblock {\em Proceedings of the American Mathematical Society},
  125(2):547--554.

\bibitem[Br\^{a}nzei and Miltersen, 2015]{Branzei2015Dictatorship}
Br\^{a}nzei, S. and Miltersen, P.~B. (2015).
\newblock {A Dictatorship Theorem for Cake Cutting}.
\newblock In {\em Proceedings of the 24th International Conference on
  Artificial Intelligence}, IJCAI'15, pages 482--488. AAAI Press.

\bibitem[Caragiannis et~al., 2011]{Caragiannis2011Towards}
Caragiannis, I., Lai, J.~K., and Procaccia, A.~D. (2011).
\newblock {Towards more expressive cake cutting}.
\newblock In {\em Proceedings of the Twenty-Second international joint
  conference on Artificial Intelligence (IJCAI'11)}, IJCAI'11, pages 127--132.
  AAAI Press.

\bibitem[Chambers, 2005]{Chambers2005Allocation}
Chambers, C.~P. (2005).
\newblock {Allocation rules for land division}.
\newblock {\em Journal of Economic Theory}, 121(2):236--258.

\bibitem[Chen et~al., 2013]{Chen2013Truth}
Chen, Y., Lai, J.~K., Parkes, D.~C., and Procaccia, A.~D. (2013).
\newblock {Truth, justice, and cake cutting}.
\newblock {\em Games and Economic Behavior}, 77(1):284--297.

\bibitem[Cohler et~al., 2011]{Cohler2011Optimal}
Cohler, Y.~J., Lai, J.~K., Parkes, D.~C., and Procaccia, A.~D. (2011).
\newblock {Optimal Envy-Free Cake Cutting}.
\newblock In {\em Proceedings of the 25th AAAI Conference on Artificial
  Intelligence (AAAI-11)}, pages 626--631.

\bibitem[Dall'Aglio and Maccheroni, 2009]{DallAglio2009Disputed}
Dall'Aglio, M. and Maccheroni, F. (2009).
\newblock {Disputed lands}.
\newblock {\em Games and Economic Behavior}, 66(1):57--77.

\bibitem[Deng et~al., 2012]{Deng2012Algorithmic}
Deng, X., Qi, Q., and Saberi, A. (2012).
\newblock {Algorithmic Solutions for Envy-Free Cake Cutting}.
\newblock {\em Operations Research}, 60(6):1461--1476.

\bibitem[Dubins and Spanier, 1961]{Dubins1961How}
Dubins, L.~E. and Spanier, E.~H. (1961).
\newblock {How to Cut A Cake Fairly}.
\newblock {\em The American Mathematical Monthly}, 68(1):1--17.

\bibitem[Hill, 1983]{Hill1983Determining}
Hill, T.~P. (1983).
\newblock {Determining a Fair Border}.
\newblock {\em The American Mathematical Monthly}, 90(7):438--442.

\bibitem[H\"{u}sseinov, 2011]{Husseinov2011Theory}
H\"{u}sseinov, F. (2011).
\newblock {A theory of a heterogeneous divisible commodity exchange economy}.
\newblock {\em Journal of Mathematical Economics}, 47(1):54--59.

\bibitem[H\"{u}sseinov and Sagara, 2013]{Husseinov2013Existence}
H\"{u}sseinov, F. and Sagara, N. (2013).
\newblock {Existence of efficient envy-free allocations of a heterogeneous
  divisible commodity with nonadditive utilities}.
\newblock {\em Social Choice and Welfare}, pages 1--18.

\bibitem[Ichiishi and Idzik, 1999]{Ichiishi1999Equitable}
Ichiishi, T. and Idzik, A. (1999).
\newblock {Equitable allocation of divisible goods}.
\newblock {\em Journal of Mathematical Economics}, 32(4):389--400.

\bibitem[Iyer and Huhns, 2009]{Iyer2009Procedure}
Iyer, K. and Huhns, M.~N. (2009).
\newblock {A Procedure for the Allocation of Two-Dimensional Resources in a
  Multiagent System}.
\newblock {\em International Journal of Cooperative Information Systems},
  18:1--34.

\bibitem[Katz, 1997]{Katz19973D}
Katz, M.~J. (1997).
\newblock {3-D vertical ray shooting and 2-D point enclosure, range searching,
  and arc shooting amidst convex fat objects}.
\newblock {\em Computational Geometry}, 8(6):299--316.

\bibitem[Kurokawa et~al., 2013]{Kurokawa2013How}
Kurokawa, D., Lai, J.~K., and Procaccia, A.~D. (2013).
\newblock {How to Cut a Cake Before the Party Ends}.
\newblock In {\em Proceedings of the Twenty-Seventh AAAI Conference on
  Artificial Intelligence}, pages 555--561.

\bibitem[Legut et~al., 1994]{Legut1994Economies}
Legut, J., Potters, J. A.~M., and Tijs, S.~H. (1994).
\newblock {Economies with Land - A Game Theoretical Approach}.
\newblock {\em Games and Economic Behavior}, 6(3):416--430.

\bibitem[LiCalzi and Nicol\`{o}, 2009]{LiCalzi2009Efficient}
LiCalzi, M. and Nicol\`{o}, A. (2009).
\newblock {Efficient egalitarian equivalent allocations over a single good}.
\newblock {\em Economic Theory}, 40(1):27--45.

\bibitem[Maccheroni and Marinacci, 2003]{Maccheroni2003How}
Maccheroni, F. and Marinacci, M. (2003).
\newblock {How to cut a pizza fairly: Fair division with decreasing marginal
  evaluations}.
\newblock {\em Social Choice and Welfare}, 20(3):457--465.

\bibitem[Manabe and Okamoto, 2010]{Manabe2010MetaEnvyFree}
Manabe, Y. and Okamoto, T. (2010).
\newblock {Meta-Envy-Free Cake-Cutting Protocols}.
\newblock In Hlin\v{e}n\'{y}, P. and Ku\v{c}era, A., editors, {\em Mathematical
  Foundations of Computer Science 2010}, volume 6281 of {\em Lecture Notes in
  Computer Science}, chapter~44, pages 501--512. Springer Berlin Heidelberg,
  Berlin, Heidelberg.

\bibitem[Mirchandani, 2013]{Mirchandani2013Superadditivity}
Mirchandani, R.~S. (2013).
\newblock {Superadditivity and Subadditivity in Fair Division}.
\newblock {\em Journal of Mathematics Research}, 5(3):78--91.

\bibitem[Nicol\`{o} et~al., 2012]{Nicolo2012Equal}
Nicol\`{o}, A., Perea, and Roberti, P. (2012).
\newblock {Equal opportunity equivalence in land division}.
\newblock {\em SERIEs - Journal of the Spanish Economic Association},
  3(1-2):133--142.

\bibitem[Nicol\`{o} and Yu, 2008]{Nicolo2008Strategic}
Nicol\`{o}, A. and Yu, Y. (2008).
\newblock {Strategic divide and choose}.
\newblock {\em Games and Economic Behavior}, 64(1):268--289.

\bibitem[Procaccia, 2015]{Procaccia2015Cake}
Procaccia, A.~D. (2015).
\newblock {Cake Cutting Algorithms}.
\newblock In Brandt, F., Conitzer, V., Endriss, U., Lang, J., and Procaccia,
  A.~D., editors, {\em Handbook of Computational Social Choice}, chapter~13,
  pages 261--283. Cambridge University Press.

\bibitem[Reijnierse and Potters, 1998]{Reijnierse1998Finding}
Reijnierse, J.~H. and Potters, J. A.~M. (1998).
\newblock {On finding an envy-free Pareto-optimal division}.
\newblock {\em Mathematical Programming}, 83(1-3):291--311.

\bibitem[Robertson and Webb, 1998]{Robertson1998CakeCutting}
Robertson, J.~M. and Webb, W.~A. (1998).
\newblock {\em {Cake-Cutting Algorithms: Be Fair if You Can}}.
\newblock A K Peters/CRC Press, first edition.

\bibitem[Saberi and Wang, 2009]{Saberi2009Cutting}
Saberi, A. and Wang, Y. (2009).
\newblock {Cutting a Cake for Five People}.
\newblock In Goldberg, A.~V. and Zhou, Y., editors, {\em Algorithmic Aspects in
  Information and Management}, volume 5564 of {\em Lecture Notes in Computer
  Science}, pages 292--300. Springer Berlin Heidelberg.

\bibitem[Sagara and Vlach, 2005]{Sagara2005Equity}
Sagara, N. and Vlach, M. (2005).
\newblock {Equity and Efficiency in a Measure Space with Nonadditive
  Preferences: The Problem of Cake Division}.
\newblock {\em Journal of Economic Literature}, 90:1.

\bibitem[Segal-Halevi, 2018]{segalhalevi2018redividing}
Segal-Halevi, E. (2018).
\newblock {Redividing the Cake}.
\newblock In {\em Proceedings of IJCAI'18}, pages 498--504.
\newblock arXiv preprint 1603.00286.

\bibitem[Segal-Halevi et~al., 2015]{SegalHalevi2015EnvyFree}
Segal-Halevi, E., Hassidim, A., and Aumann, Y. (2015).
\newblock {Envy-Free Cake-Cutting in Two Dimensions}.
\newblock In {\em Proceedings of the 29th AAAI Conference on Artificial
  Intelligence (AAAI-15)}, pages 1021--1028.
\newblock arXiv preprint 1609.03938.

\bibitem[Segal-Halevi et~al., 2016]{SegalHalevi2016Waste}
Segal-Halevi, E., Hassidim, A., and Aumann, Y. (2016).
\newblock {Waste Makes Haste: Bounded Time Algorithms for Envy-Free Cake
  Cutting with Free Disposal}.
\newblock {\em ACM Transactions on Algorithms (TALG)}, 13(1):12:1--12:32.

\bibitem[Segal-Halevi et~al., 2017]{SegalHalevi2017Fair}
Segal-Halevi, E., Nitzan, S., Hassidim, A., and Aumann, Y. (2017).
\newblock {Fair and square: Cake-cutting in two dimensions}.
\newblock {\em Journal of Mathematical Economics}, 70:1--28.

\bibitem[Segal-Halevi and Sziklai, 2018]{segalhalevi2018monotonicity}
Segal-Halevi, E. and Sziklai, B.~R. (2018).
\newblock {Monotonicity and Competitive Equilibrium in Cake-cutting}.
\newblock {\em Economic Theory}, pages 1--39.
\newblock arXiv preprint 1510.05229.

\bibitem[Steinhaus, 1948]{Steinhaus1948Problem}
Steinhaus, H. (1948).
\newblock {The problem of fair division}.
\newblock {\em Econometrica}, 16(1):101--104.

\bibitem[Stromquist, 1980]{Stromquist1980How}
Stromquist, W. (1980).
\newblock {How to Cut a Cake Fairly}.
\newblock {\em The American Mathematical Monthly}, 87(8):640--644.

\bibitem[Stromquist, 2008]{Stromquist2008Envyfree}
Stromquist, W. (2008).
\newblock {Envy-free cake divisions cannot be found by finite protocols}.
\newblock {\em Electronic Journal of Combinatorics}, 15(1):\#R11.
\newblock Research paper 11, 10 pp., 91B32.

\bibitem[Su, 1999]{Su1999Rental}
Su, F.~E. (1999).
\newblock {Rental Harmony: Sperner's Lemma in Fair Division}.
\newblock {\em The American Mathematical Monthly}, 106(10):930--942.

\bibitem[Thomson, 2007]{Thomson2007Children}
Thomson, W. (2007).
\newblock {Children Crying at Birthday Parties. Why?}
\newblock {\em Economic Theory}, 31(3):501--521.

\bibitem[Webb, 1990]{Webb1990Combinatorial}
Webb, W.~A. (1990).
\newblock {A Combinatorial Algorithm to Establish a Fair Border}.
\newblock {\em European Journal of Combinatorics}, 11(3):301--304.

\bibitem[Weller, 1985]{Weller1985Fair}
Weller, D. (1985).
\newblock {Fair division of a measurable space}.
\newblock {\em Journal of Mathematical Economics}, 14(1):5--17.

\end{thebibliography}

\end{document}